\documentclass[article]{aa}
\usepackage{txfonts}
\usepackage{graphicx}
\usepackage{natbib}

\bibpunct{(}{)}{;}{a}{}{,}

\usepackage{color}
\newcommand{\kms}{\,{\rm km\,s}^{-1}}

\newcommand{\Lsun}{L_\odot}
\newcommand{\Msun}{M_\odot}

\newcommand{\as}{\ifmmode {^{\scriptscriptstyle\prime\prime}}
        \else $^{\scriptscriptstyle\prime\prime}$\fi}

\begin{document}
\title{Chemistry in Disks X: \\ The Molecular Content of Proto-planetary Disks in Taurus
\thanks{Based on observations carried out with the IRAM 30-m radiotelescope
and Plateau de Bure interferometers.
 IRAM is supported by INSU/CNRS (France), MPG (Germany) and IGN (Spain).}
}

\author{S.Guilloteau\inst{1,2}, L.Reboussin\inst{1,2}, A.Dutrey\inst{1,2},
E.Chapillon\inst{1,2,3}, V.Wakelam\inst{1,2}, V.Pi\'etu\inst{3}, E.Di Folco\inst{1,2},
D.Semenov\inst{4}, and Th.Henning\inst{4}
} 
\institute{
Univ. Bordeaux, LAB, UMR 5804, F-33270, Floirac, France
\and
CNRS, LAB, UMR 5804, F-33270 Floirac, France\\
  \email{[guilloteau,reboussin,dutrey,wakelam,difolco]@obs.u-bordeaux1.fr}
\and
IRAM, 300 rue de la piscine, F-38406
Saint Martin d'H\`eres, France
\and{}
Max Planck Institute f\"ur Astronomie, K\"onigstuhl 17, D-69117 Heidelberg, Germany
}

\offprints{S.Guilloteau, \email{Stephane.Guilloteau@obs.u-bordeaux1.fr}}

\date{Received / Accepted } %
\authorrunning{Guilloteau et al.} %
\titlerunning{A survey of molecules in proto-planetary disks.}

\abstract
{}
{We attempt to determine the molecular composition of disks around young low-mass stars.}
{We used the IRAM 30-m radiotelescope to perform a sensitive wideband survey of 30 stars
in the Taurus Auriga region known to be surrounded by gaseous circumstellar disks.
We simultaneously observed HCO$^+$(3-2), HCN(3-2), C$_2$H(3-2), CS(5-4), and two transitions
of SO. We combine the results with a previous survey which observed $^{13}$CO (2-1), CN(2-1),
two o-H$_2$CO lines and another transition of SO. We use available interferometric data
to derive excitation temperatures of CN and C$_2$H in several sources.
We determine characteristic sizes of the gas disks and column
densities of all molecules using a parametric power-law disk model.
Our study is mostly sensitive to molecules at 200-400 au from
the stars. We compare the derived column densities to the predictions
of an extensive gas-grain chemical disk model, under conditions
representative of T Tauri disks.}
{This survey provides 20 new detections of HCO$^+$ in disks, 18 in HCN, 11 in C$_2$H, 8 in CS and
4 in SO. HCO$^+$ is detected in almost all sources, and its J=3-2 line is essentially optically thick, providing
good estimates of the disk radii. The other transitions are (at least partially) optically thin.
Large variations of the  column density ratios are observed, but do not correlate with any specific property
of the star or disk. Disks around Herbig Ae stars appear less rich in molecules than those
around T Tauri stars, although the sample remains small. SO is only found in the (presumably younger)
embedded objects, perhaps reflecting an evolution of the S chemistry due to increasing
depletion with time. Overall, the
molecular column densities, and in particular the CN/HCN and CN/C$_2$H ratios, are well reproduced
by gas-grain chemistry in cold disks.} 
{This study provides a comprehensive census of simple molecules in disks of radii $> 200-300$ au.
Extending that to smaller disks, or searching for less abundant or more complex molecules
requires a much more sensitive facility, i.e. NOEMA and ALMA.}


\maketitle{}

\section{Introduction}
\label{sec:intro}

One of the fundamental problems of modern astrophysics is to comprehend the
formation and evolution of planets and to discern their physical and chemical
structures \citep[e.g.][]{Burke+etal_2014,Marcy+etal_2014,Raymond+etal_2014} 
The initial conditions in protoplanetary disks play a crucial role in determining the
properties of the emerging planetary systems
\citep{Oberg+etal_2011,Moriarty+etal_2014,Mordasini+etal_2015,Thiabaud+etal_2015}.
While the immensely daunting goal of detecting and characterizing physical
structures and molecular contents of exoplanetary atmospheres is no longer a mere dream
\citep[e.g][]{Fraine+etal_2014,Brogi+etal_2014,Barman+etal_2015},
the relevant statistical studies will only become  doable with the next
generation of observational facilities such as the James Webb Space Telescope
and the European Extremely Large Telescope. In contrast, the situation with
accessing the physical and chemical structure of protoplanetary disks
{\em en masse} is  more favorable thanks to the emerging sensitive instruments,
like the Atacama Large Millimeter/submillimeter Array (ALMA) and
NOrthern Extended Millimeter Array (NOEMA).

Pivotal information about temperature, density and mineralogical
content of protoplanetary disks has been obtained via (mainly unresolved)
dust continuum observations \citep{Williams+Cieza_2011,Henning+Meeus_2011}.
In addition, a rapidly growing number of results comes  from the analysis
of spectrally and spatially resolved
line data \citep{Henning+Semenov_2013,Dutrey+etal_2014,Pontoppidan+etal_2014},
where molecules serve first and foremost as distinct probes of disk physics.
Since the initial discovery of $\sim 10$ molecules in protoplanetary disks
almost 20 years ago \citep{Dutrey+etal_1997,Kastner+etal_1997}, the chemical
composition of disks still remains largely a mystery.

This situation is a combination of several complicated factors.
First, protoplanetary disks are relatively small objects, with radii
$\sim 100-1\,000$~au,  hence sub-arcsec resolution is required to study their
spatial distributions. Second, disks are comprised of a small amount of
leftover material remained after the build-up of the central star(s),
$\sim 1-10 \Msun$. Given the scarcity  of detectable diagnostic species
wrt the main gas component, H$_2$, high sensitivity and long integration
times are required to  detect their weak line emission.
Third, the line emission in protoplanetary disks is observed on
top of a strong emission from dust particles, requiring high spectral
resolution and the right molecular line set-up. Therefore, previous
observational  studies have focussed on a few exceptionally bright/large
nearby disks with nearly face-on orientation that may not be representative
of their class. One notable  exception is the T~Tauri system of TW Hya,
which is the most nearby, compact disk located at
$\sim 55$~pc 
\citep[e.g.][]{Thi+etal_2004}.
Another obstacle is the potential contamination of the line emission
stemming from the disk by the surrounding molecular cloud
\citep[e.g.][]{Guilloteau+etal_2013,Guilloteau+etal_2014}. This led observers to
select preferentially isolated objects, which on average should be older
than most of the disks.

As a result of the small sizes and demanding integration times,
spatially resolved line observations of protoplanetary disks are still
relatively rare, and the radial distributions of molecules are poorly constrained
\citep[see ][]{Pietu+etal_2007,Chapillon+etal_2012,Qi+etal_2013c,Teague+etal_2015}.
Also, until a few years ago, large telescopes or interferometers had limited bandwidth,
so a full spectral survey required substantial observing time, often of order 100
of hours or more, as only a few lines could be observed simultaneously.
Nonetheless, limited spatially-resolved molecular surveys targeting a handful of
sources in lines of several simple species were performed by the
``Chemistry in Disks''
\citep[CID,][]{Dutrey+etal_2007,Schreyer+etal_2008,Henning+etal_2010,Dutrey+etal_2011,Chapillon+etal_2012,
Teague+etal_2015} and ``Disk Imaging Survey of Chemistry with SMA'' \citep[DISCS,][]{Oberg+etal_2010,Oberg+etal_2011}
consortia.

Recent increase in the telescope sensitivity thanks to the advent
of ALMA and upgrade of receivers of
the IRAM facilities resulted in  detections of new, more complex molecules
in disks, such as HC$_3$N \citep{Chapillon+etal_2012b}, cyclic C$_3$H$_2$ \citep{Qi+etal_2013b},
 and CH$_3$CN \citep{Oberg+etal_2015}.
Also, more extended surveys (either in number of targeted disks or molecules)
with single-dish radio-antennas became doable.
However, the complete, unbiased frequency coverage between 208 and 270 GHz performed
at the IRAM 30-m telescope for the LkCa~15 transition disks by \citet{Punzi+etal_2015}, did not reveal
any new molecule in this source.
\citet{Guilloteau+etal_2013} observed $^{13}$CO, CN, H$_{2}$CO, and SO in
42 disks around T~Tauri and Herbig~Ae stars of the Taurus-Auriga association
with the IRAM 30-m telescope. They found that the  $^{13}$CO
lines are often dominated by residual emission from the cloud, and that
the CN lines are better tracers of the whole gas disk. The sources with strong
SO emission are usually also strong in ortho-H$_2$CO, and this emission
most likely comes from remaining envelopes or outflows. The overall gas disk detection rate
from the all molecular tracers was $\sim 68\%$. A follow-up study for the
29 T~Tauri disks located in $\rho$~Ophiuchi and upper Scorpius was
performed by \cite{Reboussin+etal_2015}. It confirmed the widespread
contamination by molecular clouds in $^{13}$CO, and also
showed substantial contamination in the ortho-H$_2$CO emission. However,
the disk detection rate was much lower, perhaps indicating a smaller
average disk size in the $\rho$~Oph region than in the Taurus
cloud, or a different chemistry resulting from the younger ages of
$\rho$ Oph.

Still, little is known about the chemical evolution as a function
of disk age and stellar properties in a statistically significant sample
of sources.  In this study, we take advantage of the low noise, sideband separating,
dual-polarization receivers of the IRAM 30-m telescope, coupled to wide band, high spectral resolution
backends providing about 200 kHz over an instantaneous frequency coverage of
$\approx 2 \times 8$ GHz. Our main aim is to perform a sensitive
search for a number of chemically diverse, diagnostic molecules in
$\sim 30$ protoplanetary disks in the Taurus region, covering  a wide
range of stellar characteristics, and to provide a simple chemical analysis
of the observational trends.

\begin{figure*}[t]
\begin{center}
\includegraphics[height=10.0cm]{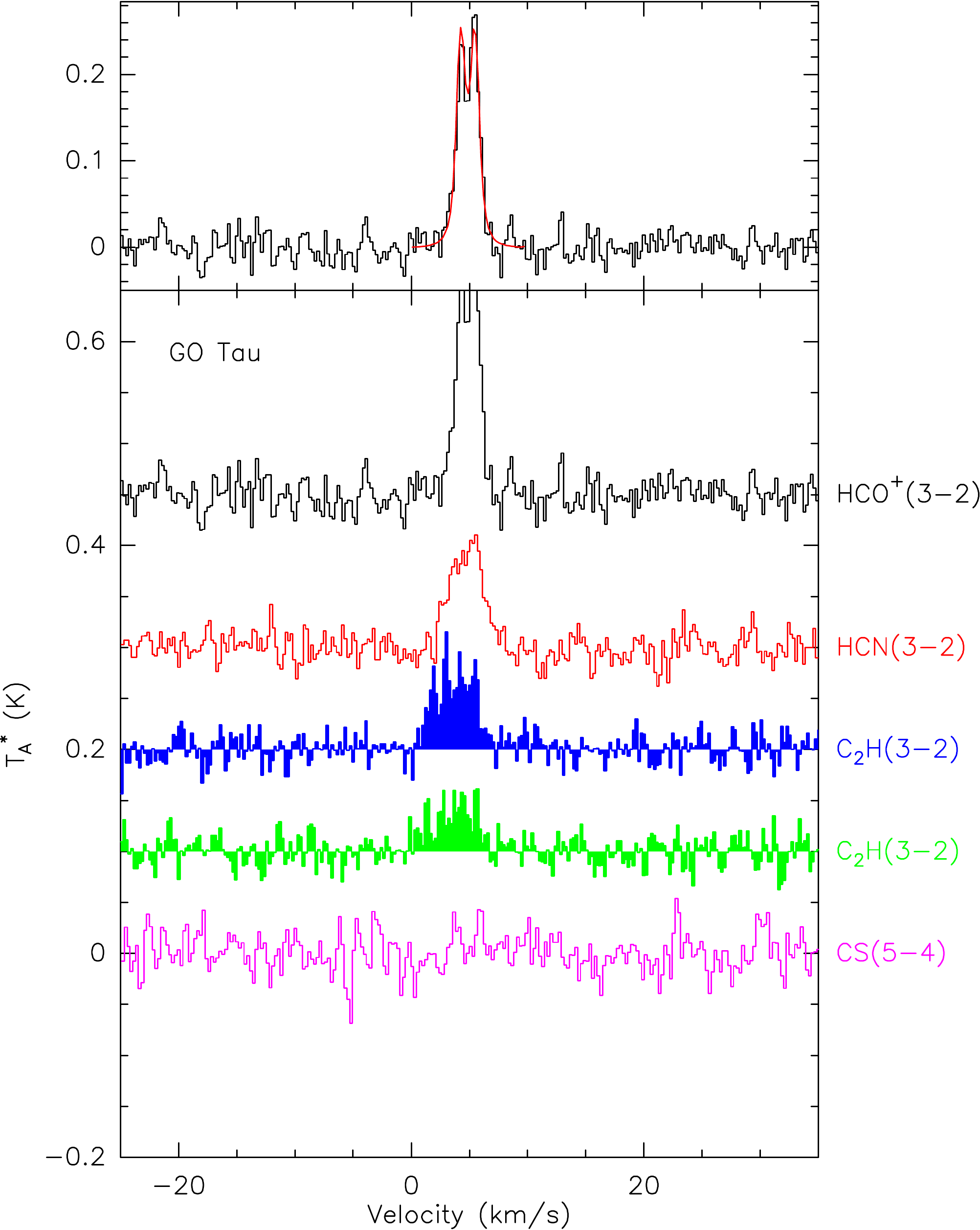}
\includegraphics[height=10.0cm]{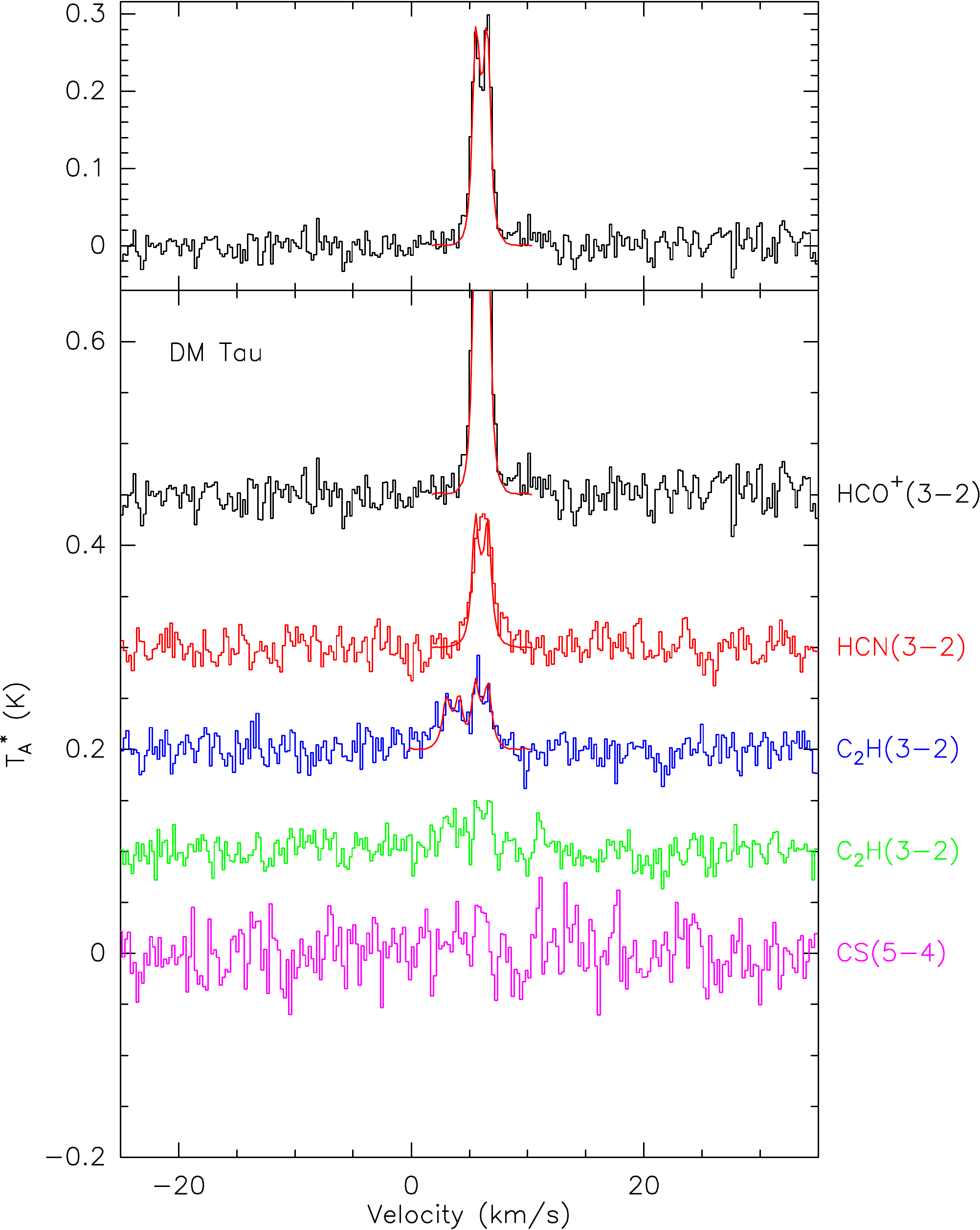}
\end{center}
\caption{Spectra of the observed transitions towards GO Tau and DM Tau.
Intensity scale is T$_A^*$; conversion to flux density can be made by multiplying by 9 Jy/K.
The thin red lines are the synthetic profiles obtained through the complete disk modelling
(Sect.\ref{sec:sub:model}).
The two groups of two hyperfine components for C$_2$H are presented in separate spectra:
J=7/2-5/2 in blue, J=5/2-3/2 in green. The upper panels present with a
different intensity scale the HCO$^+$ J=3-2 spectra
with the best fit model superimposed}.
\label{fig:gotau}
\end{figure*}

\section{Observations and Data Analysis}
\label{sec:obs}

\begin{figure*}[!ht] 
\begin{center}
\includegraphics[width=13.0cm]{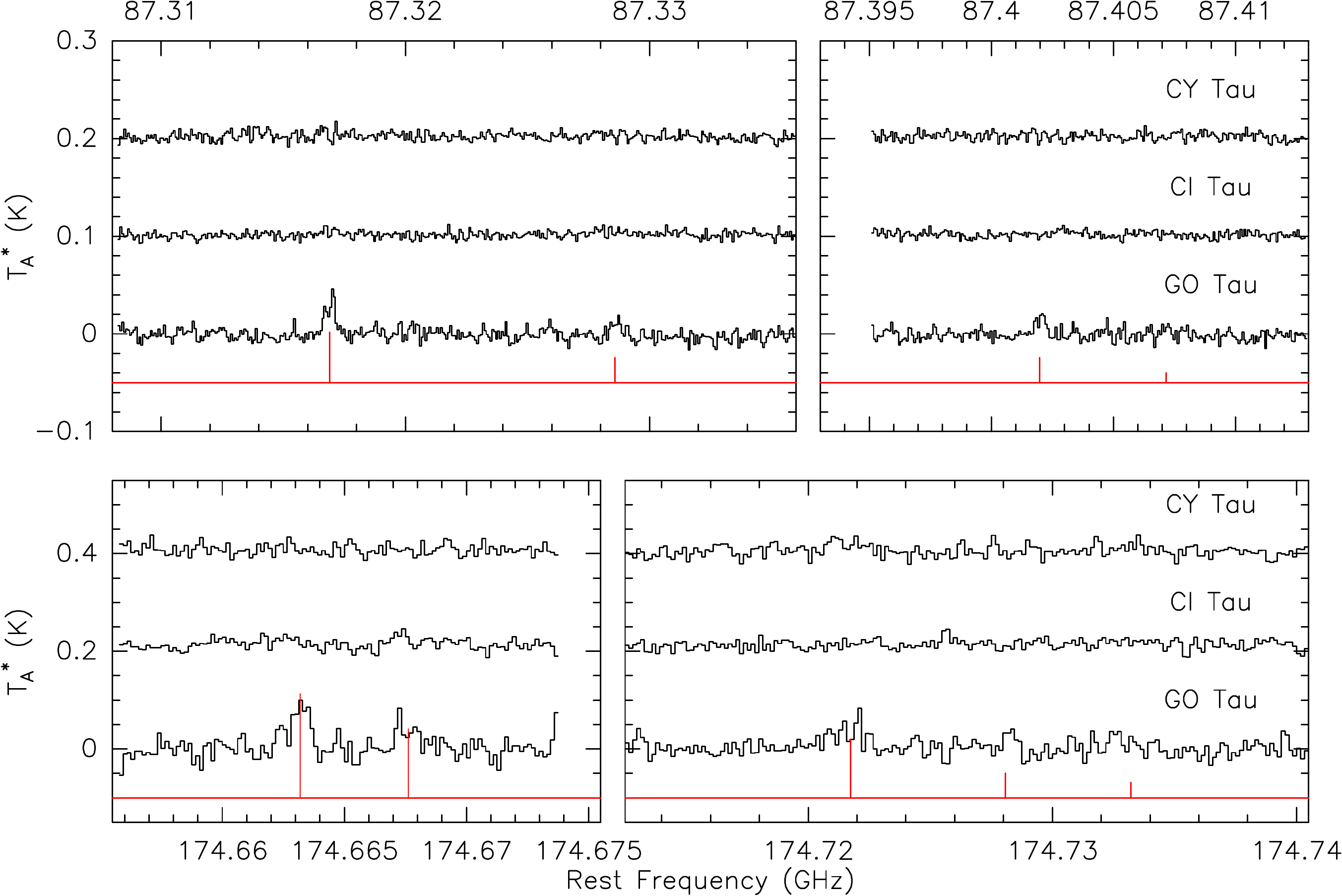}
\end{center}
\caption{N=1-0 (top) and 2-1 (bottom) C$_2$H emission towards CI Tau, CY Tau and GO Tau.
Conversion to flux density can be obtained by scaling by 6.1 Jy/K and
7.0 Jy/K respectively the observed spectra. The red ticks indicate
the expected positions and relative intensities of the hyperfine components.}
\label{fig:cch}
\end{figure*} 

\subsection{Source Sample}
\label{sec:sub:sample}
Our sample is derived from the source list of \citet{Guilloteau+etal_2013}.
It includes all sources for which a disk detection is unambiguous, but
explicitely avoids those where contamination by an outflow is suggested
by strong H$_2$CO or SO lines, such as DG Tau, T Tau or Haro 6-10. A total
of 30 disks has thus been observed, out of the 47 sources studied
in \citet{Guilloteau+etal_2013}.  We divide this sample into three
categories of sources. The first category is embedded objects, which
are probably younger than most other stars; it contains Haro 6-13, Haro 6-33,
Haro 6-5B, HH 30, and DO Tau. The second is Herbig Ae (HAe) stars, which includes
AB Aur, CQ Tau, MWC 480, and MWC 758. Although its spectral type is unknown
due to its edge-on configuration, we also include IRAS04302+2247 in this
category because its stellar mass is high, $1.7 \Msun$. It could equally
be classified in the ``embedded objects'' category without affecting our
main conclusions.  The last category is T Tauri stars.  Embedded objects
are highlighted in italics in the Tables, and HAe stars in boldface.
In Fig.~\ref{fig:correl-cd} and \ref{fig:correl-ratio-cn}-\ref{fig:correl-ratio-hco+},
embedded objects appear in red, HAe stars in blue, and T Tauri stars in black.

\subsection{30-m observations}
\label{sec:sub:obs}
Observations were carried out with the IRAM 30-m telescope, from Dec 3 to Dec 9, 2012.
Good weather conditions prevailed during the observing session.  The dual-polarization
EMIR 1.3 mm receiver was used to observe all lines simultaneously. Symmetric wobbler switching
in azimuth 60$''$ away from the targets provided very flat baselines.

Pointing was performed on 3C84. The estimated rms pointing error is $2-3''$, while
the telescope beam size is 10.5$''$ at 267 GHz. A more accurate limit can
be set from the symmetric aspect of the
double-peaked profiles of HCO$^+$(3-2), which results from the Keplerian rotation
of the disks. The largest asymmetries are obtained for LkCa15 and GG Tau, and
lead to a ratio between the two peaks on  the order
of 0.8, which indicate that the pointing offset is about $2''$ given the source size
near 8$"$ for these objects.

The EMIR receivers are equipped with sideband separating mixers, providing about 8 GHz
bandwidth in each sideband (separated by about 8 GHz).
Mean system temperatures ranged between 230 and 330 K at 267 GHz, and 160 to 260 K
at 245 GHz.  Two slightly different (overlapping) tunings were used, one covering
HOC$^+$, the other H$^{13}$CO$^+$ and CS(5-4).
In Upper side band,  this provided the best sensitivity in the [261.168-267.948] GHz
Frequency range, with typically 30 \% higher noise in the [267.948-268.948] GHz range and a factor two in the
[260.168-261.168] GHz range. In the Lower sideband, the best range is [245.487-252.267] GHz, with 20 \% higher noise in the
[252.265-253.267] GHz window, and 40 \% higher noise in the [244.487-245.287] GHz range containing CS(5-4).
The lines covered by the whole setup are indicated in Table \ref{tab:lines}. The spectral
resolution ranges from $\sim 0.24$ km.s$^{-1}$ for CS 5-4 to $\sim 0.22$ km.s$^{-1}$ for HCO$^+$(3-2).

Conversion from $T_A^*$ to flux density was performed assuming an efficiency of $J_K = 9.0$ Jy/K. The
absolute calibration may be uncertain to within 15 \%, but this affects all sources equally, and
does not impact the line ratios from source to source. Relative intensities within a source
should be more accurate given the simultaneity of the observations, as they  are only affected
by sideband gain variations across the receiver passband, and
should not exceed 10 \%. Furthermore, a comparison of the HCO$^+$(3-2) line flux with those derived by
\citet{Oberg+etal_2010} from SMA observations for DM Tau, LkCa 15, MWC 480, GM Aur, AA Tau and CQ Tau
shows agreement within the errorbars due to thermal noise, in general better than 10 \%. This
calibration uncertainty must be kept in mind when comparing the current molecules with
those detected by \citet{Guilloteau+etal_2013}.

\begin{table}[!h]
\caption{Frequencies of transitions}  
\begin{tabular}{lcr}
 Frequency & Molecule and & Detection \\
 (GHz)  &  Transition     &  \\
 \hline
 244.935609  & CS J=5-4 & Yes \\
 245.606310  & HC$_3$N  J=27-26 & \\
 251.825816  & SO ($6_5-5_4$) & Yes  \\
 260.2553390  & H$^{13}$CO$^+$ J=3-2 & (Yes) \\
 261.8344691  & C$_2$H  &  \\ 
 261.843756   & SO ($6_7-5_6$) & Yes  \\
  261.9781200 & C$_2$H  &  \\ 
  262.0042600 & C$_2$H N=3-2 J=7/2-5/2 F=4-3 & Yes \\ 
  262.0064820 & C$_2$H N=3-2 J=7/2-5/2 F=3-2 & Yes \\ 
  262.0649860 & C$_2$H N=3-2 J=5/2-3/2 F=3-2  & Yes  \\ 
  262.0674690 & C$_2$H N=3-2 J=5/2-3/2 F=2-1 & Yes \\ 
  262.0789347 & C$_2$H  &  \\ 
  262.2086143  & C$_2$H   &   \\ 
  262.2225856  & C$_2$H   &  \\ 
  262.2369577  & C$_2$H   &  \\ 
  262.2509290  & C$_2$H   &  \\ 
 263.792301 & HC$_3$N   J=29-28 & \\
 265.886436 & HCN J=3-2 & Yes \\
 267.557633 & HCO$^+$ J=3-2 & Yes \\
 268.451094 & HOC$^+$ J=3-2 &  \\
\hline
\end{tabular}
\tablefoot{Lines in the observed band. The detection column
indicates whether the line has been detected in the sample
(after source averaging for the marginal detection of H$^{13}$CO$^+$).}
\label{tab:lines}
\end{table}

Integration times ranged from 1 to 2 hours on source.
Except for the CS(5-4) line where it reaches 20 mK, the median noise
for all spectra is $\sigma_T \approx 13$ mK at the
$\delta V \sim 0.25 \kms$ spectral resolution for most spectral lines,
with relatively limited scatter (20 \%). The previous survey for
$^{13}$CO, CN and H$_2$CO reached a similar sensitivity,
although with a much larger scatter from source to source (7 to 25 mK)
\citep{Guilloteau+etal_2013}. With a characteristic line width around
$\Delta V \sim 3 \kms$, the flux sensitivity,
$\sigma_S = J_K \sigma_T \sqrt{\delta V \Delta V}$ is around 0.1 Jy\,$\kms$.
On average, our survey depth is 16 times better than those of \citet{Salter+etal_2011}
and 3 times better than those of \citet{Oberg+etal_2010,Oberg+etal_2011}.

Fig.\ref{fig:gotau} show the spectra obtained for GO Tau and DM Tau, which illustrate
the line shapes and the impact of the hyperfine structure for the C$_2$H lines.
Spectra for all sources are displayed in Appendix \ref{app:plots}.
\begin{table*}
\caption{Fit Results: HCO$^+$ and HCN}
\begin{tabular}{l rrr rrr }
\hline
Source & \multicolumn{3}{c}{HCO$^+$} & \multicolumn{3}{c}{HCN} \\
       & Area & V$_\mathrm{LSR}$ & $\Delta V$ &  Area & V$_\mathrm{LSR}$ & $\Delta V$ \\
       & Jy.$\kms$ & $\kms$ & $\kms$ & Jy.$\kms$ & $\kms$ & $\kms$ \\
\hline
\textbf{\textit{04302+2247}}  &     12.65  $\pm$      0.21  & 5.62 $\pm$ 0.04  & 4.83  $\pm$ 0.09  &
         1.81  $\pm$      0.17  &  [  5.60   ] &  [  4.80  ] \\
      AA Tau  &      1.71  $\pm$      0.21  & 7.65 $\pm$ 0.34  & 5.17  $\pm$ 0.67  &
       1.03 $\pm$      0.20   &  6.61$\pm$ 0.56   &  5.50 $\pm$ 1.25   \\
\textbf{AB Aur}  &      4.79  $\pm$      0.22  & 5.73 $\pm$ 0.07  & 2.89  $\pm$ 0.14  &
       1.59 $\pm$      0.22   &  5.30$\pm$ 0.23   &  2.96 $\pm$ 0.44   \\
      CI Tau  &      2.70  $\pm$      0.17  & 5.84 $\pm$ 0.11  & 3.17  $\pm$ 0.22  &
       0.77 $\pm$      0.16   &  6.02$\pm$ 0.32   &  2.96 $\pm$ 0.73   \\
\textbf{CQ Tau}  &      1.12  $\pm$      0.24  & 5.91 $\pm$ 0.73  & 7.19  $\pm$ 1.87  &
         0.61  $\pm$      0.19  &  [  5.90   ] &  [  7.20  ] \\
      CW Tau  &      0.46  $\pm$      0.12  & 4.16 $\pm$ 0.23  & 1.64  $\pm$ 0.38  &
    $ <       0.48 $ & & \\
      CY Tau  &      2.12  $\pm$      0.14  & 7.31 $\pm$ 0.07  & 1.98  $\pm$ 0.13  &
       0.90 $\pm$      0.14   &  7.19$\pm$ 0.21   &  2.54 $\pm$ 0.41   \\
      DL Tau  &      4.30  $\pm$      0.14  & 6.11 $\pm$ 0.05  & 2.83  $\pm$ 0.09  &
         1.02  $\pm$      0.13  &  [  6.10   ] &  [  2.80  ] \\
      DM Tau  &      5.23  $\pm$      0.13  & 6.07 $\pm$ 0.02  & 1.90  $\pm$ 0.05  &
       2.55 $\pm$      0.12   &  6.15$\pm$ 0.05   &  2.00 $\pm$ 0.12   \\
      DN Tau  &      1.34  $\pm$      0.13  & 6.23 $\pm$ 0.14  & 2.75  $\pm$ 0.28  &
       1.20 $\pm$      0.16   &  6.85$\pm$ 0.27   &  3.98 $\pm$ 0.52   \\
\textit{DO Tau}  &      0.93  $\pm$      0.12  & 6.47 $\pm$ 0.14  & 1.99  $\pm$ 0.31  &
    $ <       0.31 $ & & \\
      FT Tau  &      1.08  $\pm$      0.16  & 7.41 $\pm$ 0.20  & 2.64  $\pm$ 0.43  &
    $ <       0.37 $ & & \\
      GG Tau  &      6.59  $\pm$      0.15  & 6.58 $\pm$ 0.03  & 2.79  $\pm$ 0.07  &
       2.88 $\pm$      0.16   &  6.51$\pm$ 0.08   &  2.77 $\pm$ 0.16   \\
      GM Aur  &      4.37  $\pm$      0.16  & 5.75 $\pm$ 0.08  & 4.06  $\pm$ 0.16  &
       1.59 $\pm$      0.17   &  5.85$\pm$ 0.24   &  4.53 $\pm$ 0.56   \\
      GO Tau  &      5.39  $\pm$      0.13  & 4.94 $\pm$ 0.03  & 2.16  $\pm$ 0.06  &
       3.44 $\pm$      0.15   &  4.81$\pm$ 0.07   &  3.29 $\pm$ 0.15   \\

\textit{Haro 6-13}  &      2.28  $\pm$      0.17  & 5.18 $\pm$ 0.10  & 2.83  $\pm$ 0.25  &
         0.52  $\pm$      0.13  &  [  5.20   ] &  [  2.90  ] \\
\textit{Haro 6-33}  &      3.21  $\pm$      0.13  & 5.52 $\pm$ 0.05  & 2.43  $\pm$ 0.11  &
         0.77  $\pm$      0.11  &  [  5.50   ] &  [  2.50  ] \\
\textit{Haro 6-5B}  &      2.83  $\pm$      0.24  & 7.78 $\pm$ 0.25  & 5.92  $\pm$ 0.52  &
    $ <       0.59 $ & & \\
\textit{HH 30}  &      2.01  $\pm$      0.24  & 6.97 $\pm$ 0.18  & 3.41  $\pm$ 0.57  &
    $ <       0.49 $ & & \\
      HK Tau  &      2.74  $\pm$      0.27  & 6.14 $\pm$ 0.53  & 9.27  $\pm$ 0.91  &
         0.97  $\pm$      0.24  &  [  6.30   ] &  [  9.50  ] \\
    HV Tau C  &      2.97  $\pm$      0.24  & 5.57 $\pm$ 0.34  & 8.45  $\pm$ 0.74  &
    $ <       0.58 $ & & \\
      IQ Tau  &      2.17  $\pm$      0.22  & 5.70 $\pm$ 0.26  & 4.87  $\pm$ 0.53  &
       0.78 $\pm$      0.18   &  6.45$\pm$ 0.55   &  4.20 $\pm$ 0.82   \\

    LkH$\alpha$ 358  &      0.26  $\pm$      0.08  & 7.58 $\pm$ 0.06  & 0.42  $\pm$ 0.16  &
    $ <       0.33 $ & & \\
     LkCa 15  &      5.03  $\pm$      0.16  & 6.08 $\pm$ 0.06  & 3.67  $\pm$ 0.13  &
       4.42 $\pm$      0.18   &  6.17$\pm$ 0.08   &  4.20 $\pm$ 0.19   \\
\textbf{MWC 480}  &      4.27  $\pm$      0.14  & 5.16 $\pm$ 0.07  & 3.86  $\pm$ 0.13  &
       2.01 $\pm$      0.17   &  5.42$\pm$ 0.30   &  6.76 $\pm$ 0.61   \\
\textbf{MWC 758}  &      0.80  $\pm$      0.12  & 5.26 $\pm$ 0.20  & 2.62  $\pm$ 0.35  &
    $ <       0.31 $ & & \\
      RW Aur  &      2.67  $\pm$      0.29  & 4.34 $\pm$ 0.70  &12.78  $\pm$ 1.61  &
    $ <       0.75 $ & & \\
      RY Tau  &      1.90  $\pm$      0.33  & 5.12 $\pm$ 0.92  &10.15  $\pm$ 1.83  &
         1.26  $\pm$      0.29  &  [  5.20   ] &  [ 10.00  ] \\
      SU Aur  &      1.58  $\pm$      0.24  & 4.39 $\pm$ 0.49  & 7.27  $\pm$ 1.51  &
    $ <       0.79 $ & & \\
    UZ Tau E  &      0.99  $\pm$      0.21  & 6.72 $\pm$ 0.60  & 5.57  $\pm$ 1.36  &
         0.58  $\pm$      0.19  &  [  7.00   ] &  [  5.90  ] \\
\hline
\end{tabular}
\tablefoot{Source names are in italics for embedded sources, in boldface for HAe stars.
\textbf{Number in brackets indicate the assumed value foor fixed parameters.}}
\label{tab:hco-hcn}
\end{table*}

\subsection{Ancillary 30-m data}
In three sources, CI Tau, CY Tau, and GO Tau,
we obtained high sensitivity spectra
of the C$_2$H N=1-0 and N=2-1 transitions.
Each source was observed with the IRAM 30-m telescope in Jul 2010 for about 6 hours.
We used wobbler switching, and the stable weather conditions yielded
flat baselines. System temperatures were
around 90 -- 110 K at 3\,mm, but ranged between 300 and 500 K at 174.7 GHz
The spectra were smoothed to 0.26 $\kms$ resolution, yielding
a rms noise about 6 mK at 87 GHz for the 1-0 line, and
16 mK for the 2-1.  Spectra are displayed in Fig.\ref{fig:cch}.
Lines are clearly detected in GO Tau. Summing the spectra of CI Tau and CY Tau
(after appropriate recentering for their respective velocities) also indicates
a $5 \sigma$ detection in those sources (although each source is only detected
at the $3 \sigma$ level).

\subsection{30-m spectra analysis}
All lines were fitted by simple Gaussian profiles. Although some of them are clearly double peaked
(see Fig.\ref{fig:gotau}),
this process still gives the
appropriate total line flux within the statistical uncertainties.
Results are given in Tables \ref{tab:lines}-\ref{tab:cch-cs}.
The N=3-2 C$_2$H transition is split in 11 hyperfine components,
but 94 \% of the line flux comes in two well separated groups of two hyperfine components,
with an hyperfine splitting of about 3 km.s$^{-1}$
which is rather similar to the typical line width. We used the HFS method of the CLASS
(\footnote{see https://www.iram.fr/IRAMFR/GILDAS/})  program to
make a fit of all components together, assuming optically thin emission (total opacity $< 0.5$).
From the hyperfine line ratios, \citet{Punzi+etal_2015} find non-negligible optical
depths for CN and especially C$_2$H in LkCa 15. Their quoted values for opacities are within those allowed by
our (somewhat less sensitive) data:
our best fit total opacity for C$_2$H is 4, but with large errors ($\sim 2-3$). DM Tau may also be partially
optically thick in C$_2$H, with a total opacity of 0.8, but again a large error. All other sources
are consistent with optically thin emission. The reported flux in Table \ref{tab:cch-cs}
is the sum of the 4 strongest transitions.

\begin{table*}
\caption{Fit Results: CCH (main group of hyperfine components) and CS}
\begin{tabular}{l rrr rrr l}
\hline
Source & \multicolumn{3}{c}{C$_2$H} & \multicolumn{3}{c}{CS} \\
       & Area & V$_\mathrm{LSR}$ & $\Delta V$ &  Area & V$_\mathrm{LSR}$ & $\Delta V$ \\
       & Jy.$\kms$ & $\kms$ & $\kms$ & Jy.$\kms$ & $\kms$ & $\kms$ \\
\hline

\textbf{\textit{04302+2247}}&
    $ <       0.79 $ & & &
       3.03 $\pm$      0.50   &  6.02$\pm$ 0.35   &  4.38 $\pm$ 0.89   \\
AA Tau&
      1.10  $\pm$      0.35  & 6.70 $\pm$ 0.71  & 4.30  $\pm$ 1.47  &
       0.97 $\pm$      0.25   &  5.77$\pm$ 0.40   &  2.96 $\pm$ 0.76   \\
\textbf{AB Aur} &
      1.31  $\pm$      0.43  &  [  5.30   ] &  [  2.96   ] &
\multicolumn{3}{c}{no data} \\
CI Tau&
      1.97  $\pm$      0.48  & 5.90 $\pm$ 0.53  & 4.91  $\pm$ 1.54  &
         1.42  $\pm$      0.37  &  [  5.80   ] &  [  3.20  ] \\
\textbf{CQ Tau} &
    $ <       0.94 $ & & &
    $ <       0.98 $ & & \\
CW Tau&
    $ <       0.61 $ & & &
    $ <       0.71 $ & & \\
CY Tau&
      2.30  $\pm$      0.33  & 7.36 $\pm$ 0.24  & 2.57  $\pm$ 0.32  &
       1.05 $\pm$      0.26   &  7.04$\pm$ 0.52   &  3.76 $\pm$ 0.85   \\

DL Tau&
      1.22  $\pm$      0.30  & 6.17 $\pm$ 0.37  & 2.55  $\pm$ 0.60  &
    $ <       1.06 $ & & \\
DM Tau&
      3.88  $\pm$      0.27  & 6.02 $\pm$ 0.07  & 1.98  $\pm$ 0.15  &
         0.61  $\pm$      0.18  &  [  6.06   ] &  [  1.90  ] \\
DN Tau&
      1.58  $\pm$      0.34  & 6.54 $\pm$ 0.40  & 3.51  $\pm$ 0.77  &
    $ <       0.32 $ & & \\
\textit{DO Tau} &
    $ <       0.59 $ & & &
       1.24 $\pm$      0.24   &  4.38$\pm$ 0.32   &  2.99 $\pm$ 0.50   \\

FT Tau&
      1.00  $\pm$      0.24  &  [  7.40   ] &  [  2.50   ] &
    $ <       0.68 $ & & \\

GG Tau&
      2.89  $\pm$      0.29  & 6.13 $\pm$ 0.19  & 2.99  $\pm$ 0.28  &
       2.61 $\pm$      0.40   &  6.71$\pm$ 0.24   &  2.91 $\pm$ 0.39   \\
GM Aur&
    $ <       0.68 $ & & &
       1.18 $\pm$      0.19   &  5.63$\pm$ 0.28   &  3.25 $\pm$ 0.45   \\
GO Tau&
      5.66  $\pm$      0.30  & 4.80 $\pm$ 0.09  & 2.56  $\pm$ 0.12  &
    $ <       0.47 $ & & \\

\textit{Haro 6-13} &
    $ <       0.68 $ & & &
    $ <       0.98 $ & & \\
\textit{Haro 6-33} &
    $ <       0.59 $ & & &
       1.87 $\pm$      0.16   &  5.60$\pm$ 0.06   &  1.51 $\pm$ 0.14   \\
\textit{Haro 6-5B} &
    $ <       0.90 $ & & &
    $ <       0.87 $ & & \\
\textit{HH 30} &
    $ <       0.65 $ & & &
    $ <       0.64 $ & & \\
HK Tau&
    $ <       1.00 $ & & &
    $ <       1.80 $ & & \\
HV Tau C&
    $ <       0.87 $ & & &
    $ <       0.83 $ & & \\
IQ Tau&
      1.56  $\pm$      0.40  & 4.99 $\pm$ 0.51  & 4.06  $\pm$ 1.12  &
       1.27 $\pm$      0.27   &  5.42$\pm$ 0.41   &  3.56 $\pm$ 0.72   \\
LkH$\alpha$ 358&
    $ <       0.30 $ & & &
    $ <       0.50 $ & & \\
LkCa 15&
      2.69  $\pm$      0.30  & 5.77 $\pm$ 0.16  & 2.62  $\pm$ 0.29  &
       1.41 $\pm$      0.32   &  6.08$\pm$ 0.38   &  4.18 $\pm$ 1.47   \\
\textbf{MWC 480}&
      2.07  $\pm$      0.23  &  [  5.42   ] &  [  6.76   ] &
    $ <       0.70 $ & & \\
\textbf{MWC 758}&
    $ <       0.61 $ & & &
    $ <       0.58 $ & & \\
RW Aur&
    $ <       1.21 $ & & &
    $ <       1.02 $ & & \\
RY Tau&
      1.46  $\pm$      0.41  &  [  5.20   ] &  [ 10.00   ] &
    $ <       1.43 $ & & \\
SU Aur&
    $ <       0.87 $ & & &
    $ <       1.56 $ & & \\
UZ Tau E&
    $ <       0.83 $ & & &
    $ <       1.46 $ & & \\
\hline
\end{tabular}
\label{tab:cch-cs}
\end{table*}

\begin{figure}[!hb] 
\begin{center}
\includegraphics[width=0.8\columnwidth]{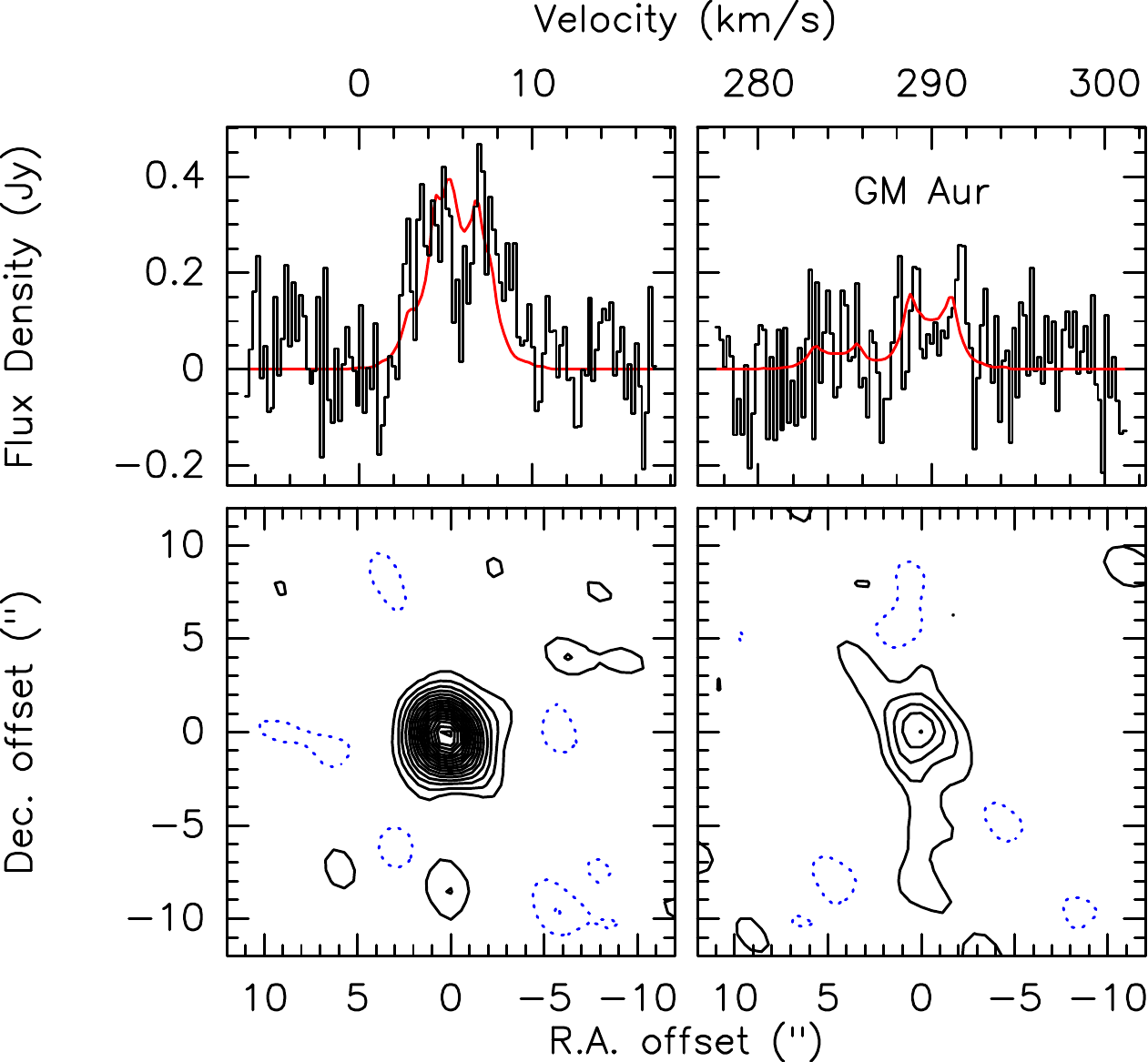}
\end{center}
\caption{Integrated spectra of the two groups of hyperfine
components for CN N=2-1 line, and maps of the signal-to-noise
ratio 
 towards GM Aur (contours are in steps of
$2 \sigma$). The profile of the best fit disk model
(see Sec.\ref{sec:sub:model}) is shown in red.}
\label{fig:gmaur:pdbi}
\end{figure} 

\subsection{Interferometer data}
To better constrain the disk properties, we use the CN and CO IRAM
Plateau de Bure interferometer (PdBI) data reported in \citet{Guilloteau+etal_2014}.
In addition, we also analyzed in a similar way unpublished observations
of GM Aur in CN, performed in compact
configuration on 29 Oct, 2007, which yielded an angular resolution
of $2.8\times 2.1''$.
Figure \ref{fig:gmaur:pdbi} shows the resulting integrated spectra
and integrated intensity maps of the two sets of hyperfine components.

Disk properties, specifically source
velocity, disk inclination and outer radius, and star dynamical
mass, are derived from these data sets, and completed by results
from \citet{Simon+etal_2000,Pietu+etal_2007,Schaefer+etal_2009} when needed
for higher accuracy.

Similarly, the C$_2$H PdBI data reported by \citet{Henning+etal_2010}
for DM\,Tau, LkCa\,15 and MWC\,480 has been reanalyzed to provide a
complementary view on the C$_2$H distribution.

\section{Results}
\begin{figure}[!hb] 
\begin{center}
\includegraphics[width=0.8\columnwidth]{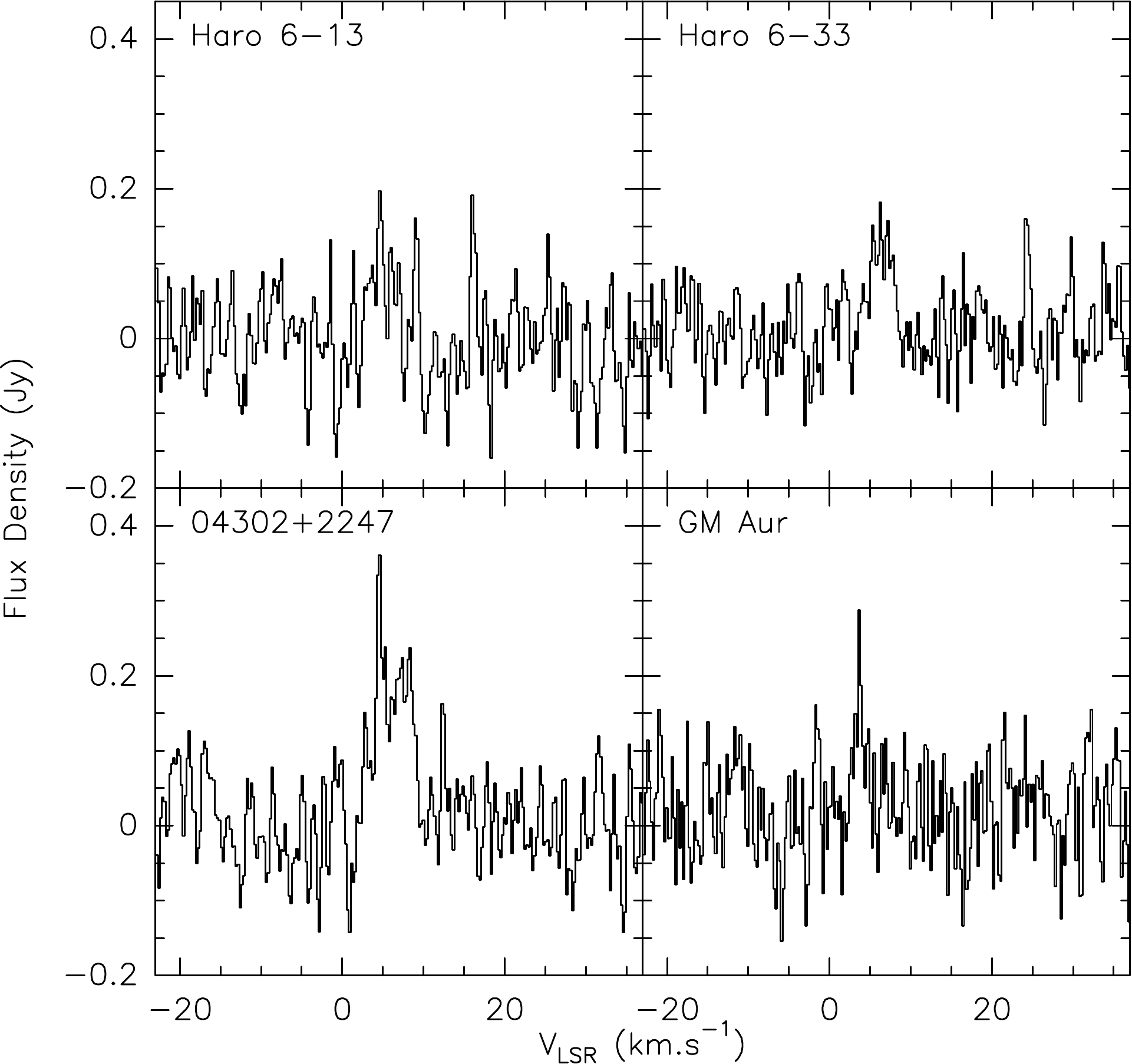}
\end{center}
\caption{Averages of the SO($6_5-5_4$) and  SO($6_7-5_6$) spectra resampled
to a common velocity scale.}
\label{fig:so}
\end{figure} 

\begin{table*}
\caption{HCO$^+$ outer radii and line detections}
\begin{tabular}{l rrrr rrr rr}
\hline
Source & R(HCO$^+$) & $T_0$ & Flux & Line width & \multicolumn{3}{c}{Detection ?} & R$_\mathrm{inter}$ & Ref \\
  & (au) & (K) & (Jy\,$\kms$) & $\kms$ & HCN & CCH & CS  &  (au) \\
(1) & (2) & (3) & (4) & (5) & (6) & (7) & (8) & (9) & (10) \\
\hline
  \textbf{\textit{04302+2247}} &
  860 &  30 &    12.65 &       4.83 $\pm$       0.09
 &  10.5  &  &   6.1 & 750$\pm$56 & CN, 2 \\

      AA Tau &
   440 &  15 &     1.71 &       5.17 $\pm$       0.67
 &   5.1 &   3.1 &   3.9  & $\approx 500$ & CO, 13 \\
      \textbf{AB Aur} &
   520 & 30 &      4.79 &       2.89 $\pm$       0.14
 &   7.2 &   3.0 &   - & 890$\pm$10 & $^{13}$CO, 14 \\

      CI Tau &
   560 &  15 &     2.70 &       3.17 $\pm$       0.22
 &   4.8 &   4.1 &   3.8 & 520$\pm$13 & CN, 2 \\
      \textbf{CQ Tau} &
   250 & 30 &      1.12 &       7.19 $\pm$       1.87
 &   3.2  & & & 200$\pm$20 & CO, 16 \\
      CW Tau &
   230 &    15 &    0.46 &       1.64 $\pm$       0.38
 &  &  & & 210$\pm$7  & $^{13}$CO, 1 \\
      CY Tau &
   490 &   15 &    2.12 &       1.98 $\pm$       0.13
 &   6.3  &   7.0  &   4.0 & 295$\pm$11 & CN, 2 \\

      DL Tau &
   700 & 15 &      4.30 &       2.83 $\pm$       0.09
 &   8.1 &   4.1 &  & 463$\pm$6 & CO, CN, 2 \\
      DM Tau &
   780 & 15 &      5.23 &       1.90 $\pm$       0.05
 &  20.4 &  14.4 &   3.4 & 800$\pm$30 & HCO$^+$ 1-0, 11 \\
      DN Tau &
   390 &  15 &     1.34 &       2.75 $\pm$       0.28
 &   7.6 &   4.7 &  & 241$\pm$7 & CN, 2 \\
      \textit{DO Tau} &
   330 &  15 &      0.93 &       1.99 $\pm$       0.31
 & &  &   5.1 & \textit{300} & \textit{CN, 17} \\

      FT Tau &
   350 &  15 &     1.08 &       2.64 $\pm$       0.43
 &  &   4.2 & & $> 60$ & cont, 5 \\
      GG Tau &
   610 & 30 &      6.59 &       2.79 $\pm$       0.07
 &  18.1 &   9.9  &   6.6 & 800 & $^{13}$CO, 9 \\
      GM Aur &
   710 & 15 &      4.37 &       4.06 $\pm$       0.16
 &   9.3  & &   6.3 & 525$\pm$20 & CO, 12 \\
      GO Tau &
   790 &  15    & 5.39 &       2.16 $\pm$       0.06
 &  23.2 &  19.2 & & 587$\pm$55 & CN, 2 \\

   \textit{Haro 6-13} &
   510 &  15 &      2.28 &       2.83 $\pm$       0.25
 &   4.0  & &  & $> 180$, \textit{280} & CO, 6; \textit{CN, 17} \\
   \textit{Haro 6-33} &
   610 & 30 &       3.21 &       2.43 $\pm$       0.11
 &   6.7 & &  11.9 & 300$\pm$10, \textit{260$\pm$30} & CO, 6; \textit{CN, 17} \\
   \textit{Haro 6-5B} &
   570 &  15 &     2.83 &       5.92 $\pm$       0.52
 &  &  &  & $\approx 600$, \textit{310} & $^{13}$CO 4; \textit{CN, 17} \\
       \textit{HH 30} &
   480 & 15 &      2.01 &       3.41 $\pm$       0.57
 & & & &  420$\pm$25  & $^{13}$CO, 7 \\
      HK Tau &
   400 & 30 &      2.74 &       9.27 $\pm$       0.91
 &   4.1 & & & $> 200$ & CO, 8 \\
    HV Tau C &
   410 &  30 &     2.97 &       8.45 $\pm$       0.74
 & & & & 256$\pm$51 & CN, 2 \\

      IQ Tau &
   500 &  15 &     2.17 &       4.87 $\pm$       0.53
 &   4.3 &   3.9 &   4.6 & 220 $\pm$15 & CN, 2 \\
    LkH$\alpha$ 358 &
   170 & 15 &     0.26 &       0.42 $\pm$       0.16
 & & & & 150$\pm$17 & CO, 6 \\
     LkCa 15 &
   760 & 15 &      5.03 &       3.67 $\pm$       0.13
 &  24.7 &   9.0 &   4.4 & 660$\pm$60 & HCO$^+$ 1-0, 11 \\

     \textbf{MWC 480} &
   490 & 30 &      4.27 &       3.86 $\pm$       0.13
 &  11.6 &   9.1 & & 520$\pm$50 & HCO$^+$ 1-0, 11 \\
     \textbf{MWC 758} &
   210 & 30 &      0.80 &       2.62 $\pm$       0.35
 & & & & 270$\pm$15 & CO, 16 \\

      RW Aur &
   390 & 30 &      2.67 &      12.78 $\pm$       1.61
 & & & & 50 & CO, 15 \\
      RY Tau &
   330 &  30 &     1.90 &      10.15 $\pm$       1.83
 &   4.4  &   3.6  &  & $\approx 400$ & CO, 3 \\
      SU Aur &
   300 & 30 &      1.58 &       7.27 $\pm$       1.51
 & & & & $> 150$ & $^{13}$CO, 1 \\ 
    UZ Tau E &
   240 &  30 &     0.99 &       5.57 $\pm$       1.36
 &   3.0 & & & 300$\pm$20 & CO, 10 \\
\hline
\end{tabular}
\label{tab:radii}
\tablefoot{HCO$^+$ outer radii (Col.2) for the assumed temperature (Col.3). Cols.4-5 are
the line flux and width.
Cols.6-8 indicate the signal to noise for line detections. Col.9 indicates interferometric
radii R$_\mathrm{inter}$ (or, when in italics, R$_\mathrm{single}$ from reference [17]),
for the molecules and references given in Col.10. References are:
[1] \citep{Pietu+etal_2014},
[2] \citep{Guilloteau+etal_2014},
[3] \citep{Coffey+etal_2015},
[4] \citep{Yokogawa+etal_2002},
[5] \citep{Guilloteau+etal_2011},
[6] \citep{Schaefer+etal_2009},
[7] \citep{Pety+etal_2006},
[8] \citep{Jensen+Akeson_2014},
[9] \citep{Guilloteau+etal_1999},
[10] \citep{Simon+etal_2000},
[11] \citep{Pietu+etal_2007},
[12] \citep{Dutrey+etal_1998},
[13] Kessler-Silacci, 2004, PhD Thesis,
[14] \citep{Pietu+etal_2005},
[15] \citep{Cabrit+etal_2006},
[16] \citep{Chapillon+etal_2008},
\textit{[17] \citep{Guilloteau+etal_2013}, single-dish estimate
from CN line flux}
}
\end{table*}
%

Combined with the previous study by \citet{Guilloteau+etal_2013}, our
survey provides an important quantitative step in number of molecules detected
in disks, with 20 new detections of HCO$^+$, 18 of HCN, 19 in CN, 11 in C$_2$H,
10 in H$_2$CO, 8 in CS, 4 in SO and 4 in C$^{17}$O in a sample of 30 disks,
of which only 5 had already been studied extensively.

HCO$^+$ is detected in practically all sources. Detection is marginal in two
sources, CW Tau and LkH$\alpha$ 358, in which confusion with the molecular cloud
makes  the interpretation ambiguous (see Sect.\ref{sec:obj:cw_tau} and \ref{sec:obj:lkha_358}).
All strong lines are double peaked, as expected from Keplerian rotation
in circumstellar disks.  This also indicates that there is little contamination
from the surrounding clouds (or outflows) for these sources. A few sources have very wide lines,
in good agreement with the results from CN measurements of \citet{Guilloteau+etal_2013}.
A peculiar case is RW Aur, where the HCO$^+$ line is extremely wide, and may not originate
from a circumstellar disk (see Sect.\ref{sec:obj:rw_aur}).
HCN is detected in 20 (although marginally in CQ Tau and UZ Tau E) out of 30 sources.
C$_2$H is detected in 13 (ignoring AB Aur) out of 30 sources, and CS in 10 out of 29 sources (AB Aur
was not observed in CS). Both lines are weak (peak flux 0.3-0.5 Jy).

Several lines of SO appear in the observed band, two of them being detected: the SO($6_5-5_4$) line
at 251.825816 GHz (in Haro 6-33 and IRAS04302+2247, the Butterfly star, hereafter 04302+2247)
and the SO($6_7-5_6$) at 261.843756 GHz (in Haro 6-13, Haro 6-33
and 04302+2247, see Table \ref{tab:so}). Combined spectra obtained by averaging
the two lines are displayed in Fig.\ref{fig:so}. These spectra also
suggest a marginal detection in GM Aur.

No other molecule is detected in the band, which included lines of HOC$^+$,
two lines of HC$_3$N and the H$^{13}$CO$^+$ J=3-2 transition. The latter provides
upper limits to the HCO$^+$ opacity (see Section\ref{sec:sub:hco} for details).

\begin{table}
\caption{SO detections}
\label{tab:so}
\begin{tabular}{c|rr}
\hline
Transition         & SO $6_5-5_4$ & SO $6_7-5_6$ \\
Source Name        &  \multicolumn{2}{c}{(Jy\,km\,s$^{-1}$)} \\
\hline
IRAS04302+2247 & $0.90 \pm 0.13$ & $1.60 \pm 0.18$ \\
Haro 6-13      & $0.26 \pm 0.10$ & $0.80 \pm 0.17$ \\
Haro 6-33      & $0.43 \pm 0.09$ & $0.48 \pm 0.13$ \\
GM Aur         & $0.30 \pm 0.13$ & $0.48 \pm 0.13$ \\
\hline
\end{tabular}
\tablefoot{Integrated line flux for the SO line detections}
\end{table}

\section{Analysis Method}

\subsection{Definition of disk radii}
\label{sec:sub:def}

In the following sections, we introduce several radii
with the goal of characterizing the size of the observed disks
in a consistent way. In Sect.\ref{sec:sub:flux}, we first use a simple approximation to
derive the outer radius of the molecular distribution of HCO$^+$,
R(HCO$^+$), using the same approach as \citet{Guilloteau+etal_2013}
for CN.
In Sect.\ref{sec:sub:profile}, we also show that an estimate of
the outer radius can be obtained from the overall shape of the line profile,
when the stellar mass is known.
Finally, Sect.\ref{sec:sub:model} derives radii based on a detailed
modelling of the line emission, using a 3-D ray tracing code to
simulate the emission from a disk with a power distribution of
molecules and temperatures. The derived outer radius of that distribution
is called $R_\mathrm{single}$ if only the line profile can be fit,
and $R_\mathrm{inter}$ when we can model resolved images of the line
emission.
Based on these independent estimates, we show that a single radius
is sufficient to characterize the HCO$^+$ and CN emission, as well
as that of C$_2$H when the later method can be applied to it. This
\textit{disk radius} $R_\mathrm{disk}$ is then used to derive the molecular column
densities from the optically thinner emission of other molecules.

Although real molecular distributions will not be truncated
power laws, this model can still be a reasonable approximation.
This happens because the density in disks is falling rapidly
with radius,
(e.g. as $1/r^3$ for a surface density as $1/r^{1.5}$ and isothermal
disk, and even more steeply for exponentially
tapered profiles predicted by viscous evolution models). This
implies a sharp transition beyond which the observed spectral
lines become unexcited. Only CO, because of its low critical
density, may behave differently and have a larger outer radius.

\subsection{Radii from line flux}
\label{sec:sub:flux}
Line formation in a Keplerian disk is largely dominated
by the Keplerian shear \citep{Horne+Marsh_1986}. We use
the approach described in \citet[][Appendix D]{Guilloteau+etal_2013}, to which
we refer for additional details.
The integrated line flux is given by
\begin{equation}
\int S_{\nu} d\mathrm{v} = B_{\nu}(T_0) (\rho \Delta V) \pi R_{\mathrm{out}}^2 / D^2 {\rm cos}(i)
\label{eq:flux}
\end{equation}
for inclinations $i<60-70^\circ$.
$T_0$ is the average disk temperature, $\Delta V$ the local
linewidth, $R_\mathrm{out}$ the disk outer radius, and $D$ the source distance.
$\rho$ \textbf{is a dimensionless factor which} depends on the line opacity $\tau_l$ \citep[see][their Fig.\,4]{Guilloteau+Dutrey_1998}.
$\rho \approx \tau_l$ for optically thin lines, saturating as
$\propto \log(\tau_l)$ for large optical depths.
In the optically thin regime, $\tau_l$ is inversely proportional
to $\cos(i)$ and $\Delta V$ and proportional to the molecular column density, so the dependency on
inclination and line width cancels, and the flux just
scales with the total number of molecules. For optically thick, nearly edge-on objects,
where the above formula breaks down, \citet{Beckwith+Sargent_1993}
have shown that the expected flux is not a strong function of inclination,
so using $\cos{i} = 0.5$ remains appropriate.
Inverting Eq.\ref{eq:flux}, the outer radius is given by
\begin{equation}
R_\mathrm{out}  =  D \left( \frac{\int S_{\nu} d\mathrm{v} }{B_{\nu}(T_0) (\rho \Delta V) \pi {\rm cos}(i)} \right) ^{1/2}
\label{eq:main-rout}
\end{equation}

In disks, the local line width is typically around $0.2 \kms$ \citep[e.g.][]{Pietu+etal_2007},
so the outer radius can be derived provided $\rho$ and $T_0$ can be estimated.
In \citet{Guilloteau+etal_2013}, we used the hyperfine structure of CN to provide
an estimate of the line opacity, and thus of $\rho$, and simply assumed $T_0=15$\,K for
large disks around T Tauri stars, and 30\,K for small disks, or those around HAe stars.
We show in the next paragraph that using $\rho=1$ is appropriate for HCO$^+$.

\subsection{HCO$^+$ line flux}
\label{sec:sub:hco}

\paragraph{Simple Approximation: $\rho$(HCO$^+)=1$\\} \hspace{2.0ex} Some upper limit on the HCO$^+$ J=3-2 line opacity
comes from the non (or at best marginal) detection of the H$^{13}$CO$^+$ J=3-2 transition. For the strongest HCO$^+$
J=3-2 emitter, 04302+2414, H$^{13}$CO$^+$ J=3-2 is at least 20 times fainter.  Combining the signals from the stronger
sources (re-centered on the source velocity and weighted by the HCO$^+$ J=3-2 detected line intensity, but not
re-scaled in terms of line width), we obtain a marginal detection of H$^{13}$CO$^+$ J=3-2 at a level 20 times smaller
than that of the main isotopologues. So the average opacity of the HCO$^+$ J=3-2 line does not exceed a few. We
thus use $\rho = 1$ in Eq.\ref{eq:main-rout}. As in \citet{Guilloteau+etal_2013}, we also assume $\Delta V = 0.2
\kms$, and $T_0 = 15$ (for T Tauri stars) or 30 K (for F and A stars, or small disks) (see Table \ref{tab:radii} Col.3
for the value used for each source).

The derived outer radii, called $R(\mathrm{HCO}^+)$, are given in Table \ref{tab:radii}, and are in general quite
consistent with prior knowledge coming from  resolved images made with interferometers (see Table \ref{tab:radii}
Col.10 and associated references) despite our very crude assumption about the disk temperature and HCO$^+$
opacity. For 17 sources, the radii match to better than 20\%, while for the 10 remaining ones, $R(\mathrm{HCO}^+)$ is
larger (3 sources have no other reliable radius).

If the HCO$^+$(3-2) was optically thin, lower values of $\rho$ should be used, which would
imply \textit{larger} disk radii, becoming inconsistent with the CN radii, and even exceeding
the radii derived from CO.
This simple exercice shows that the HCO$^+$ J=3-2 transition is mostly optically thick in all disks.
As a consequence, it implies that the HCO$^+$ content cannot be simply determined from this transition
(which only sets a lower limit),
and that direct comparison of intensity ratios between HCO$^+$(3-2) and other transitions will be meaningless.
A more complete disk model, as done in Sec.\ref{sec:sub:model}
is required for HCO$^+$.
On the contrary, all other transitions in general have significantly smaller flux
(\footnote{except for HCN in LkCa 15 and perhaps DN Tau}),
and thus must be optically thin. Note that for C$_2$H, one should
consider for this the line flux per hyperfine component, which is only 0.3 times the integrated line flux given in
Table \ref{tab:hco-hcn}. The same remark also applies to CN lines.

\begin{table*}[!t]
\caption{Interferometric Modeling of CN N=2-1}
\label{tab:inter}
\begin{tabular}{rrr rrrrrr}
\hline
Source & R$_\mathrm{inter}$ &  i         & T$_{100}$ & $p$ & $\Sigma$ (CN)         & Data \\
       & (au)             & ($^\circ$) & (K)       &     & ($10^{12}$ cm$^{-2}$ )& reference \\
\hline
\textbf{\textit{04302+2247}} & 
  $ > 700$    & 
  65 $\pm$ 2 & 
  [17]   & 
  0.8 $\pm$ 0.4 & %
  6 $\pm$ 2 & 1 \\ 
CI Tau & 
      530 $\pm$ 20 & 
       50 $\pm$ 2  & 
     11.6 $\pm$ 1.6 & 
      1.7 $\pm$ 0.2 & 
      24 $\pm$  6 & 1   \\ 
CY Tau & 
  290  $\pm$ 20 & 
    22 $\pm$ 2 & 
  17.1 $\pm$ 1.8 & %
   2.0 $\pm$ 0.1 & %
   9.2 $\pm$ 2.6 & 1 \\
DL Tau   & 
  460 $\pm$ 6 & 
   43 $\pm$ 2 & 
 13.6 $\pm$ 1.2 & 
  0.5 $\pm$ 0.1 & 
   24 $\pm$ 6 & 1 \\ 
DM Tau $^{(a)}$ & 
  620 $\pm$ 15 &
  35 $\pm$ 5 &
  13 $\pm$ 1 &
  0.6 $\pm$ 0.1 &
  58 $\pm$ 5 & 2 \\
DN Tau  & 
   230 $\pm$ 10  & %
   30 $\pm$ 3 & 
  11.5 $\pm$ 0.9 & 
   1.3 $\pm$ 0.3 & 
    17 $\pm$ 4 & 1 \\ 
GM Aur & 
       370 $\pm$       80  & 
        43 $\pm$        8 & 
      17.5 $\pm$       3.5 & 
       1.0 $\pm$       0.3 & 
        20 $\pm$ 10  & 3 \\ 
GO Tau & 
  590 $\pm$ 60 & 
   55 $\pm$ 1  & 
 11.4 $\pm$ 0.3 & 
  0.8 $\pm$ 0.1 & 
   22 $\pm$ 5 & 1 \\ 
HV Tau C & 
 290 $\pm$ 25 & 
  87 $\pm$ 2 & 
  13 $\pm$ 2 & 
  1.6 $\pm$ 0.2 & 
  9 $\pm$ 3 & 1 \\ 
IQ Tau & 
  220 $\pm$ 20  & 
   56 $\pm$ 4   & 
 13.2 $\pm$ 0.8 & 
  0.8 $\pm$ 0.4 & 
   35 $\pm$ 9 & 1 \\ 
LkCa15 & 475 $\pm$ 15 & 50 $\pm$ 1 & 15.5 $\pm$ 1 & 1.10 $\pm$ 0.07 & 51 $\pm$ 6  & 2 \\
LkCa15 $^{(b)}$ & 
  630 $\pm$ 35 &
  55 $\pm$ 5 &
  13 $\pm$ 2 &
   0.8 $\pm$ 0.1 &
   58 $\pm$5 & 2 \\
 \textbf{MWC 480}  & %
  545 $\pm$ 35 &
    37 &
   22 $\pm$ 4 &
   2.1 $\pm$ 0.1 &
   10.4 $\pm$ 0.9 & 2 \\ 
\hline
\end{tabular}
\tablefoot{Column densities at 300 au are in units of $10^{12}$cm$^{-2}$.
Temperatures are given at 100 au. (a) exponent of temperature law
was fitted as $q=0.60 \pm 0.05$, (b) exponent of temperature law
was fitted as $q=0.95 \pm 0.05$. References: 1) \citet{Guilloteau+etal_2014}, 2) \citet{Chapillon+etal_2012},
3) this work.}
\end{table*}

\paragraph{Discrepant radii: warmer and/or younger sources ?\\}
\hspace{2.0ex} Where a difference exists, the HCO$^+$-based radius is larger than the interferometric measurement.
Measurements based on CN only give a lower limit to the disk radius, because the CN N=2-1 line
may be sub-thermally excited in the outer disk, while HCO$^+$ is more easily thermalized.
However, in some cases, the HCO$^+$-based radius is larger than the radius measured in CO.
To reconcile both values, we can thus
either (1) increase $\rho \Delta V$ or (2) increase the temperature $T_0$ in Eq.\ref{eq:main-rout}.
$R_\mathrm{out}$ would
only scale as $1/\log(\tau)$ in case (1), so the required increase in $\tau$ might
conflict with the upper limit on H$^{13}$CO$^+$. Thus case (2), where $R_\mathrm{out}$
scales as $1/\sqrt(T)$,
is more likely.  In fact, such differences are most prominent on smaller disks, which
are probably on average warmer than larger sources, as the temperature drops with radius roughly
as $r^{-q}$ with $q \approx 0.5 - 0.6$ in disks. This could apply to
DL Tau and CY Tau, which would need to be a factor 2 warmer to reconcile radii.
It could be also that the local line width $\Delta V$ is larger in smaller disks.

In all cases, this is insufficient for Haro 6-33, where we already assumed a relatively
warm disk, or Haro 6-13 where a factor 4 in temperature would be needed.
We note in this respect that both sources are heavily confused, so it is possible
that the CO radius derived by \citet{Schaefer+etal_2009} is biased towards low values
because one cannot get the full disk extent near the systemic velocity.

It is also worth noting that Haro 6-33 has the strongest CS/CN line ratio compared to any other
source, reflecting perhaps an unusual chemistry. Its relatively large content in S-bearing
molecules is also attested by the detection of SO lines. Contamination by an outflow
is not fully excluded.

\subsection{Radii from line profiles}
\label{sec:sub:profile}
In any case, the radii given in Table \ref{tab:radii} are only
very simple estimates. Another approach consists in deriving the outer
radii from the separation of the two velocity peaks in the line profile,
$\Delta V_\mathrm{peak}$.
Because of the Keplerian velocity field, this separation is
\begin{equation}
\Delta V_\mathrm{peak} = 2 \sqrt{\frac{G M_*}{R_\mathrm{out}}} \sin{i}
\label{eq:vout}
\end{equation}
so if the stellar mass $M_*$ and disk inclination $i$ are
reasonably well known, the outer radius can be estimated. Instead
of using  Eq.\ref{eq:vout}
directly, we develop in the next section an analysis based on a
complete disk modelling to simulate the line profiles. This is possible
for sources for which sufficient interferometric data (or other
ancillary information) allow to constrain the basic disk geometry,
as detailed below.

\begin{table*}
\caption{Modeled column densities from the 30-m data}
\label{tab:cd}
%
\begin{tabular}{rrr rrrrrrr}
\hline
Source & R$_\mathrm{disk}$ & T$_{100}$ & HCO$^+$ & HCN & CN & CCH & CS & SO & H$_2$CO\\
       & (au) &  (K) & \multicolumn{7}{c}{$(10^{12}$ cm$^{-2}$)} \\
\hline
\textbf{\textit{04302+2247}}  &       500. &        25. &
       5.6 $\pm$       0.6 &
      0.73 $\pm$      0.05 &
       5.7 $\pm$       0.5 &
       2.6 $\pm$       1.2 &
       6.7 $\pm$       0.9 &
      16.0 $\pm$       1.0 &
       6.6 $\pm$       0.3 \\ 

AA Tau &       350. &        13. &
       1.4 $\pm$       0.1 &
       1.1 $\pm$       0.2 &
      10.5 $\pm$       0.7 &
      13.4 $\pm$       2.0 &
       9.4 $\pm$       3.1 &
 $<$        7.0 &
       0.9 $\pm$       0.3 \\ 

\textbf{AB Aur} &       600. &        30. &
       0.9 $\pm$       0.0 &
       0.4 $\pm$       0.0 &
       1.8 $\pm$       0.4 &
       6.3 $\pm$       1.4 &
       6.3 $\pm$       1.4 &
       2.8 $\pm$       0.9 &
       1.6 $\pm$       0.2 \\ 

CI Tau &       520. &        12. &
       3.9 $\pm$       0.4 &
       0.7 $\pm$       0.1 &
      13.6 $\pm$       0.7 &
      20.4 $\pm$       2.0 &
      20.9 $\pm$       5.9 &
       9.0 $\pm$       3.7 \\ 

\textbf{CQ Tau} &       200. &        50. &
       0.35 $\pm$      0.07 &
       0.27 $\pm$      0.10 &
       4.2 $\pm$       1.1 &
 $<$        7.5 &
 $<$        1.8 &
 $<$        3.1 &
       3.3 $\pm$       0.5 \\ 

CW Tau \\

CY Tau &       300. &        18. &
       1.7 $\pm$       0.1 &
       0.5 $\pm$       0.1 &
      11.9 $\pm$       0.7 &
      18.9 $\pm$       2.0 &
       2.8 $\pm$       0.6 &
 $<$        2.9 \\ 

DL Tau &       460. &        16. &
       3.0 $\pm$       0.1 &
       0.7 $\pm$       0.1 &
      10.4 $\pm$       0.5 &
       9.7 $\pm$       1.1 &
       4.2 $\pm$       1.3 &
 $<$        4.3 &
       1.9 $\pm$       0.2 \\ 
DM Tau &       600. &        13. &
      10.5 $\pm$       0.8 &
       2.6 $\pm$       0.2 &
      43.7 $\pm$       2.0 &
      43.1 $\pm$       1.9 &
       3.7 $\pm$       1.2 &
 $<$        4.6 \\ 
DN Tau &       230. &        12. &
       2.2 $\pm$       0.5 &
       1.9 $\pm$       0.4 &
      18.4 $\pm$       1.7 &
      29.8 $\pm$       3.5 &
 $<$        5.0 &
 $<$        8.7 &
       1.5 $\pm$       0.5 \\ 
\textit{DO Tau} &       350. &        15. &
       0.5 $\pm$       0.1 &
       0.2 $\pm$       0.1 &
       5.3 $\pm$       0.7 &
 $<$        4.1 &
 $<$        3.0 &
 $<$        4.2 &
 $<$        0.9 \\ 

FT Tau &       200. &        20. &
       0.7 $\pm$       0.1 &
       0.2 $\pm$       0.1 &
       6.6 $\pm$       0.8 &
      10.7 $\pm$       2.5 &
 $<$        3.0 &
 $<$        3.4 &
 $<$        0.7 \\ 

GG Tau &       500. &        25. &
       2.7 $\pm$       0.1 &
       1.7 $\pm$       0.1 &
      17.3 $\pm$       1.1 &
      22.1 $\pm$       1.1 &
       8.6 $\pm$       1.0 &
 $<$        2.6 &
       3.0 $\pm$       0.5 \\ 
GM Aur &       400. &        20. &
       2.0 $\pm$       0.1 &
       0.8 $\pm$       0.1 &
       8.0 $\pm$       0.3 &
       4.5 $\pm$       1.1 &
       4.3 $\pm$       0.5 &
       7.4 $\pm$       1.2 \\ 
GO Tau &       600. &        13. &
      19.3 $\pm$       1.7 &
       3.3 $\pm$       0.3 &
      35.1 $\pm$       1.4 &
      85.3 $\pm$       3.2 &
       4.4 $\pm$       0.9 &
 $<$        5.6 \\ 
\textit{Haro6-13 } &       640. &        18. &
       0.55 $\pm$       0.06 &
       0.20 $\pm$       0.04 &
       1.2 $\pm$       0.2 &
 $<$        2.6 &
       1.9 $\pm$       0.8 &
       7.6 $\pm$       0.9 &
       1.7 $\pm$       0.2 \\ 
\textit{Haro6-33 } &       300. &        18. &
       4.4 $\pm$       0.6 &
       0.5 $\pm$       0.1 &
       5.1 $\pm$       0.6 &
 $<$        4.4 &
      19.6 $\pm$       3.9 &
       9.3 $\pm$       1.4 &
       2.5 $\pm$       0.4 \\ 
\textit{Haro6-5b } &       300. &        18. &
       2.3 $\pm$       0.2 &
       0.3 $\pm$       0.1 &
       4.3 $\pm$       0.5 &
       7.1 $\pm$       2.2 &
 $<$        2.5 &
 $<$        4.5 &
       1.0 $\pm$       0.2 \\ 
\textit{HH 30} \\
HK Tau$^{(a)}$ &       100. &        25. &
      15.0 $\pm$       0.1 &
       2.8 $\pm$       0.9 &
      48.6 $\pm$       8.2 &
 $<$       25.4 &
 $<$       23.8 &
 $<$       10.0 &
       4.4 $\pm$       2.5 \\ 
HV Tau C &       300. &        13. &
       9.6 $\pm$       2.4 &
       0.4 $\pm$       0.2 &
      12.3 $\pm$       1.5 &
       8.7 $\pm$       2.9 &
 $<$        4.9 &
 $<$        8.7 &
 $<$        1.2 \\ 
IQ Tau &       220. &        13. &
      12.8 $\pm$       4.1 &
       1.0 $\pm$       0.2 &
      23.9 $\pm$       2.4 &
      27.0 $\pm$       4.1 &
 $<$       65.8 &
 $<$        9.5 &
       1.4 $\pm$       0.6 \\ 
LkH$\alpha$ 358 \\
LkCa15 &       500. &        13. &
       9.0 $\pm$       0.8 &
       7.8 $\pm$       0.7 &
      82.0 $\pm$       7.7 &
      32.4 $\pm$       1.9 &
      13.6 $\pm$       1.7 &
 $<$        9.0 \\ 
\textbf{MWC 480} &       450. &        23. &
       1.8 $\pm$       0.1 &
       0.5 $\pm$       0.1 &
      12.1 $\pm$       0.5 &
      10.2 $\pm$       1.0 &
       1.2 $\pm$       0.4 &
 $<$        2.1 &
 $<$        0.5 \\ 
\textbf{MWC 758} &       250. &        30. &
       0.25 $\pm$      0.03 &
       0.13 $\pm$      0.05 &
 $<$        3.7 &
 $<$        3.7 &
 $<$        1.2 &
 $<$        1.9 &
 $<$        1.4 \\ 

RW Aur \\
RY Tau &       210. &        30. &
       0.9 $\pm$       0.1 &
       1.0 $\pm$       0.2 &
      29.4 $\pm$       5.5 &
      16.4 $\pm$       3.4 &
 $<$        3.0 &
 $<$        5.7 &
 $<$        0.5 \\ 

SU Aur \\
UZ Tau E &       210. &        23. &
       0.7 $\pm$       0.1 &
       0.5 $\pm$       0.1 &
      10.5 $\pm$       1.1 &
       4.7 $\pm$       2.2 &
       5.2 $\pm$       2.6 &
 $<$        3.8 &
 $<$        1.0 \\ 

\hline
\end{tabular}
%
\tablefoot{Column densities at 300 au are in units of $10^{12}$cm$^{-2}$.
Disk radii (see Sect.\ref{sec:sub:def})} are in au, and assumed temperatures at 100 au in K.
(a) using the upper limit from H$^{13}$CO$^+$: the HCO$^+$ line is
completely optically thick
\end{table*}

\begin{figure*}[!ht]
\begin{center}
\includegraphics[height=22.0cm]{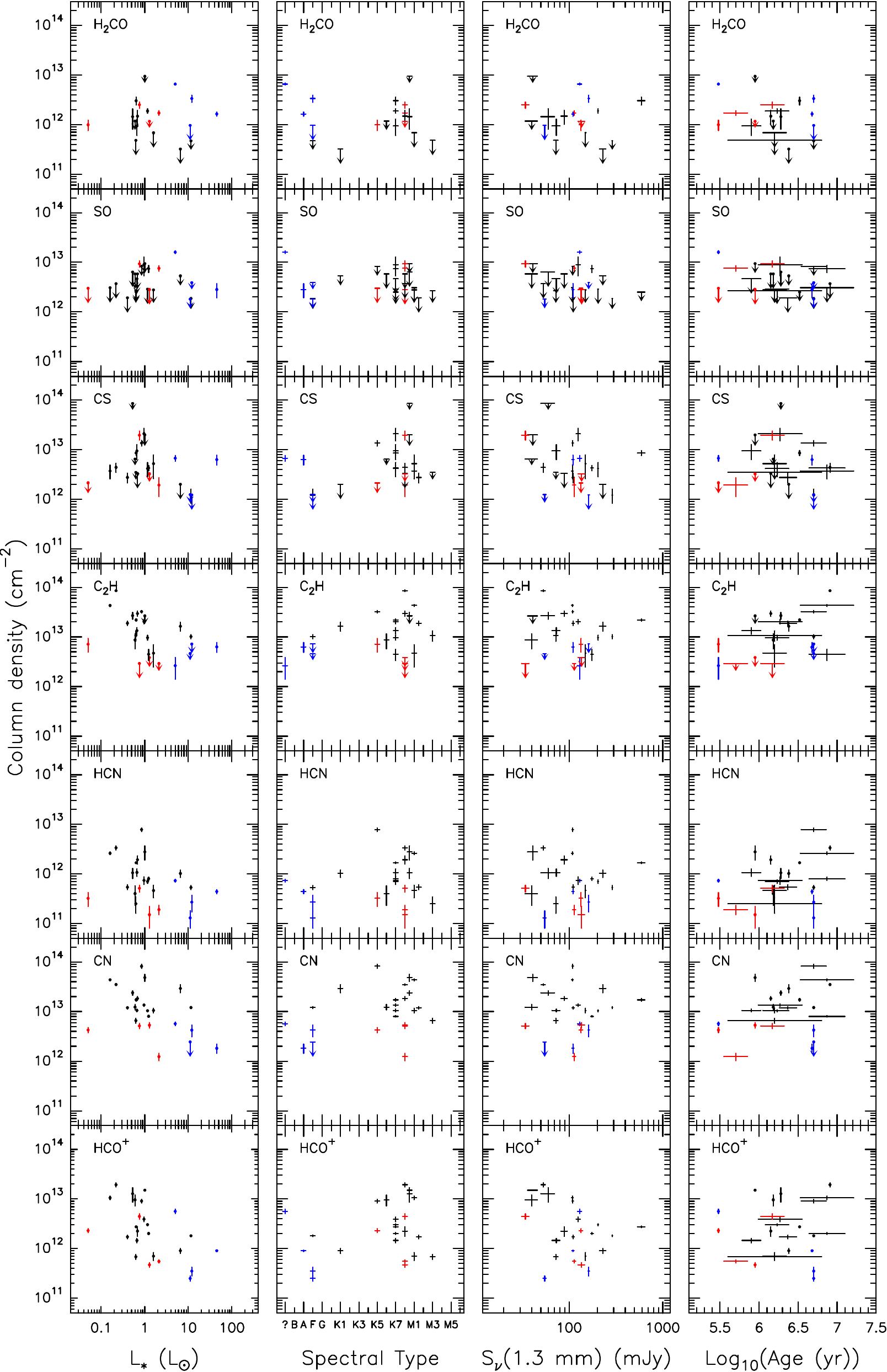}
\end{center}
\caption{Correlation plot of the column density of molecules
at 300 au with stellar luminosity, Spectral Type, 1.3\,mm flux
density and stellar age. Embedded objects appear in red,
HAe stars in blue, and T Tauri stars in black. Errorbars are $\pm 1 \sigma$,
with limits at $2 \sigma$ indicated by arrows.}
\label{fig:correl-cd}
\end{figure*}

\subsection{From Flux Densities to Molecular Column Densities}
\label{sec:sub:model}

When the basic disk geometry: source velocity, disk inclination and
star dynamical mass is known, a simple power law disk model can
allow us to retrieve other disk parameters, such as the (excitation)
temperature radial profile (for mostly optically thick lines)
or the molecular surface density radial profile (for optically
thin lines). The assumed profiles are
\begin{equation}
\Sigma(r) = \Sigma_{300} (r/300\,\mathrm{au})^{-p}
\end{equation}
\begin{equation}
T(r) = T_{100} (r/100\,\mathrm{au})^{-q}
\end{equation}
for the molecular surface density and (rotation) temperature respectively.
The choice of the pivot, 300 and 100 au, is made
to minimize the errors on the values at this radius, as discussed
in \citet{Pietu+etal_2007}.

Even in case of low S/N, usefull limits on
molecular column densities can be obtained by assuming, for example,
that the (excitation/rotation) temperature of the molecule is given from
other measurements, either CO or HCO$^+$ transitions. Some assumption
about the local line width is required, but it only
matters for substantially optically thick lines.

  We use the DiskFit tool \citep{Pietu+etal_2007}
to adjust the observed line profiles, following the method used
by \citet{Dutrey+etal_2011} for the study of S-bearing molecules.
DiskFit minimizes a $\chi^2$ function in the $uv$ plane,
comparing the measured visibilities (or line flux in case
of single-dish profiles) to those predicted by the ray tracing code.
Each visibility is weighted according to the theoretical noise
derived from system temperature, channel width and integration time.
The minimizations are performed using a modified Levenberg-Marquardt
and the errors derived from the covariance matrix.

We assume the exponent of the temperature profile is $q=0.4$.
The local line width is set to 0.16 or 0.2 $\kms$; HK Tau required
a larger value to best fit the profiles, but this local line
width has no impact on the derived column densities.
In a first step, to determine the systemic properties (LSR
velocity, inclination $i$ and stellar mass $M_*$), we used the interferometric data,
leaving $p$, the temperature $T$ and the outer radius $R_\mathrm{out}$
as free parameters. Results obtained from CN are given in Table \ref{tab:inter}.

The derived values of $i,T,M_*$ and
$R_\mathrm{out}$ were used for single-dish analysis, for which we fix $p=1$.
The exponent $p$ has little impact on
the column densities at 300 au. $p=1$ is
within the errors of the values found from the interferometer
data: the weighted mean of values in Table \ref{tab:inter} is $1.15 \pm 0.16$.
The temperature value $T$
also has limited influence, because the 2-1 and 3-2 transitions of
the observed molecules are
mostly sensitive to the column density (because of the compensation
between excitation temperature and partition function for
temperatures between 10 and 25 K \citep[see e.g.][their Fig.4]{Dartois+etal_2003}).
A possible exception is the CS J=5-4 line which becomes sensitive to
the temperature below 15-20 K. In this respect, the small (factor 2)
difference between the column densities given for DM Tau by \citet{Dutrey+etal_2011}
from the J=3-2 line and found here could be due to a somewhat larger
(rotation) temperature than assumed.

The single-dish spectra are then constrained by a single free parameter:
the molecular column density at 300 au, $\Sigma$. When no CN interferometric
data is available, the temperature is taken from CO interferometric data
if possible (which may overestimate the actual rotation temperature
of most molecules, because of vertical temperature gradients in disks),
or simply assumed. For disks without CN interferometric data,
$R_\mathrm{out}$ is also adjusted to best fit the HCO$^+$ spectrum.

The sources which are well suited for this study are
\object{CY Tau}, \object{HV Tau C}, \object{IQ Tau}, \object{DL Tau}, \object{DM Tau},
\object{GG Tau}, \object{UZ Tau}, \object{CI Tau}, \object{DN Tau}, \object{LkCa 15},
\object{GO Tau}, \object{GM Aur}, \object{MWC 480}, the Butterfly star, \object{04302+2247}.
\object{Haro 6-13} and \object{Haro 6-33} can be studied with less precision, because
of the limited accuracy on their disk parameters from the CO data
of \citet{Schaefer+etal_2009}.  A similar caution should be applied
to \object{AB Aur} because of the source complexity \citep{Pietu+etal_2005}.
The small disk sizes of \object{CQ Tau} and \object{MWC 758}
also result in limited accuracy.

Despite a lack of accurate sizes from interferometric data, we also consider RY Tau, HK Tau, AA Tau, Haro 6-5b, and DO
Tau because all have sufficiently well constrained inclinations to allow some estimate of disk sizes from the HCO$^+$
spectra. We assumed a stellar mass of $2.2 \Msun$ for RY Tau and a systemic velocity of $6.75 \kms$. For HK Tau, we
assume all the emission is coming from HK Tau B, because it is the most massive and most inclined disk of the system,
and thus better fits the wide lines which are detected.

We exclude CW Tau, HH 30 and LkH$\alpha$ 358 from the analysis because of contamination by the cloud, and also SU Aur
and RW Aur due to their peculiar nature and very small disks. Little can be said for very small disks ($\leq 150$ au),
as only substantially optically thick lines are detectable in this case.


The fitted lines include: HCO$^+$(3-2), HCN(3-2), CN(2-1), C$_2$H(3-2), CS(5-4)
plus the SO and (ortho)-H$_2$CO lines when detected. We do not consider
$^{13}$CO(2-1) for two reasons. First, for many sources, the IRAM 30-m telescope spectra
are affected by strong contamination from the surrounding molecular cloud.
Second, when interferometric data are available (DM Tau, LkCa 15 and MWC\,480,
\citet{Pietu+etal_2007}, GM Aur, \citet{Dutrey+etal_2011}),
the exponent of the surface density law, $p$, is often much steeper
($p > 2-3$) than for CN or HCO$^+$. This steep exponent probably reflects
only the very outer parts of the disks, where the $^{13}$CO lines are
optically thin, but transitions from other molecules are no longer excited.

Results for column densities are given in Tables \ref{tab:cd}
and ratios in Table \ref{tab:ratio}.

\subsection{C$_2$H multi-transition analysis}

For C$_2$H, three sources have been previously imaged with
the IRAM PdBI array in the  N=1-0 and 2-1 lines at
$\sim 3''$ resolution: DM Tau,
LkCa 15 and MWC\,480 by \citet{Henning+etal_2010}.
For DM Tau and LkCa 15 our column densities
derived from the N=3-2 lines under the simple hypothesis above are a
factor 2 larger than those of \citet{Henning+etal_2010}. This probably
reflects a somewhat larger excitation temperature than assumed.
The results, provided by DiskFit as described in Sect.\ref{sec:sub:model},
using this data combined with the N=3-2 spectra are given in Table \ref{tab:cch},
assuming $q = 0.4$. The derived temperatures, 15 and 16 K, are indeed slightly
above those quoted by \citet{Henning+etal_2010}, 11 and 10 K respectively
for DM Tau and LkCa 15. In MWC 480, the interferometric data has
a low Signal to Noise ratio and thus a limited influence on the global fit.

In addition, the N=1-0 and 2-1 transitions of C$_2$H were observed at the
IRAM 30-m telescope for CI Tau, CY Tau and GO Tau. Lines in GO Tau are well detected,
but those in CI Tau and CY Tau only marginally. These spectra
were used in a multi-line fit to derive the C$_2$H column density
in these three sources, with results given  in Table \ref{tab:cch}.
Overall, the fit results are in remarkable agreement with those
of the CN interferometer data, in terms of disk radii but also
excitation temperatures. The excitation temperatures
derived from CN or C$_2$H also agree with the values assumed
for HCO$^+$. This gives credit to our simple
assumption for temperatures, and thus to the disk radii derived solely from HCO$^+$
in other sources.

We note that excitation temperatures and column densities
for C$_2$H in LkCa15 found in this detailed disk modelling are
quite different from those derived by \citet{Punzi+etal_2015}
from the N=3-2 IRAM 30-m telescope spectra only. This is presumably because
\citet{Punzi+etal_2015} assumed a uniform beam filling factor
in the 30-m telescope beam, of order 1 given the outer radius of the disk,
$\sim 500$\,au, and thus derive very low excitation temperatures (as the opacity
of each of the detected hyperfine component is $\sim 0.5$
from the hyperfine ratios). Consequently, these underestimated temperatures
result in high column densities because of the partition function at very low
temperatures.

However, in a Keplerian disk, the filling
factor is limited to less than $\sim \delta v/V_\mathrm{kep}(R_\mathrm{out})$
by the Keplerian shear, and is affected by the $\cos(i)$ factor due to
the inclination \citep[e.g.][their Appendix D]{Horne+Marsh_1986,Guilloteau+etal_2013}.
In the case of LkCa 15, this filling factor is at most  $\sim 0.05$.
The full multi-line disk modelling explicitely accounts for the details
of the line formation mechanism, and provides reliable estimate of the column
density and excitation (though of course under the power-law approximation).

\begin{table*}
\caption{Multi-transition modeling of C$_2$H}
\label{tab:cch}
\begin{tabular}{rcc ccc}
\hline
Source & R$_\mathrm{inter}$ &  i & T$_{100}$ & $p$ & $\Sigma$ (C$_2$H)  \\
      & (au)           & ($^\circ$) & (K) &  & $10^{12}$\,cm$^{-2}$ \\
\hline
\hline
\multicolumn{6}{c}{Combined Interferometric \& 30-m multi-transition analysis} \\
\hline
DM Tau (a) & 
  560 $\pm$ 40 &
  35 &
  14.0 $\pm$ 1.1 &
  1.5 $\pm$ 0.2 &
  13 $\pm$ 2 \\
LkCa15 & 
  600 $\pm$ 40 &
  52 &
  16 $\pm$ 2 &
   0.7 $\pm$ 0.3 &
   18 $\pm$2 \\
MWC 480 &
  400 $\pm$ 40 &
  37 &
  14 $\pm$ 4 &
   1 &
   9.3 $\pm$ 1.8 \\
%
\hline
\hline
\multicolumn{6}{c}{30-m multi-transition analysis} \\
\hline
      & R$_\mathrm{single}$ \\ 
\hline
CI Tau &
  520  &
  51 &
  13 $\pm$ 1 &
   1 &
  14  $\pm$ 2 \\
CY Tau &
  300  &
  21 &
  27 $\pm$ 4 &
   1 &
  14  $\pm$ 4 \\
GO Tau &
  800 $\pm$ 40  &
  55 &
  12.1 $\pm$ 0.3 &
   1 &
  74  $\pm$ 4 \\
\hline
\end{tabular}
\tablefoot{C$_2$H column densities at 300 au are in units of $10^{12}$cm$^{-2}$.
Temperatures are given at 100 au: note that \citet{Henning+etal_2010} gave temperatures
at 300 au, which are smaller by a factor 1.55 since we assume $q=0.4$.}
\end{table*}

\begin{table*}
\caption{Column density ratios at 300 au}
\label{tab:ratio}
\begin{tabular}{r rrrrrr}
\hline
Source & HCO$^+$/CN & HCN/CN & CCH/CN & CS/CN & SO/CN & H$_2$CO/CN\\
\hline
\hline
\textbf{\textit{04302+2247}} &
      0.99 $\pm$       0.18 &
     0.129 $\pm$     0.019 &
       0.5 $\pm$        0.3 &
       1.2 $\pm$        0.3 &
       2.8 $\pm$        0.4 &
      1.16 $\pm$       0.15 \\ 
AA Tau &
      0.14 $\pm$       0.02 &
      0.10 $\pm$       0.02 &
       1.3 $\pm$        0.3 &
       0.9 $\pm$        0.4 &
 $<$       0.44 &
     0.09 $\pm$     0.04 \\ 
\textbf{AB Aur} &
      0.50 $\pm$       0.12 &
      0.24 $\pm$       0.08 &
       3.5 $\pm$        1.5 &
 $<$      0.000 &
       1.5 $\pm$        0.8 &
       0.9 $\pm$        0.3 \\ 
CI Tau &
      0.29 $\pm$       0.05 &
     0.054 $\pm$     0.013 &
       1.5 $\pm$        0.2 &
       1.5 $\pm$        0.5 &
       0.7 $\pm$        0.3 \\ 

\textbf{CQ Tau} &
      0.08 $\pm$       0.04 &
      0.06 $\pm$       0.04 &
       0.5 $\pm$        0.7 &
 $<$       0.29 &
       0.4 $\pm$        0.4 &
       0.8 $\pm$        0.3 \\ 
CW Tau \\
CY Tau &
     0.144 $\pm$     0.019 &
     0.045 $\pm$     0.008 &
       1.6 $\pm$        0.3 &
      0.23 $\pm$       0.06 &
 $<$       0.16 \\ 
DL Tau &
      0.29 $\pm$       0.03 &
     0.067 $\pm$     0.009 &
      0.93 $\pm$       0.15 &
      0.40 $\pm$       0.14 &
 $<$       0.28 &
      0.18 $\pm$       0.03 \\ 
DM Tau &
      0.24 $\pm$       0.03 &
     0.059 $\pm$     0.007 &
      0.99 $\pm$       0.09 &
      0.08 $\pm$       0.03 &
 $<$       0.07 \\ 
DN Tau &
      0.12 $\pm$       0.04 &
      0.10 $\pm$       0.03 &
       1.6 $\pm$        0.3 &
 $<$       0.18 &
 $<$       0.32 &
      0.08 $\pm$       0.03 \\ 
\textit{DO Tau} &
      0.09 $\pm$       0.02 &
     0.029 $\pm$     0.017 &
       0.2 $\pm$        0.3 &
       0.2 $\pm$        0.2 &
 $<$       0.53 &
      0.11 $\pm$       0.07 \\ 

FT Tau &
      0.10 $\pm$       0.02 &
     0.037 $\pm$     0.018 &
       1.6 $\pm$        0.6 &
      0.23 $\pm$       0.18 &
      0.06 $\pm$       0.18 &
 $<$       0.07 \\ 
GG Tau &
     0.158 $\pm$     0.016 &
     0.096 $\pm$     0.010 &
      1.27 $\pm$       0.14 &
      0.49 $\pm$       0.09 &
      0.04 $\pm$       0.05 &
      0.17 $\pm$       0.04 \\ 
GM Aur &
     0.250 $\pm$     0.020 &
     0.099 $\pm$     0.011 &
      0.56 $\pm$       0.16 &
      0.53 $\pm$       0.08 &
      0.92 $\pm$       0.18 \\ 
GO Tau &
      0.55 $\pm$       0.07 &
     0.095 $\pm$     0.012 &
      2.43 $\pm$       0.19 &
      0.13 $\pm$       0.03 &
 $<$       0.11 \\ 

\textit{Haro6-13 } &
      0.44 $\pm$       0.10 &
      0.15 $\pm$       0.05 &
       1.0 $\pm$        0.9 &
       1.6 $\pm$        0.9 &
       6.1 $\pm$        1.8 &
       1.4 $\pm$        0.4 \\ 
\textit{Haro6-33 } &
       0.9 $\pm$        0.2 &
      0.10 $\pm$       0.03 &
 $<$       0.57 &
       3.8 $\pm$        1.2 &
       1.8 $\pm$        0.5 &
      0.49 $\pm$       0.13 \\ 
\textit{Haro6-5b } &
      0.54 $\pm$       0.11 &
      0.08 $\pm$       0.03 &
       1.7 $\pm$        0.7 &
       0.1 $\pm$        0.2 &
 $<$       0.70 &
      0.23 $\pm$       0.08 \\ 
\textit{HH 30} \\
HK Tau &
       2.5 $\pm$        1.5 &
      0.06 $\pm$       0.03 &
       0.2 $\pm$        0.2 &
      0.08 $\pm$       0.18 &
      0.06 $\pm$       0.08 &
     0.10 $\pm$     0.05 \\ 
HV Tau-C &
       0.8 $\pm$        0.3 &
     0.032 $\pm$     0.018 &
       0.7 $\pm$        0.3 &
      0.26 $\pm$       0.16 &
 $<$       0.47 &
      0.03 $\pm$       0.04 \\ 

IQ Tau &
       0.5 $\pm$        0.2 &
     0.044 $\pm$     0.015 &
       1.1 $\pm$        0.3 &
       1.7 $\pm$        1.1 &
 $<$       0.27 &
      0.06 $\pm$       0.03 \\ 
LkH$\alpha$ 358 \\

LkCa15 &
     0.110 $\pm$     0.020 &
     0.095 $\pm$     0.017 &
      0.39 $\pm$       0.06 &
      0.17 $\pm$       0.04 &
      0.03 $\pm$       0.04 \\ 

\textbf{MWC 480} &
     0.150 $\pm$     0.011 &
     0.044 $\pm$     0.006 &
      0.84 $\pm$       0.12 &
      0.10 $\pm$       0.04 &
      0.03 $\pm$       0.06 &
     0.013 $\pm$     0.014 \\ 
\textbf{MWC 758} \\ 

RW Aur \\
RY Tau &
     0.031 $\pm$     0.009 &
     0.035 $\pm$     0.012 &
       0.6 $\pm$        0.2 &
 $<$       0.07 &
      0.05 $\pm$       0.07 &
 $<$      0.011 \\ 

SU Aur \\

UZ Tau E &
     0.067 $\pm$     0.019 &
     0.043 $\pm$     0.017 &
       0.4 $\pm$        0.3 &
       0.5 $\pm$        0.3 &
      0.02 $\pm$       0.12 &
 $<$       0.06 \\ 
\hline
\end{tabular}
\end{table*}

\section{Individual Sources}

This section summarizes specific information required for several sources
to interpret the results of Tables \ref{tab:lines}-\ref{tab:cch-cs}, and
of the detailed analysis presented in Sect.\ref{sec:sub:model}.
It also provides new informations on previously unknown or
poorly characterized gas disks.

\subsection{\object{CW Tau}}
\label{sec:obj:cw_tau}
From the $^{13}$CO PdBI data of \citet{Pietu+etal_2014}, the systemic velocity is 6.45 km/s,
and the best fit model leads to an integrated line flux of 2 Jy.km/s. These values are consistent
with those derived from the IRAM 30-m telescope measurements of \citet{Guilloteau+etal_2013}, who showed that
contamination by the molecular cloud appears in CN and H$_2$CO at a systemic velocity of 6.80 km/s.
Here, the HCO$^+$(3-2) spectrum is consistent
with the redshifted part being obscured by the molecular cloud, leaving only the blue-shifted
wing detected at the 4 $\sigma$ level.

\subsection{DO Tau}
\label{sec:obj:do_tau}
The properties of \object{DO Tau} are poorly known. \citet{Koerner+Sargent_1995} detected CO emission,
but with a very broad blueshifted
wing which probably traces an outflow, and may indicate that this source is quite young.
Indeed, it is surrounded by an optical nebulosity along a redshifted optical jet \citep[see][and references
therein]{Itoh+etal_2008}. \citet{Kwon+etal_2015} obtained a measurement of the disk inclination $i = 32 \pm2^\circ$ through high
resolution imaging with CARMA. The stellar mass is not known: we assumed $M_* = 0.7 \Msun$ in our
analysis. The HCO$^+$  and CN line profiles are then consistent with an outer disk radius of 350 au
and a systemic velocity of 6.5 $\kms$. The CS data indicate a possible $\sim 5 \sigma$ detection,
but the derived velocity, $\sim 4.3 \kms$, does not agree with the better determination from
HCO$^+$ or CN, so we consider this as noise.

\subsection{\object{RY Tau}}
\label{sec:obj:ry_tau}
The HCO$^+$ detection is convincing (6 $\sigma$ level), but the line is very broad. The
centroid velocity and line width are consistent with the tentative detection of CN by
\citet{Guilloteau+etal_2013}. Assuming similar line parameters, the HCN detection is
marginal (4 $\sigma$). CS is not detected. CO observations at the IRAM
interferometer \citep{Coffey+etal_2015} indicate a LSR velocity of 6.75 $\kms$, and
the system is highly inclined (70$^\circ$). The large line width is consistent
with a 2 $\Msun$ star indicated by the spectral type F8, provided the disk
is not too large ($< 300$ au).

\subsection{Haro 6-5 B}
\label{sec:obj:haro6-5b}
This nearly edge-on object is very young, and still embedded in a substantial envelope.
The HCO$^+$ line width is somewhat larger than the CN line width, which may indicate
that part of the HCO$^+$ emission comes from an outflow, as \object{Haro 6-5 B} drives a powerful
optical jet \citep{Liu+etal_2012}. \citet{Yokogawa+etal_2002} suggested that star is
a quite low-mass object, $0.25 \Msun$, based on $5''$ resolution images of $^{13}$CO, but our HCO$^+$ spectrum,
if arising from a disk only, indicates a $1 \Msun$ star.

\subsection{LkH$\alpha$ 358}
\label{sec:obj:lkha_358}
There is a tentative HCO$^+$ detection of a very narrow line, which is not compatible
with the compact CO disk detected by \citet{Schaefer+etal_2009}. There is strong confusion
in CO isotopologues (including C$^{17}$O) towards this line of sight.
The HCO$^+$ profiles however matches the $^{13}$CO profile
obtained by \citet{Guilloteau+etal_2013}. $^{13}$CO and HCO$^+$ lines may originate from
a disk, with the inner parts of the spectra hidden by strong confusion due to the cloud.
\object{LkH$\alpha$ 358} is located deep in the Taurus cloud, near HL Tau and \object{HH 30}.
An origin from the cloud is also possible: the detection of C$^{17}$O at a LSR
velocity of $7.05 \kms$ (very close to the disk \textbf{systemic} velocity, $6.8 \kms$)
and with a linewidth of $0.45 \kms$
indicates a large column density, and the positive signal in $^{13}$CO and HCO$^+$ may
just be the wings of the cloud emission profile.

\begin{figure}[!h] 
\includegraphics[width=0.8\columnwidth]{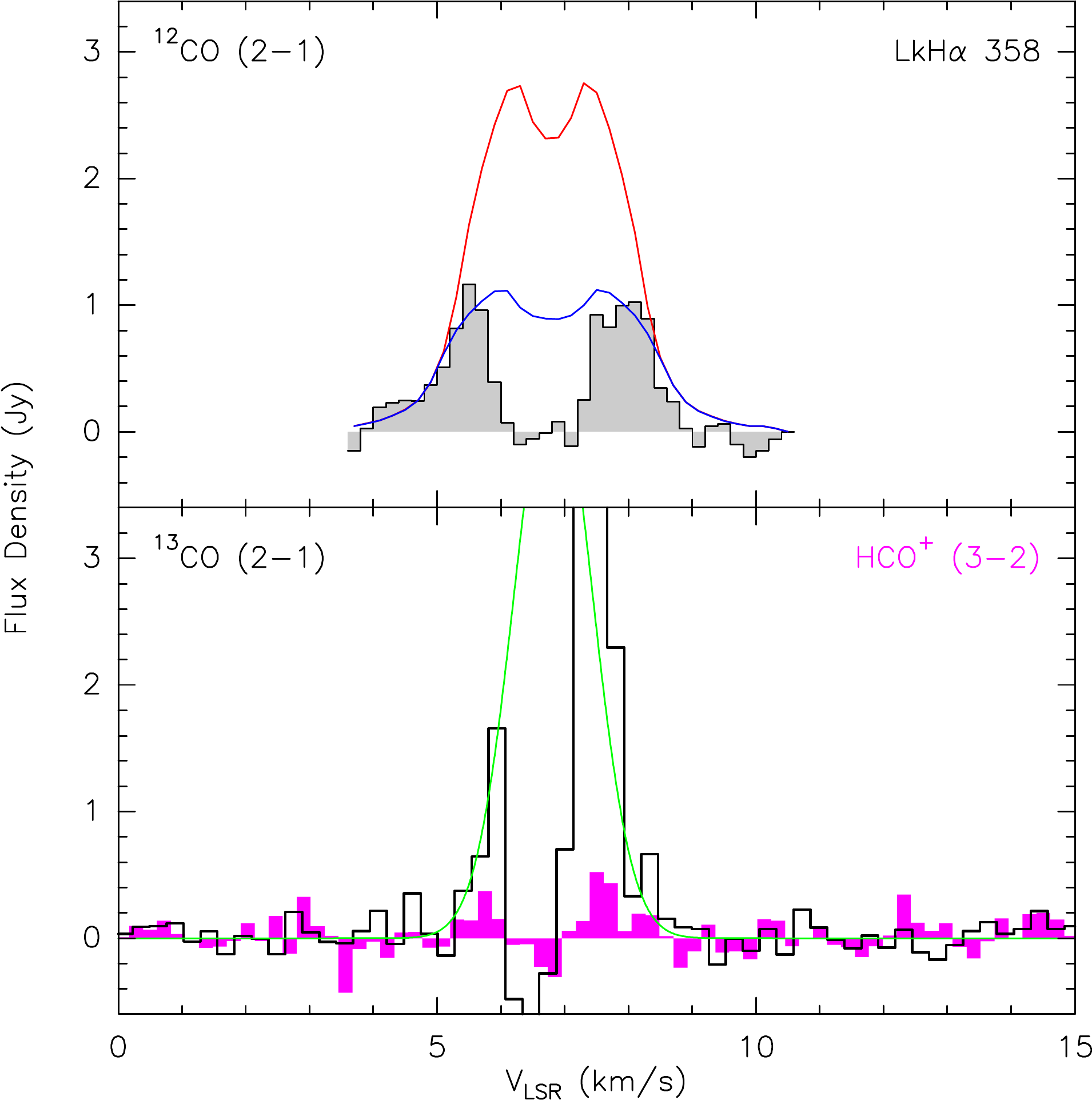}
\caption{Spectra of $^{12}$CO, $^{13}$CO and HCO$^+$ towards LkH$\alpha$ 358.
Top: $^{12}$CO from the interferometric measurement of \citet{Schaefer+etal_2009}.
The blue curve is the fitted profile based on the disk analysis made in the $uv$ plane.
The red curve uses the same disk parameters, except for the outer radius
which is here 250 au instead of 170 au. Bottom: $^{13}$CO(2-1) spectrum (histogram)
and HCO$^+$(3-2) (filled histogram). The green curve is the tentative
fit of a disk component from \citet{Guilloteau+etal_2013}.
}
\label{fig:lkha358}
\end{figure}

Figure \ref{fig:lkha358} displays the relevant spectra. It shows that the tentative
fit of a disk component in $^{13}$CO by \citet{Guilloteau+etal_2013} most likely overestimated
the disk contribution. A similar fit to the HCO$^+$ profile (masking the same
velocity range for both lines, 6.0 to $7.5 \kms$) yield an integrated
line flux of 0.15 Jy\,km\,s$^{-1}$, with
a 5$\sigma$ significance. Assuming the same excitation temperature as for
CO \citep[25 K as measured by][]{Schaefer+etal_2009}, such a flux would be compatible
with optically thick HCO$^+$ emission out to 130 au, consistent with the disk
size in CO. Note that LkH$\alpha$ 358 is one of our lowest mass star in
the sample: $0.33 \pm 0.04 \Msun$ using the inclination of $i=52\pm2^\circ$ found
by \citet{ALMA+HLTau_2015}, and the $v \sin(i) = 1.35 \pm 0.04 \kms$
from \citet{Schaefer+etal_2009}.

\subsection{GG Tau}
\label{sec:obj:gg_tau}
\object{GG Tau} is a hierarchical quintuple system \citep{Difolco+etal_2014}. The
disk is around the central triple star.
Molecules in GG Tau were already detected by \citet{Dutrey+etal_1997}.
We obtain here better signal to noise ratios, and detect C$_2$H for the
first time. For the
disk analysis, we assume an inner radius of 180 au to account for the
large cavity sculpted by the tidal interactions with the binary
\citep{Dutrey+etal_2014b}, as only $^{12}$CO has been detected from
within the cavity so far.

\subsection{AA Tau}
\label{sec:obj:aa_tau}
Emission from the disk of \object{AA Tau} is known to be affected by contamination due
to the molecular cloud, especially in CN \citep{Guilloteau+etal_2013}.
Contamination may also affect the HCO$^+$ spectrum, which displays a similar velocity
than CN, but different from those of HCN, $^{13}$CO and C$_2$H.  CS is also apparently detected.
In the analysis, we also used the integrated profiles obtained with the SMA
by \citet{Oberg+etal_2010} for CN, HCN and HCO$^+$ to improve the sensitivity
in the fit.

\subsection{\object{SU Aur}}
\label{sec:obj:su_aur}
The origin of the HCO$^+$ emission is unclear. The (partially confused) emission
detected in $^{13}$CO by \citet{Guilloteau+etal_2013} with the IRAM 30-m telescope closely
agrees with the interferometric result of \citet{Pietu+etal_2014}.
The HCO$^+$ spectrum displays in addition to this disk component a prominent
blue-shifted wing, which may be the trace of an outflow. No other molecule is detected.

\subsection{RW Aur}
\label{sec:obj:rw_aur}
\object{RW Aur} is a binary system that was imaged in $^{12}$CO by \citet{Cabrit+etal_2006}, who
find evidence for a very small (50 au radius) disk, but did not
detect $^{13}$CO emission. The rather strong, but very broad, HCO$^+$ line may come from
the tidal arm detected in CO by \citet{Cabrit+etal_2006}, rather than from a very compact disk.
An origin in a disk would imply very hot gas, and a very massive system.

\subsection{HK Tau}
\label{sec:obj:hk_tau}
\object{HK Tau} is a binary system, and it is possible that the line emission originates in
part from both stars, as they have similar disks in CO \citep{Jensen+Akeson_2014}.
However, the HCO$^+$ line width is quite large, and given the estimated stellar
masses \citep[0.6 and $1.0 \Msun$ for A and B respectively,][]{Jensen+Akeson_2014},
we tentatively assign it to the edge-on disk HK Tau B. The large line width then implies
an origin in a rather compact disk, of characteristic outer radius 100 au,
and a substantial optical depth is required to provide enough flux implying
$\rho \approx$ a few in Eq.\ref{eq:main-rout}. The modelling of the line profile indeed
requires a small, warm, disk, with higher local line width than in other sources.
However, contamination by the disk of HK Tau A may also partly explain the
relatively high flux of the HCO$^+$(3-2) line.

\subsection{FT Tau}
\label{sec:obj:ft_tau}
\begin{figure}[!h]
\includegraphics[width=0.8\columnwidth]{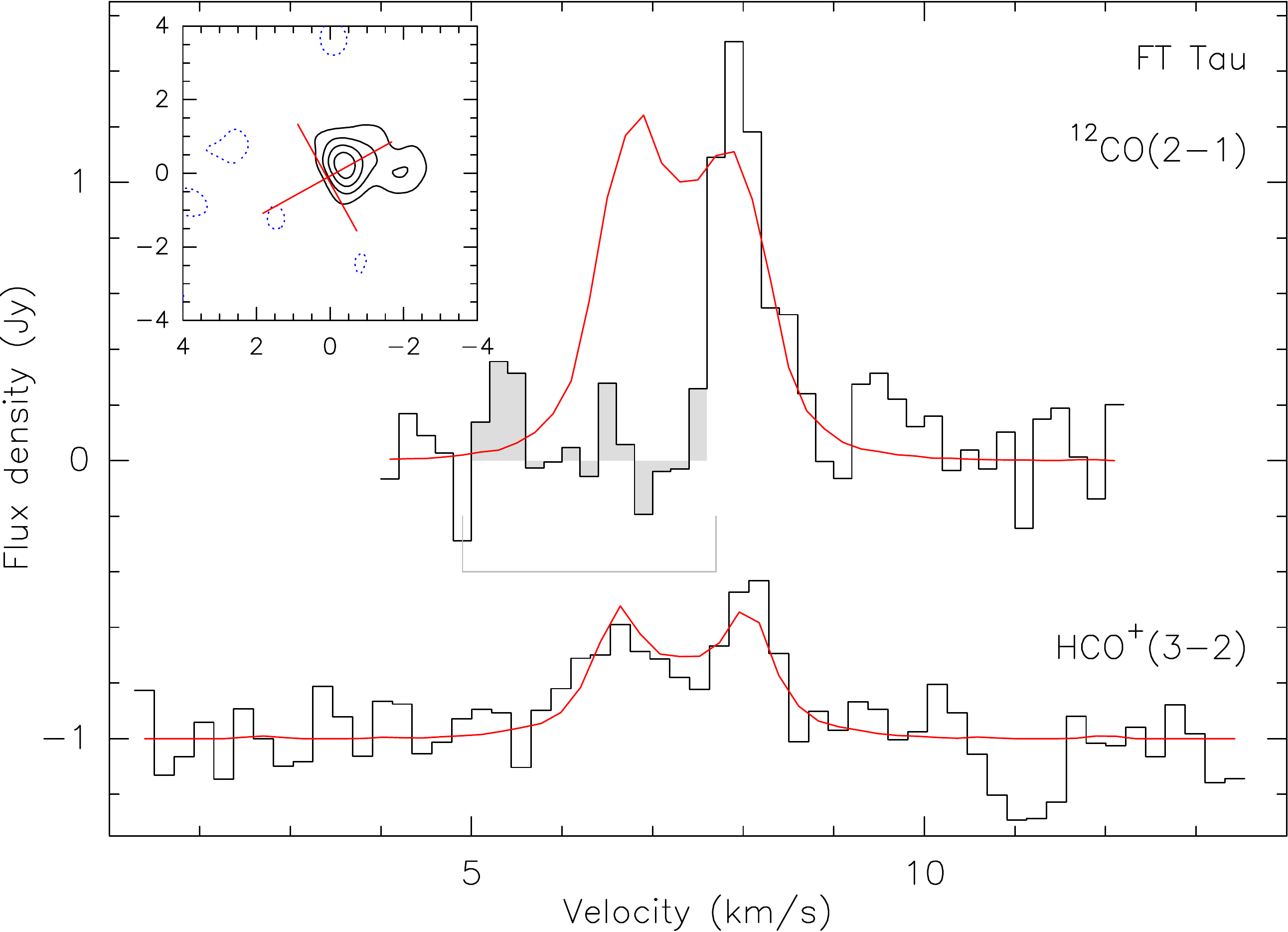}
\caption{Observed spectra and fitted profiles for FT Tau. The grey area
is the range affected by confusion from the molecular cloud, ignored
in the CO analysis. The insert shows the integrated CO emission, with
contour spacings of 50 mJy$\kms$  (angular offsets in arcsec).
The cross indicates the position
and orientation of the disk (derived from the continuum
emission and kinematics pattern).}
\label{fig:ft_tau}
\end{figure}

CO detection of \object{FT Tau} was reported by \citet{Guilloteau+etal_2011}, who found
a disk inclination around $30^\circ$, but with large uncertainty
despite the high angular resolution ($0.5 \times 0.3''$), because the disk
is small, both in line and continuum, and the CO emission severely
contaminated by a molecular cloud between 5 and $\approx 8\kms$.
Although \citet{Garufi+etal_2014} argue for higher inclination,
based on unresolved SED and line emission modelling, a recent measurement
of the dust disk inclination using CARMA gave $i = 34\pm2^\circ$ \citep{Kwon+etal_2015}.
The detection of several transitions (HCO$^+$, CN and marginally C$_2$H)
indicates a systemic velocity
of $7.3-7.4 \kms$, confirming that the blue shifted part of the disk is hidden
by the cloud in CO. To best fit all available lines, we need a small disk,
orbiting a $M_* = 0.35 \Msun$ for $i=34^\circ$. Higher inclinations would
require an even less massive star (as $v \sin(i)$ must be preserved),
and are thus unlikely. The disk radius must be around 200 au:
larger disks produce too much flux.
Figure \ref{fig:ft_tau} shows the fitted profiles for CO and HCO$^+$, and
the CO integrated area map using data from \citet{Guilloteau+etal_2011}.
The small disk size makes FT Tau similar to the ``mm-faint'',
compact disks studied by \citet{Pietu+etal_2014}. The detection of molecules
there thus indicates that the chemistry on scales 100-200 au is not
widely different from what happens in bigger disks at radii 200-500 au.

\section{Discussion}

\begin{figure*}[!th]
\begin{center}
\includegraphics[width=15.0cm]{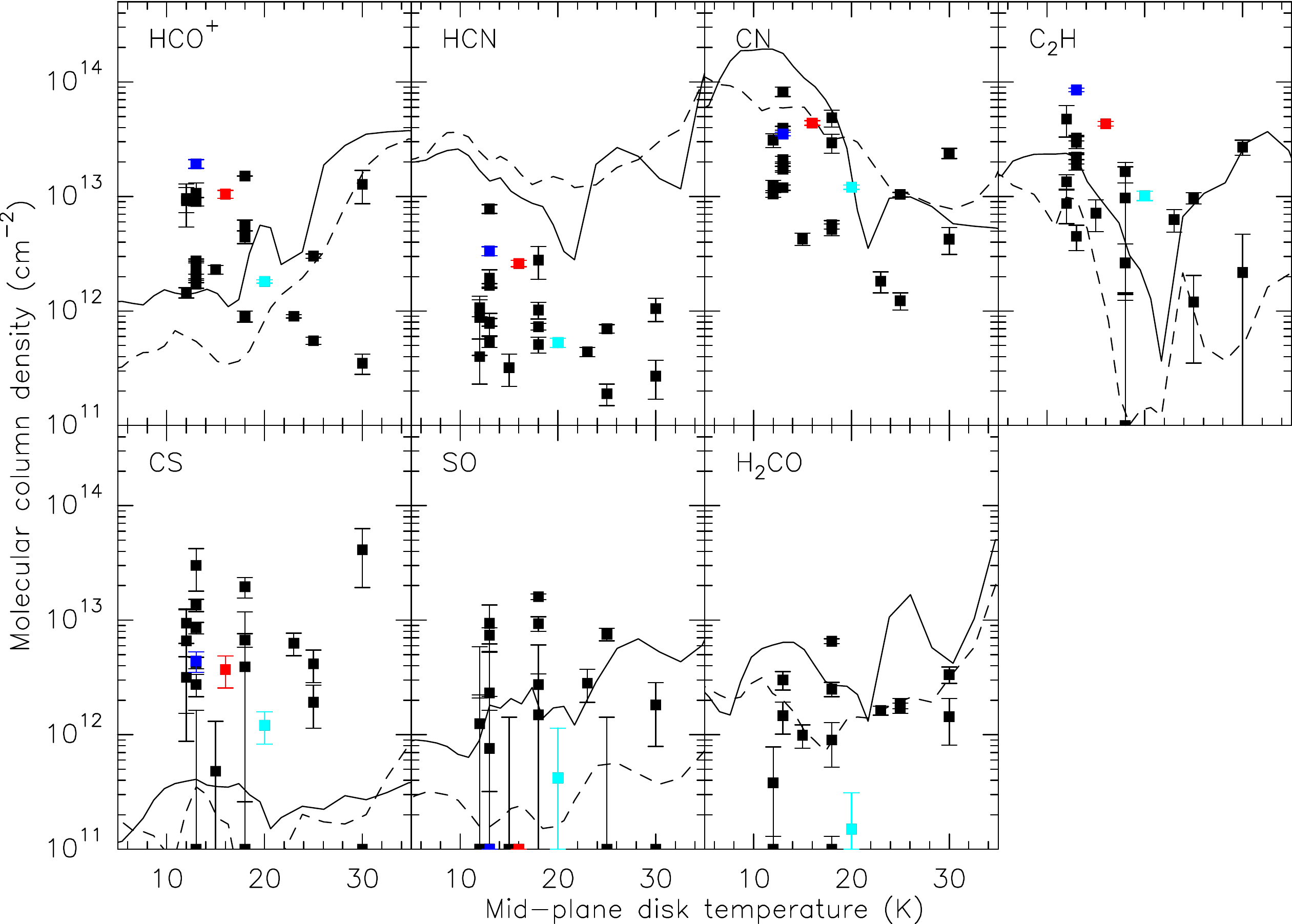}
\end{center}
\caption{Predicted column densities at 300 au
as a function of disk mid-plane temperature, with observed
points superimposed. Color markers are DM Tau (red), GO Tau (blue)
and MWC 480 (cyan). The continuous line is for ``standard low metal''
abundances, while the dashed line assumes an additional depletion
of C and O by a factor 10.}
\label{fig:chem}
\end{figure*}

\subsection{Molecular Sample}

A comparison with the unbiased frequency survey of \citet{Punzi+etal_2015} show
that, although covering only half of the 64 GHz between 206 and 270 GHz,
our study only misses 3 detectable transitions:
$^{12}$CO J=2-1 (which would be severely contaminated by molecular cloud emission
in most cases), C$^{18}$O J=2-1 and the $3_{0,3}-2_{0,2}$
transition of para H$_2$CO at 218.222 GHz.
The only other significant molecules missed by both studies, because of their upper
frequency cutoff of 270 GHz, are HNC and N$_2$H$^+$, whose 3-2 rotational lines
are at 271.981 GHz and around  279.5 GHz respectively.

Less abundant molecules, such as the deuterated species DCO$^+$ and DCN, or
the more complex ones like HC$_3$N, C$_3$H$_2$, fall in our observed band,
but require much deeper integration to be detectable.

We have for the first time  discovered several molecules in
several disks of radii $< 250$ au (RY Tau, FT Tau, IQ Tau, DN Tau, UZ Tau E and HK Tau).
This demonstrates that the TW Hya disk is not specific in this respect, and
shows that the chemistry of these regions is not fundamentally different from that
of much larger disks like DM Tau or GO Tau.

Our data includes stars with spectral types from M3 to A0, stellar masses
from 0.3 to 2.2 $\Msun$, luminosities from 0.1 to $40\,\Lsun$ and
ages ranging from $<0.5$ to more than 4 Myr. A summary of star
and disk properties is given in Table \ref{tab:sample} in Appendix
\ref{app:sample}. The column densities of the detected molecules are displayed as a function
of stellar ages, luminosities and spectral type, and also
1.3\,mm continuum flux in Fig.\ref{fig:correl-cd}.
The 1.3\,mm flux is often used as a possible proxy for the disk masses,
but temperature and opacity effects both play a non negligible role
in the derivation of masses from continuum flux, as discussed in Appendix \ref{app:sample}.
Although disk masses span the range from 0.003 to $0.1 \Msun$,
they cluster around $0.015-0.020 \Msun$.

Only two main trends are visible in Fig.\ref{fig:correl-cd}: a general
decrease of molecular column densities as a function
of stellar luminosity (except for H$_2$CO), and an overall increase
in CN, C$_2$H and HCN as a function of stellar ages.
These two trends may provide a natural explanation for the
lower CN emission found by \citet{Reboussin+etal_2015} in disks in
the younger association $\rho$ Oph, as stars are (on average) both
younger and more luminous there than in the Taurus region.
The statistical significance of this trend remains low, and the dispersion
quite large. A log-log correlation analysis of N(HCO$^+$) versus $L_*$ yields
a correlation coefficient of $r = -0.48$ ($p$ value of 0.02). A Spearman test gives
a similar result ($r = -0.505$ and $p = 0.014$), and all other correlations
are less significant.


On another hand, a statistical analysis reveals substantial
correlations between HCO$^+$, CN, C$_2$H, and HCN surface densities.
The strongest correlations are for CN with HCN, and CN with C$_2$H.
We find a log-log correlation coefficient of $r = 0.84$
(Rank-Spearman $p=7\,10^{-7}$) between the column
densities of CN and C$_2$H, similar to that for CN and HCN ($r = 0.86, p=2\,10^{-7}$).
Correlation between these molecules and HCO$^+$ come next with $r$ between 0.56 and 0.70.
Line ratios may thus be better diagnostic tools than line
intensities. These ratios are
given in Table \ref{tab:ratio} and are displayed as a function of disk/star
properties in Fig.\ref{fig:correl-ratio-cn}-\ref{fig:correl-ratio-hco+} in Appendix.
Although ratios vary by a factor $> 10$ from source to source, no obvious correlation
emerges from this analysis. For T Tauri stars, the [CN/HCN] and [C$_2$H/HCN] ratios are
higher than in typical molecular clouds \citep[e.g. for TMC-1 (CP)][]{Ohishi+etal_1992}, confirming the results found
previously on other sources \citep{Dutrey+etal_1997,Kastner+etal_2014}.
This is suggestive of photodissociation processes, for which we would
expect CN and C$_2$H to be correlated as in our data. However, the
[CN/C$_2$H] ratio still varies substantially from source to source, possibly indicating
other chemical processes at work.

We also caution that despite the substantially larger number of sources
compared to previous studies, each category still only contains a few
sources. For example, the only ``warm'' disk being well probed
by our data, MWC 480, is a heavily self-shadowed disk with little
flaring, whose properties are very close to those of T Tauri disks.
AB Aur is still embedded in an envelope \citep{Tang+etal_2012}, and
has a large (100 au radius) inner cavity \citep{Pietu+etal_2005}, so may
be atypical. The warmer disks around Group I \citep{Meeus+etal_2001} HAe stars,
like CQ Tau and MWC 758, are unfortunately too small
for precise measurements. Nevertheless, these two stars are
among those with the lowest column densities of HCO$^+$, CN, HCN, and C$_2$H,
but are strong in CO isotopologues.
In fact, the molecule-to-CO ratio is smaller in MWC\,480 than in
any T Tauri disk. A high CO content for  MWC\,480 (as well as for AB Aur
and 04302+2247) is attested by the detection of C$^{17}$O J=2-1 transition,
which is below our detection threshold for all T Tauri disks \citep{Guilloteau+etal_2013}.
A comparison with the other small disks also indicates
that HCO$^+$ and CN are substantially less abundant in CQ Tau and MWC 758
than in the T Tauri stars. This confirms the overall defficiency in molecules
in HAe disks found in the case of AB Aur by \citet{Schreyer+etal_2008},
which they explain as a result of enhanced
photodissociation due to the large UV flux from the host star
(note that such an enhancement is not included in Fig.\ref{fig:chem}.

Another noticeable trend is more S-bearing molecules and less CN in
the embedded, presumably younger, objects like Haro 6-10, 6-33, 6-5B,
DO Tau and the Butterfly star. These young sources also have
more H$_2$CO than more evolved T Tauri disks. A higher H$_2$CO content
is also found in warm disks around Herbig Ae stars,  as indicated
by AB Aur and (tentatively) CQ Tau.

\subsection{Comparison with chemical modelling}

\begin{figure}[!t]
\begin{center}
\includegraphics[width=0.8\columnwidth]{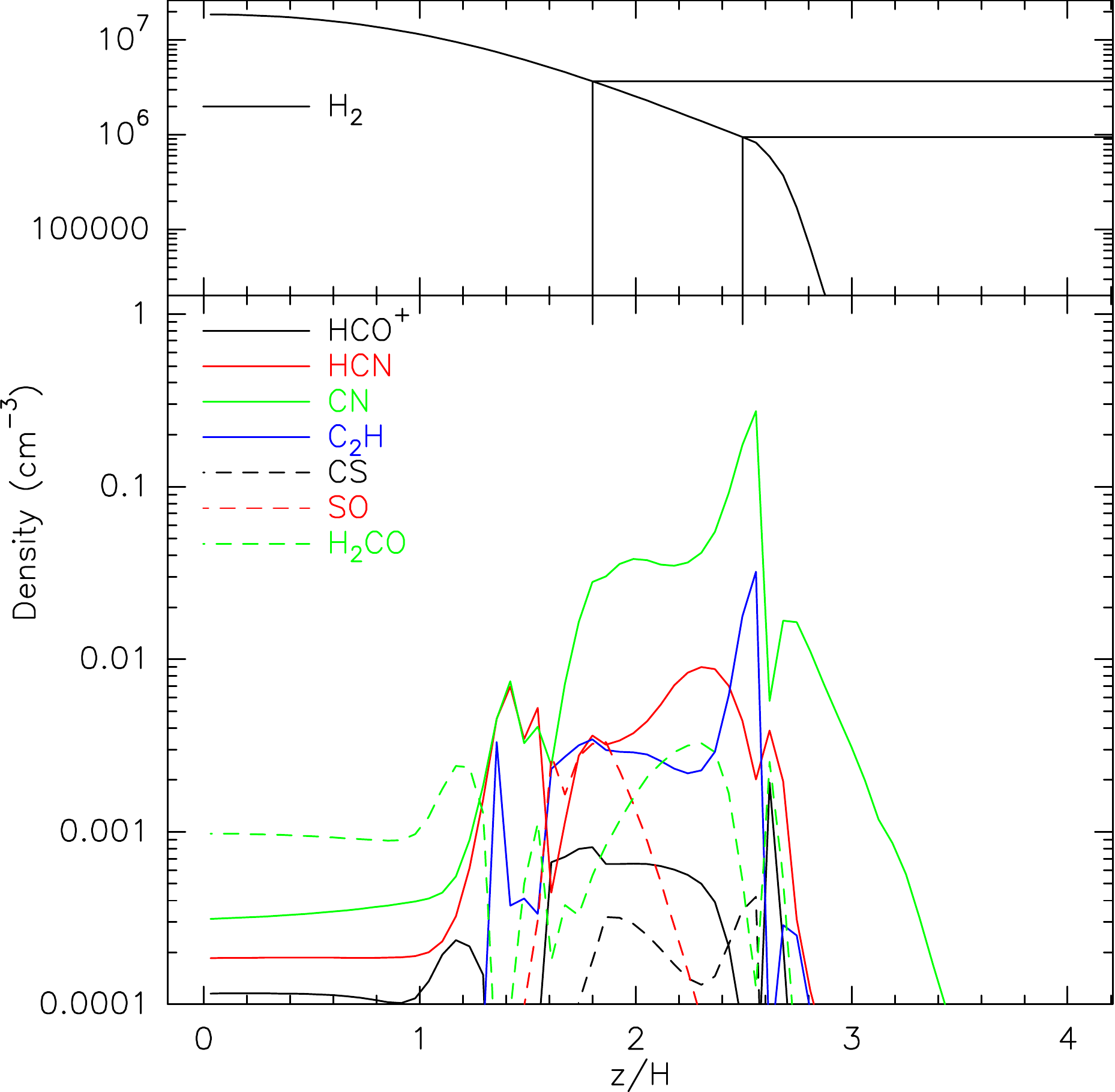}
\end{center}
\caption{Predicted molecular densities at 300 au
as a function of height above the disk mid-plane (in scale
heights, defined as $c_s/\Omega$, where $c_s$ is the
sound speed, and $\Omega$ the Keplerian angular frequency). The upper panel shows
the H$_2$ density, with the two horizontal and vertical lines delimiting
the range where most molecules are abundant.}
\label{fig:dens}
\end{figure}

A full chemical modelling of the observed disks is beyond the scope of
this work. We however show some results using the disk chemical model described
in \citet{Reboussin+etal_2015}.
We use here a density distribution with a $p=1$ exponent for the surface
density, and a disk mass of $0.03 \Msun$ for a disk extending
up to 800 au (these are roughly the expected disk parameters for
DM Tau and GO Tau). The thermal structure is not
derived self-consistently. Instead, the mid-plane temperature is
taken as a free parameter, and a vertical gradient is
included (with temperatures at 2 scale
heights typically higher by 10 K than in the mid-plane). The density
profile is derived from hydrostatic equilibrium.
The UV flux impinging on the disk is representative of T Tauri stars.

The model solves the time-dependent chemistry at each
radius independently, using a complex gas + grain chemical network,
involving reactions on dust grains.
The grain size distribution
is equivalent to 0.1$\mu$m grains.
Photodesorption was shown by \citet{Reboussin+etal_2015}
to be negligible in these conditions, and is thus not included.
The disk age is assumed to be
1 Myr. Column densities of several
molecules as a function of the disk mid-plane temperature are displayed
in Fig.\ref{fig:chem}.
We have overplotted the observed points, assuming
the apparent excitation temperature of the observed transitions is
representative of the mid-plane kinetic temperature (see justification
below).
The model roughly reproduces the observed
surface densities of CN and C$_2$H for disk mid-plane temperatures below
15 K, consistent with the fitted temperatures derived from the CN
interferometric data. The CN/HCN ratio also agrees with the measured
ranges for cold disks.

The simple combination of surface reactions
(leading to more complex molecules on grains) and thermal and non-thermal
desorption mechanisms are sufficient to produce C$_2$H and CN.
CN and C$_2$H, as well as all other
molecules, are approximately
co-located at around 2 scale heights (density 7 times below the disk
plane density) above the disk (Fig.\ref{fig:dens}). At 300 au, the density in this region
is in the range $1-4\,10^6$\,cm$^{-3}$, providing enough collisional
excitation for the observed transitions, but not a complete thermalization.
We thus expect \protect{$T_\mathrm{ex}\,\sim0.7\,T_k(z=2H)$} for most transitions, except perhaps
for HCN(3-2) which has the highest critical density of all observed molecules.
However, this region is slightly warmer than the disk mid-plane, so both differences
compensate to first order, and $T_\mathrm{ex}$ will be close to the mid-plane
temperature $T_k(z=0)$.

The model does not include X-ray ionization, so predictions for
HCO$^+$ should be considered with more caution. \citet{Henning+etal_2010}
also showed that C$_2$H could be produced in substantial quantity
close to the disk plane in presence of X-ray ionization at radii
$< 100$ au but our observations are not sensitive to these
small radii. In our chemical model, C$_2$H behaves differently at 100 au
than at 300 au, perhaps explaining why the CN/C$_2$H
ratio varies from source to source.

HCN is overpredicted by the chemical model by a factor of $\sim 5$
on average. However, part of this may be an artefact of our assumed
excitation temperature
for this line, because HCN J=3-2 is the transition with the highest
critical density, and thus the most susceptible to sub-thermal
excitation. Indeed, based on resolved observations of the HCN J=1-0
line, which is largely optically thin as attested by the hyperfine
component relative intensities, \citet{Chapillon+etal_2012} found
column densities of 7 and 11\,$10^{12}$\,cm$^{-2}$ for DM Tau and LkCa 15
respectively, a factor 2 and 1.5 above our derived values.
Non-LTE effects may become relatively more important in lower mass disks.

Finally, as in most other chemical
models, the S-bearing molecules are not well
reproduced: CS is underpredicted and SO overpredicted.
\citet{Dutrey+etal_2011} showed that in particular H$_2$S
is not detected in disks, which suggests that current chemical models
lack a major process of chemical evolution of S bearing molecules on grains.

Recent works comparing the observed O{\small I}, C{\small I}, C$^+$ and high
J CO line intensities to predictions of thermo-chemical models \citep{Chapillon+etal_2010,Bruderer+etal_2012,Du+etal_2015}
suggest that C and O are depleted in the upper
layers of the disk. A possible mechanism to obtain such an elemental depletion
resides in the disk history coupled to the dust grain evolution in the
cold mid-plane layers of the disk.  Carbon and Oxygen nuclei get trapped
in the form of CO or CO$_2$ on grains around the mid-plane, as shown
by \citet{Reboussin+etal_2015}. Simultaneously, as turbulent mixing drags
the gas and smaller grains up and down, the upper layers can progressively be
depleted in C and O nuclei in the gas phase. Grain growth in the disk
mid-plane could potentially accelerate this overall gas phase depletion.
We have computed a second set of chemical models simply assuming C and O
are depleted by a factor 10. The results are shown as the dashed lines
in Fig.\ref{fig:chem}. We note a decrease in C$_2$H surface densities.
HCN and CN are less affected by this important reduction of the
number of available C nuclei, but the predictions remain in broad
agreement with the observations (CS excluded as before).
Another interesting effect is the substantial reduction of SO, which practically
scales as the O depletion factor. A progressive reduction in elemental
abundances in the gas phase with time would thus naturally lead to stronger
SO emission in younger objects, perhaps explaining why SO is only detected
in embedded sources. Note however that \citet{Semenov+Wiebe_2011} found SO to
be sensitive to turbulent transport in the disk.

\section{Conclusion}

\begin{itemize}
\item We have performed with the IRAM 30-m telescope a (spatially unresolved) survey of 30 disks in the Taurus region,
with sufficient sensitivity to detect HCO$^+$, CN, HCN, C$_2$H, H$_2$CO,
CS, SO and C$^{17}$O, allowing us over 100 new detections of individual spectral lines
from these 8 molecules. The detection rates range from $\sim 100$ \% (for HCO$^+$) down to
$< 15 \%$ (for SO and C$^{17}$O).

\item A comparison with \citet{Punzi+etal_2015} on LkCa 15, one of the two
strongest sources in our survey, shows that, except for C$^{18}$O and one
para H$_2$CO line, our
study covers \textit{all} lines detectable in the 206 - 270 GHz range with
integration times of $\sim 2-8$ hours per source using a telescope
as sensitive as the IRAM 30-m.

\item The HCO$^+$ line flux is consistent with moderately optically thick emission
extending out to radii comparable to the outer radius derived from
CN or CO interferometric observations, which thus define ``the disk radius''.
All other lines have lower opacities.

\item The survey is complemented by resolved observations of CN (or CO)
in many sources, and with (resolved or unresolved) multi-line observations of C$_2$H
in 6 sources, allowing an estimate of disk radii and excitation conditions.
The typical excitation temperatures derived for CN and C$_2$H range from 8-10 K at 300 au to
15 to 30 K at 100 au.

\item 
The overall consistency between HCO$^+$, CN and C$_2$H suggests that non-LTE effects
are limited for these molecules (although
it is still possible to have $T_\mathrm{ex} \approx 0.7 T_k$ for most
observed transitions). In particular, we do not confirm the very low
excitation temperatures suggested for CN and C$_2$H in LkCa15 by
\citet{Punzi+etal_2015} (and consequently neither the very high column
densities for these two molecules).

\item Practically all T Tauri stars share very similar characteristics, with
enhanced CN/HCN and C$_2$H/CN ratios compared to molecular clouds.
The detection of several
molecules in small disks (radii $\sim 200$ au) indicates the chemistry in
these regions is not radically different from that of bigger disks, as already
hinted at by the only known detailed case so far, TW Hya \citep{Thi+etal_2004,
Kastner+etal_2014}.

\item As suggested by the previous study of AB Aur \citep{Schreyer+etal_2008},
disks around HAe stars appear less rich in molecules. Unfortunately
two of the five studied disks around HAe stars are small, and the current sensitivity level is
insufficient to be quantitative on the abundance of molecules.
More generally, the number of molecules appear to decrease with stellar
luminosity.

\item The content in CN and C$_2$H seems to increase with stellar age, a
result corroborating the lower CN line flux found by \citet{Reboussin+etal_2015}
in the younger $\rho$ Oph association.

\item We performed a simple chemical modelling showing that gas-grain chemistry
predicts molecular column densities in reasonable agreement with the
observations (with the noticeable, but well known exception of S-bearing molecules).

\item Progressive depletion of C and O, due to build up of CO and CO$_2$
ices on dust grains and turbulent mixing may be a clue to explain the
higher SO content of young disks.
\end{itemize}

These observations pushed the IRAM 30-m telescope to its ultimate capabilities for sensitive
spectral line surveys of many (unresolved) sources. Disks smaller than about 150 au are
no longer detectable in reasonable integration times.  The success of
our strategy however yields good promises for interferometric surveys.
For example, NOEMA will be able to provide the same frequency coverage
in just 2 setups, and would be sensitive enough to go a factor 4 to 6 times deeper
in the same integration time.  The spectral capabilities of the even more
sensitive ALMA array are less flexible, but still allows several
combinations of 3 reasonably strong lines from the above molecules to be
be covered simultaneously in Band 6 or 7.

\begin{acknowledgements}
This work was supported by ``Programme National de Physique Stellaire'' (PNPS) and ``Programme
National de Physique Chimie du Milieu Interstellaire'' (PCMI) from INSU/CNRS.
WV's research is funded by an ERC starting grant (3DICE, grant agreement 336474).
DS acknowledges support by the Deutsche Forschungsgemeinschaft
through SPP 1385: “The first ten million years of the solar system
- a planetary materials approach” (SE 1962/1-3)"
This research has made use of the SIMBAD database,
operated at CDS, Strasbourg, France.
\end{acknowledgements}

\bibliography{mays-cn}
\bibliographystyle{aa}

\clearpage

\appendix

\section{}
\label{app:sample}
Table \ref{tab:sample} summarizes the properties of the studied disks and host stars.
For simplicity and ease of comparison with other studies, the disk
mass $M_d$ is derived under the assumption of a uniform, optically thin, disk
$$
S_\nu = \frac {B_\nu(T_d) M_d}{d^2  \zeta \kappa_\nu}
$$
where $T_d$ is the mean dust temperature, $D$ the distance, $\zeta$
the dust to gas ratio, and $\kappa_\nu$ the dust emissivity at frequency $\nu$.
This can be expressed as in \citet{Andrews+etal_2013}, Eq.2 (using
quantities in SI units
$$
\log (M_d) = \log (S_\nu) + 2 \log (d) - \log (\zeta.\kappa_\nu) - \log (B_\nu(T_d))
$$
We use $\zeta = 0.01$ and $\kappa_\nu = 2.3$\,cm$^2$g$^{-1}$ at 1.3\,mm,
with the scaling $T_d = 25 (L_*/\Lsun)^{1/4}$ as in their study.
Note that \citet{Pietu+etal_2014} show that because of the radial temperature
gradient, $T_d$ is also expected to depend on the disk radius; the effect
can be significant for very small disks.  Also, the above formula does not
account for opacity, so the derived values underestimate the disk masses.
The effect is especially pronounced for bright sources (such as GG Tau) or
egde-on objects. For example,  \citet{Graefe+etal_2013} find $0.09 \Msun$
for 04302+2247, while the formula leads to $0.01 \Msun$.

The disk velocities and radii are derived from our study, using the CN, C$_2$H
and HCO$^+$ data. Typical errors are 0.05 $\kms$ on velocities, and 10\% on disk
radii. Apparent disk radii in CO molecules are expected to be slightly larger.
We assume $d=140$ pc for all sources.

\begin{table*}[!ht]
\caption{Summary of stars and disk properties}
\begin{tabular}{llllrrrrrr}
\hline
Name & RA & Dec & Spectral & $L_*$ & $S_\nu$(1.3 mm) & $M_d$ & Incli. & $V_\mathrm{LSR}$ & $R_\mathrm{disk}$ \\
     & J2000.0 & J2000.0 & Type & ($\Lsun$) & mJy & ($0.001 \Msun$) & ($^\circ$) & ($\kms$) & (au) \\
\hline
\textbf{\textit{04302+2247}} & 04:33:16.2 & 22:53:20.0 & - & 5 &  130 & $> 10$ & 70 & 5.97 & 500  \\ 
      AA Tau & 04:34:55.42 &   24:28:53.1 & K7 & 0.66 &   73 & 11 & 70 $\pm$ 5  & 6.45 & 350 \\ 
\textbf{AB Aur} & 04:55:45.80 & 30:33:04.0 & A0/B9 & 44.86 &  110 & 5 &  20-30 & 5.85 & 600 \\ 
%
\textbf{CQ Tau} & 05:35:58.485 & 24:44:54.19 & A8/F2 & 12.00 &  162 & 10 & 29 $\pm$ 2 & 6.20 & 200 \\ 
      CW Tau & 04:14:17.0 & 28:10:56.51 & K3 & 0.68 &   59 & 8 & $64 \pm 2$ & 6.42 & 210 \\ 
      CI Tau & 04:33:52.014 & 22:50:30.06 & K7 & 0.96 &  125 & 16 &  51 $\pm$ 3 & 5.74 & 520 \\ 
      CY Tau  & 04:17:33.729 & 28:20:46.86 & M1.5 & 0.40 &  111 & 19 & 24 $\pm$ 2 & 7.25 & 300 \\ 
      DL Tau & 04:33:39.077 & 25:20:38.10 & K7 & 1.16 &  204 & 25 & 43 $\pm$ 3  & 6.10 & 460 \\ 
      DM Tau & 04:33:48.70 & 18:10:10.6 & M1 & 0.16 &  109 & 25 & 35 $\pm$ 1 & 6.05 & 600 \\ 
      DN Tau & 04:35:27.38 &   24:14:58.9   & M0 & 0.68 &   89 & 13 & 30 $\pm$ 3 & 6.38 & 235 \\ 
\textit{DO Tau} & 04:38:28.59 &   26:10:49.5   & M0 & 1.29 &  136 & 16 & 32 $\pm$ 2 & 6.50 & 350  \\ 

      FT Tau & 04:23:39.188 & 24:56:14.28 & M3 & 0.63 &   73 & 11 & 34 $\pm$ 5 & 7.34 & 200 \\ 

      GG Tau & 04:32:30.34 & 17:31:40.5 & K7 & 0.64 &  593 & 86 & 37 $\pm$ 1& 6.45 & 500 \\ 
      GM Aur  & 04:55:10.98 & 30:21:59.5 & K7 & 1.23 &  176 & 21 &  50 $\pm$ 1 & 5.64 & 400 \\ 
      GO Tau & 04:43:03.050 & 25:20:18.80 & M0 & 0.22 &   53 & 11 & 52 $\pm$ 1 & 4.87 & 600 \\

\textit{Haro 6-13} & 04:32:15.419 & 24:28:59.47 & M0 & 2.11 &  114 & 12 & 40 $\pm$ 3 & 5.10 & 640 \\ 
\textit{Haro 6-33} & 04:41:38.827 & 25:56:26.68 & M0 & 0.76 &   34 & 5 & 52 $\pm$ 5 & 5.30 & 300 \\ 
\textit{Haro 6-5B} & 04:22:01.00 &   26:57:35.5 & K5 & $>0.05$ &  134 & 20 & $~75$ & 7.57 & 300 \\ 
\textit{HH 30} & 04:31:37.468 & 18:12:24.21 & M0 & 0.2 - 0.9 &   20 & $\sim 3$ &  83 $\pm$ 2 & 7.25 & 420 \\ 
      HK Tau & 04:31:50.58 &   24:24:17.9   & M0.5 & 1.00 &   41 &  $> 5$ & $\sim 85$ & 6.44 & 100 \\ 
    HV Tau C & 04:38:35.31 &   26:10:38.5   & K6 & 0.60 &   40 & $>6$ &  $\sim 85$ & 6.40 & 300 \\ 

      IQ Tau & 04:29:51.56 &   26:06:44.9   & M0.5 & 0.53 &   60 & 9  & 56 $\pm$ 4 & 5.47 & 220 \\ 
    LkH$\alpha$ 358 & 04:31:36.15 & 18:13:43.1 & K8 & 0.09 &   17 & 5 &  52 $\pm$ 2 & 6.80 & 170-250 \\ 
     LkCa 15 & 04:39:17.76 & 22:21:03.7 & K5 & 0.85 &  110 & 15 & 52 $\pm$ 1 & 6.30 & 500 \\ 
\textbf{MWC 480} & 04:58:46.27 & 29:50:37.0 & A2 & 11.50 &  289 & 18 & 37 $\pm$ 1 & 5.08 & 450 \\ 
\textbf{MWC 758} & 05:30:27.51 & 25:19:58.4 & A3 & 11.00 &   55 & 3 & 20 & 5.80 & 250 \\
      RW Aur &  05:07:49.56 & 30:24:05.1 & K3 & 1.72 &   42 & 4.5 & 45 $\pm$ 5 & 6.85 & - \\ 
      RY Tau & 04:21:57.42 & 28:26:35.6 & K1 & 6.59 &  229 & 17 & 66 $\pm$ 3  & 6.75 & 210 \\ 
      SU Aur(*) & 04:55:59.4 & 30:34:01.39 & G2 & 9.29 &    23 & 1.5 & $\sim 40$ & 6.8 \\ 
    UZ Tau E & 04:32:43.071 & 25:52:31.07 & M1 & 0.9-1.6 &  150 & 18 & 56 $\pm$ 2 & 5.70 & 210 \\ 
\hline
\end{tabular}
%
\tablefoot{
Disk properties for SU Aur and CW Tau are taken from \citet{Pietu+etal_2014}, and for HH 30
from \citet{Pety+etal_2006}. See \citet{Guilloteau+etal_2013}
for references for spectral types and luminosities \citep[(except for FT Tau, see][]{Garufi+etal_2014}.
Luminosity for 04302+2247 is from \citet{Graefe+etal_2013}.}
\label{tab:sample}
\end{table*}

\begin{figure*}
\begin{center}
\includegraphics[width=8.5cm]{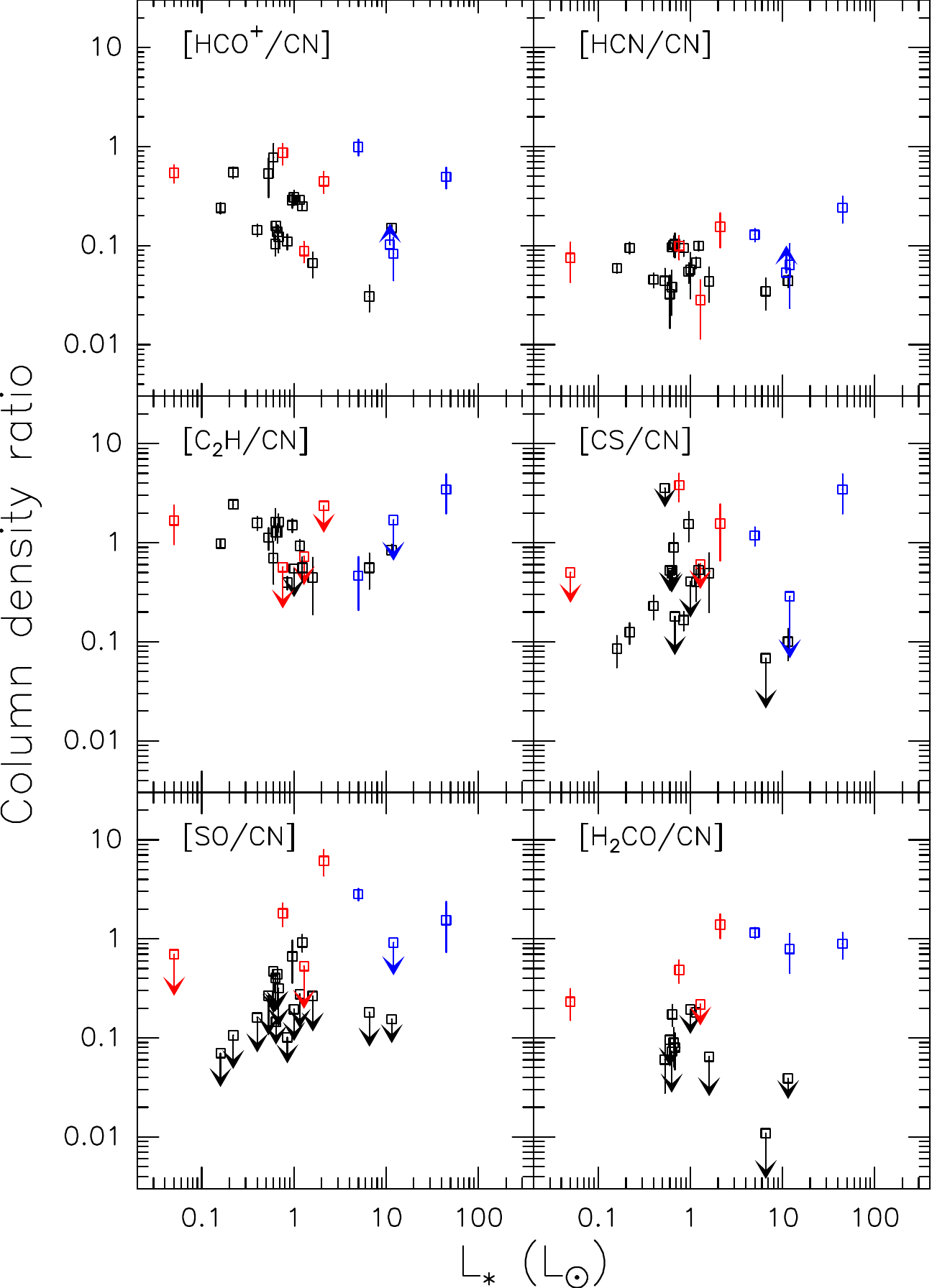}
\includegraphics[width=8.5cm]{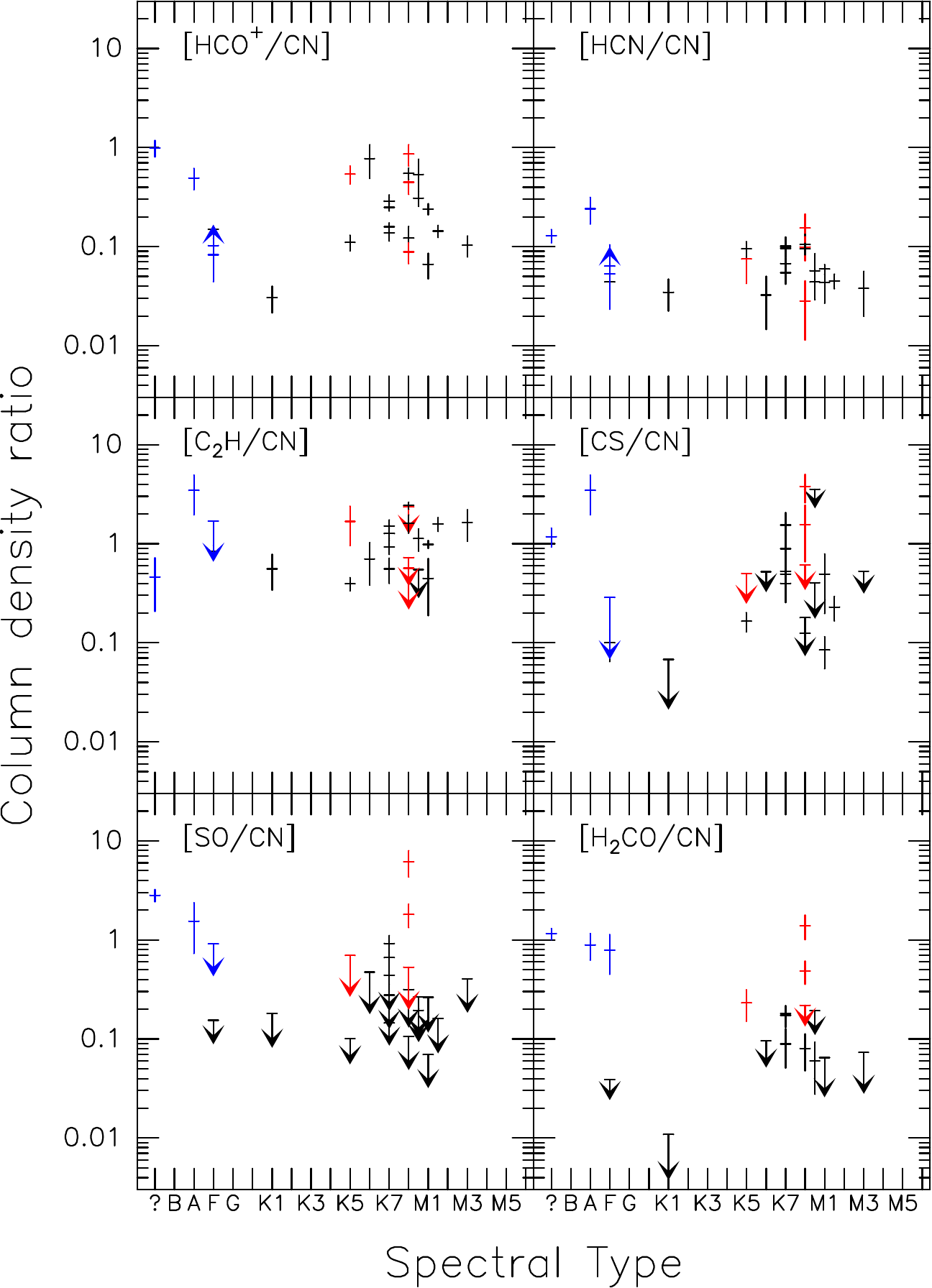}
\includegraphics[width=8.5cm]{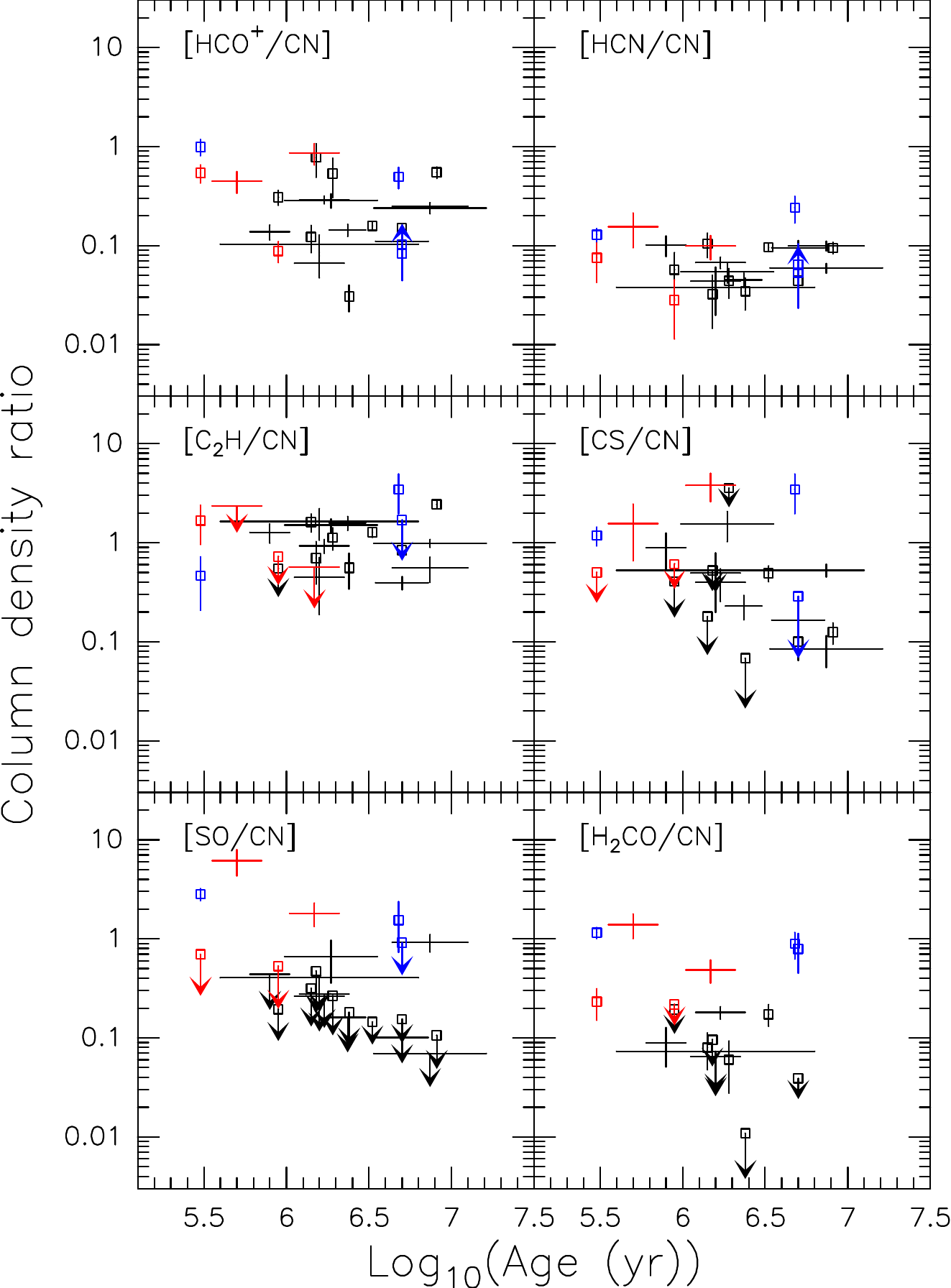}
\includegraphics[width=8.5cm]{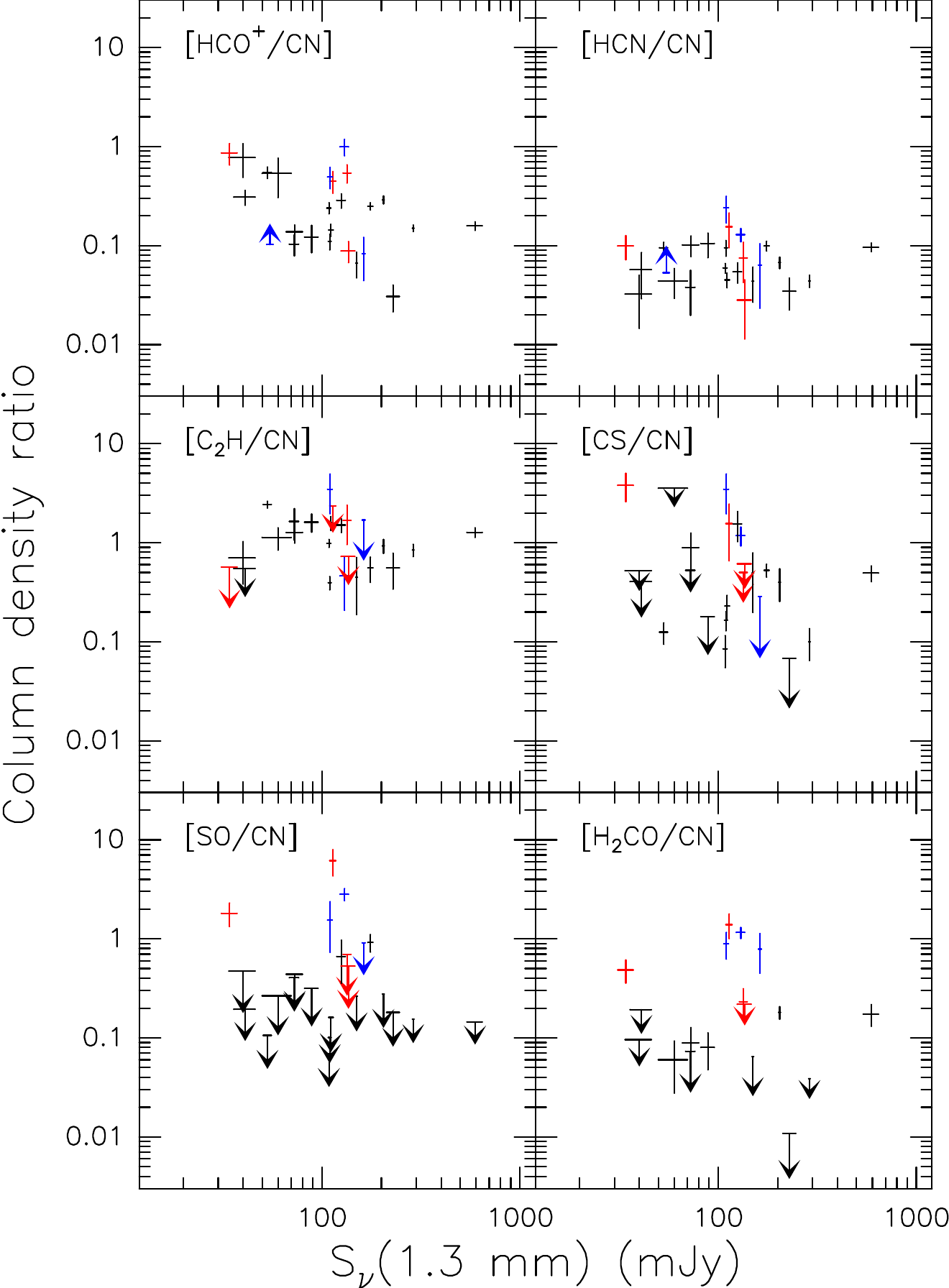}
\end{center}
\caption{Correlation plot of the column density ratios of molecules
at 300 au with stellar luminosity and Spectral Type,
age and 1.3 mm flux density. CN used as reference.}
\label{fig:correl-ratio-cn}
\end{figure*}

\begin{figure*}
\begin{center}
\includegraphics[width=8.5cm]{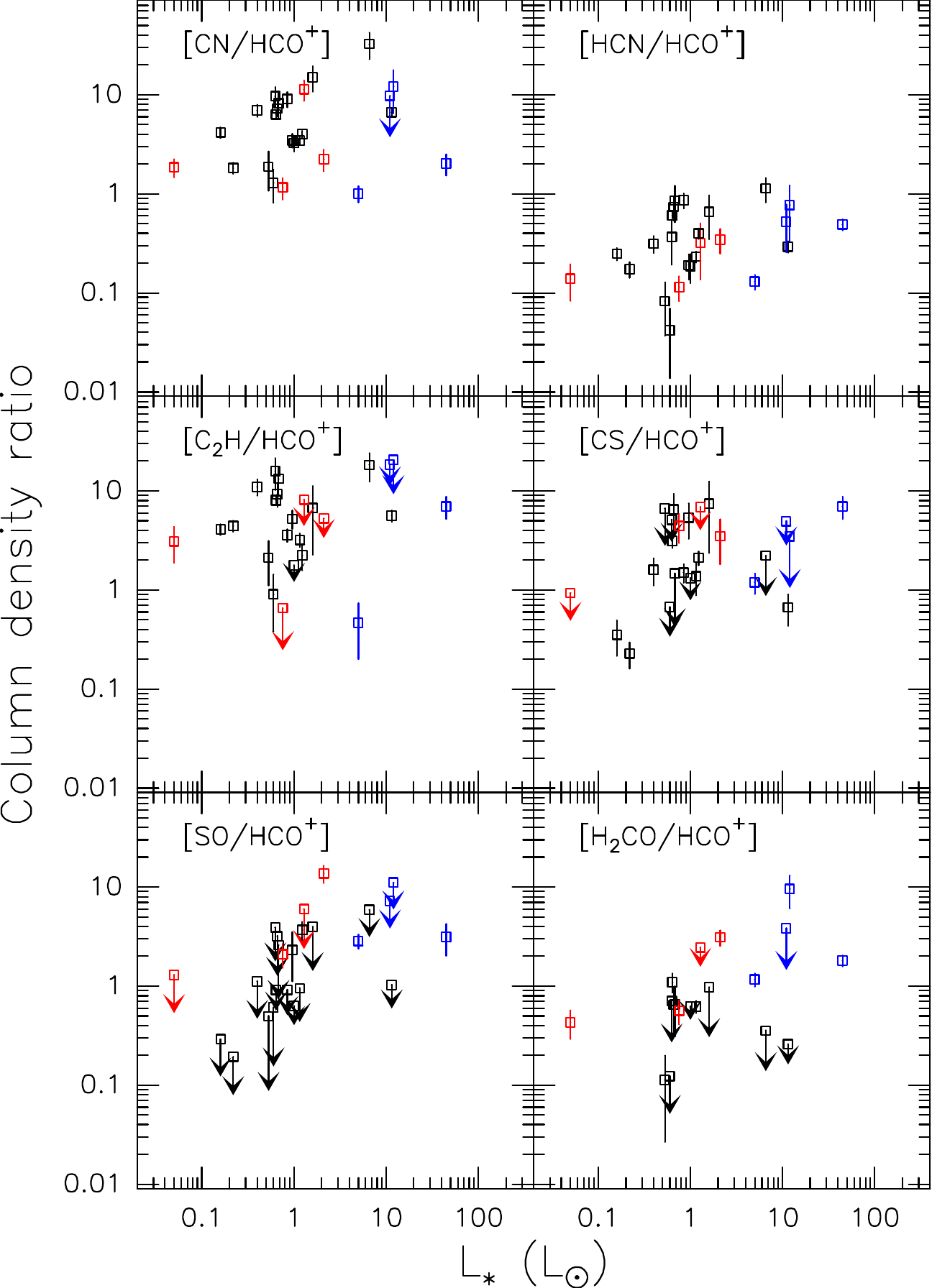}
\includegraphics[width=8.5cm]{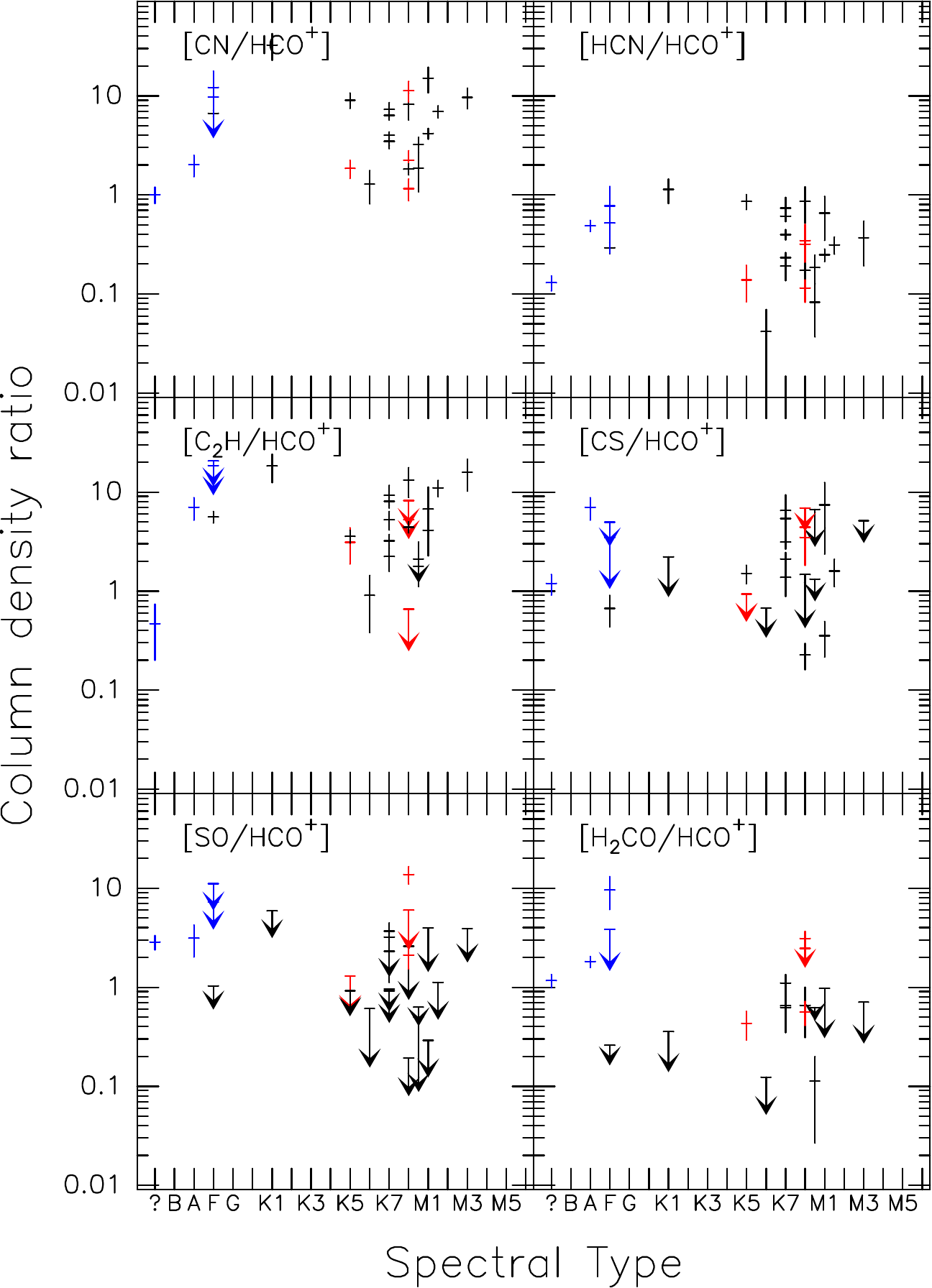}
\includegraphics[width=8.5cm]{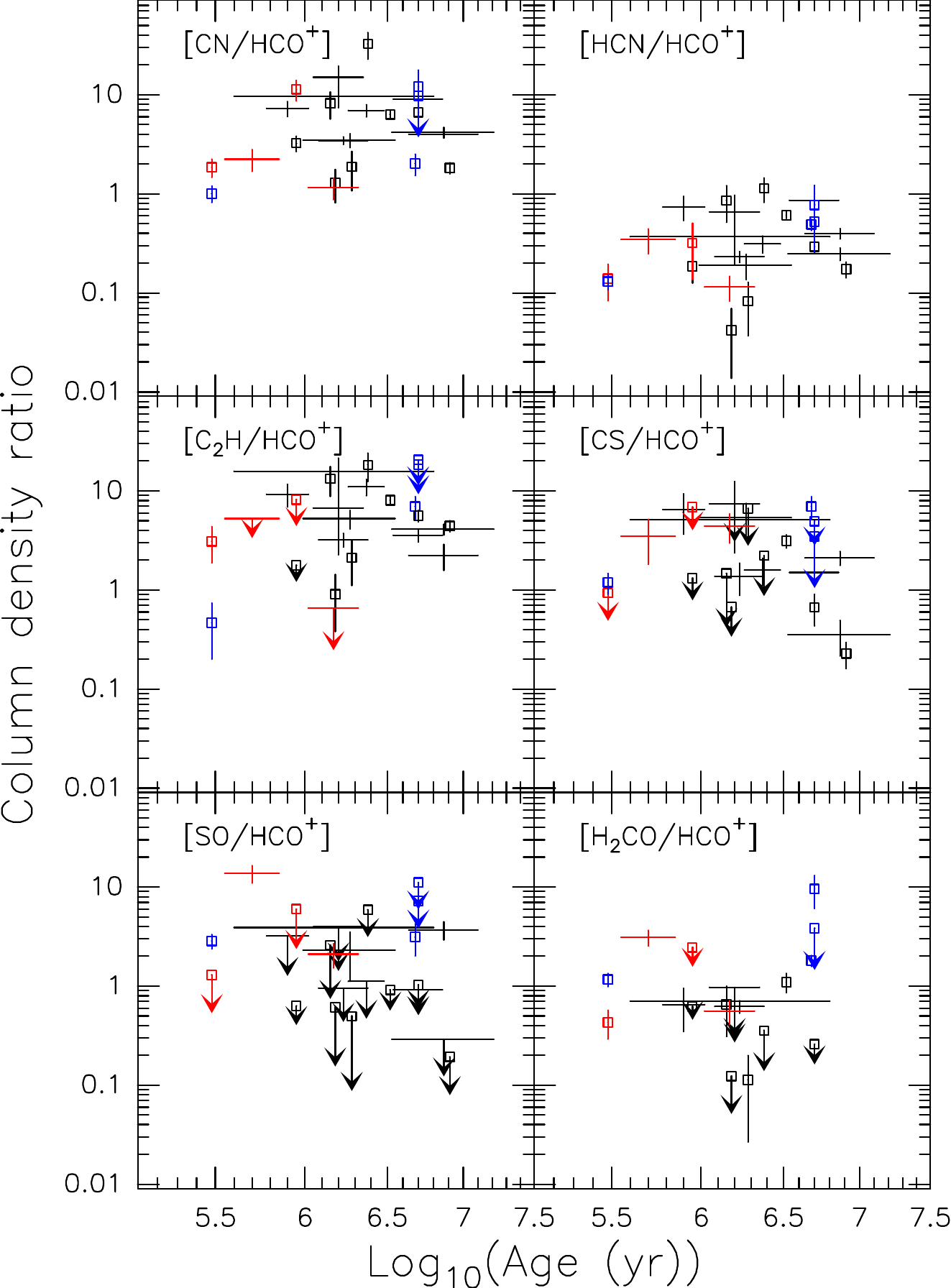}
\includegraphics[width=8.5cm]{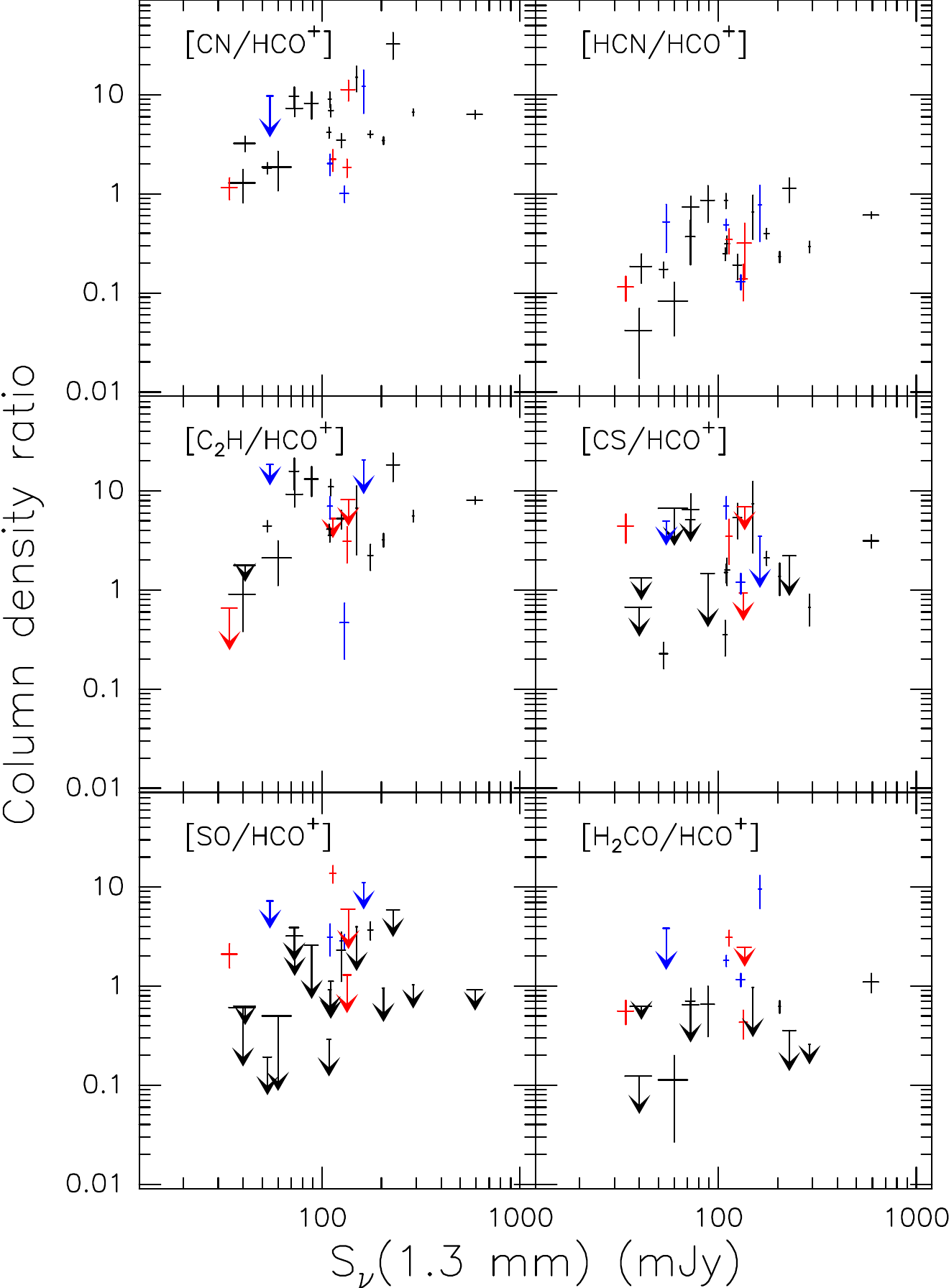}
\end{center}
\caption{Correlation plot of the column density ratios of molecules
at 300 au with stellar luminosity and Spectral Type,
age and 1.3 mm flux density. HCO$^+$ used as reference.}
\label{fig:correl-ratio-hco+}
\end{figure*}

Correlation plots of the column density ratios are displayed
in Figs.\ref{fig:correl-ratio-cn}-\ref{fig:correl-ratio-hco+}.

\clearpage

\section{ }
\label{app:plots}
This appendix displays the spectra for all observed
sources on a common velocity scale. For each source,
the top panel shows the HCO$^+$ spectrum with the best
fit profile from the disk modelling superimposed when available. The
lower panels show all spectral lines (with the two groups
of two hyperfine components for C$_2$H presented as
separate spectra) on a common intensity scale.

\begin{figure}
\includegraphics[height=11.5cm]{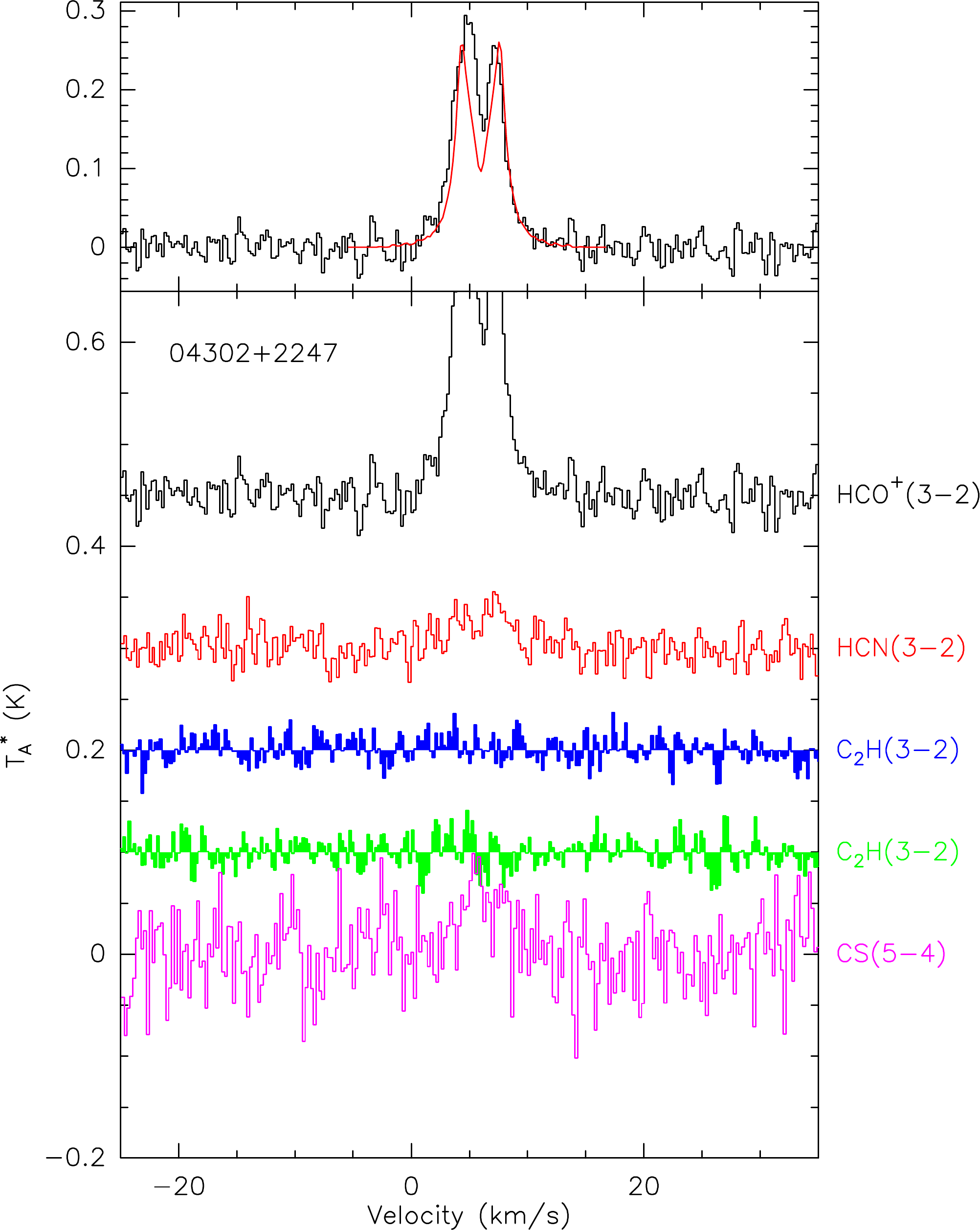}
\caption{Spectra of the observed transitions towards 04302+2247}
\label{fig:04302+2247}
\end{figure}
\begin{figure}
\includegraphics[height=11.5cm]{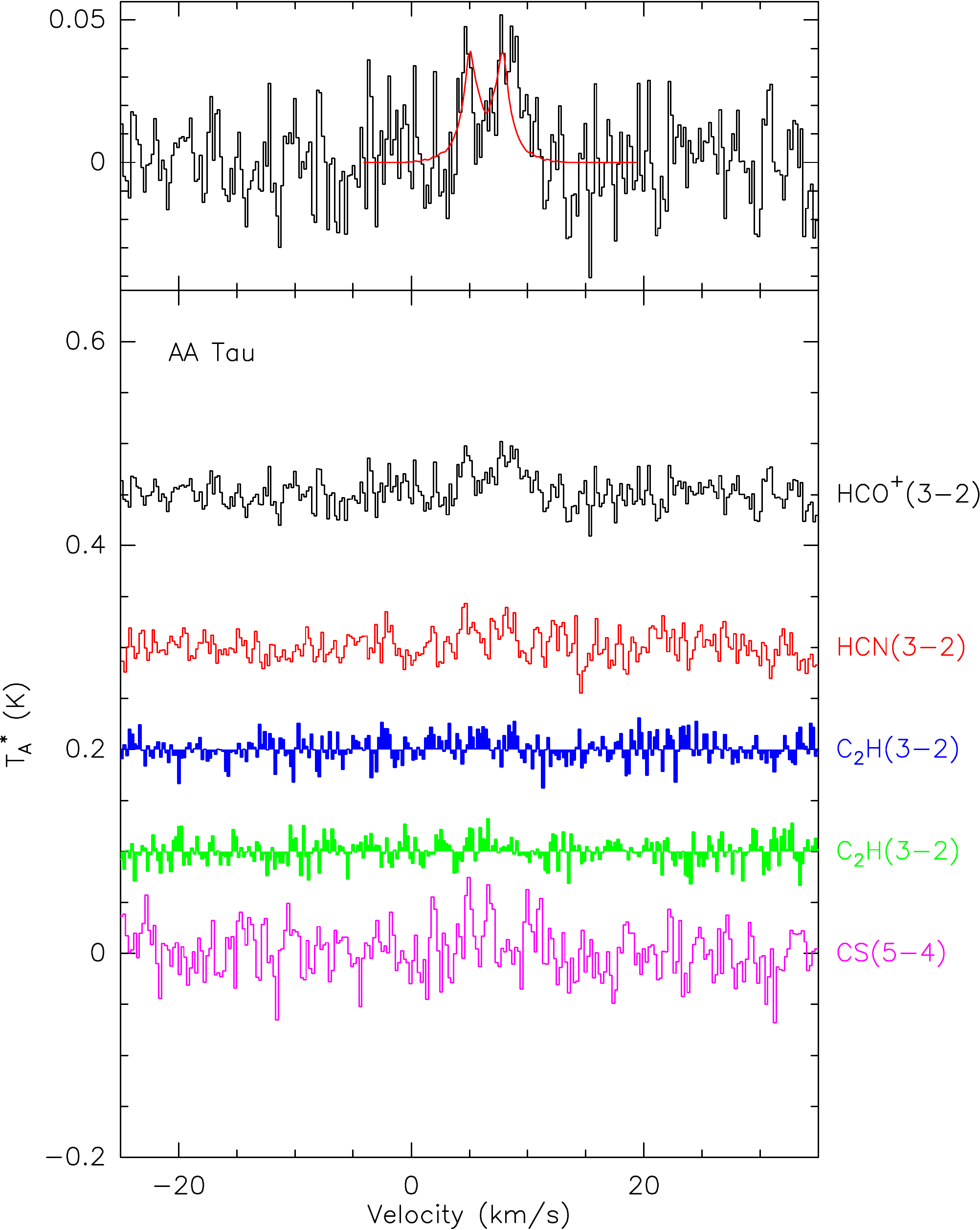}
\caption{Spectra of the observed transitions towards AA Tau}
\label{fig:AA_TAU}
\end{figure}
\clearpage
\begin{figure}
\includegraphics[height=11.5cm]{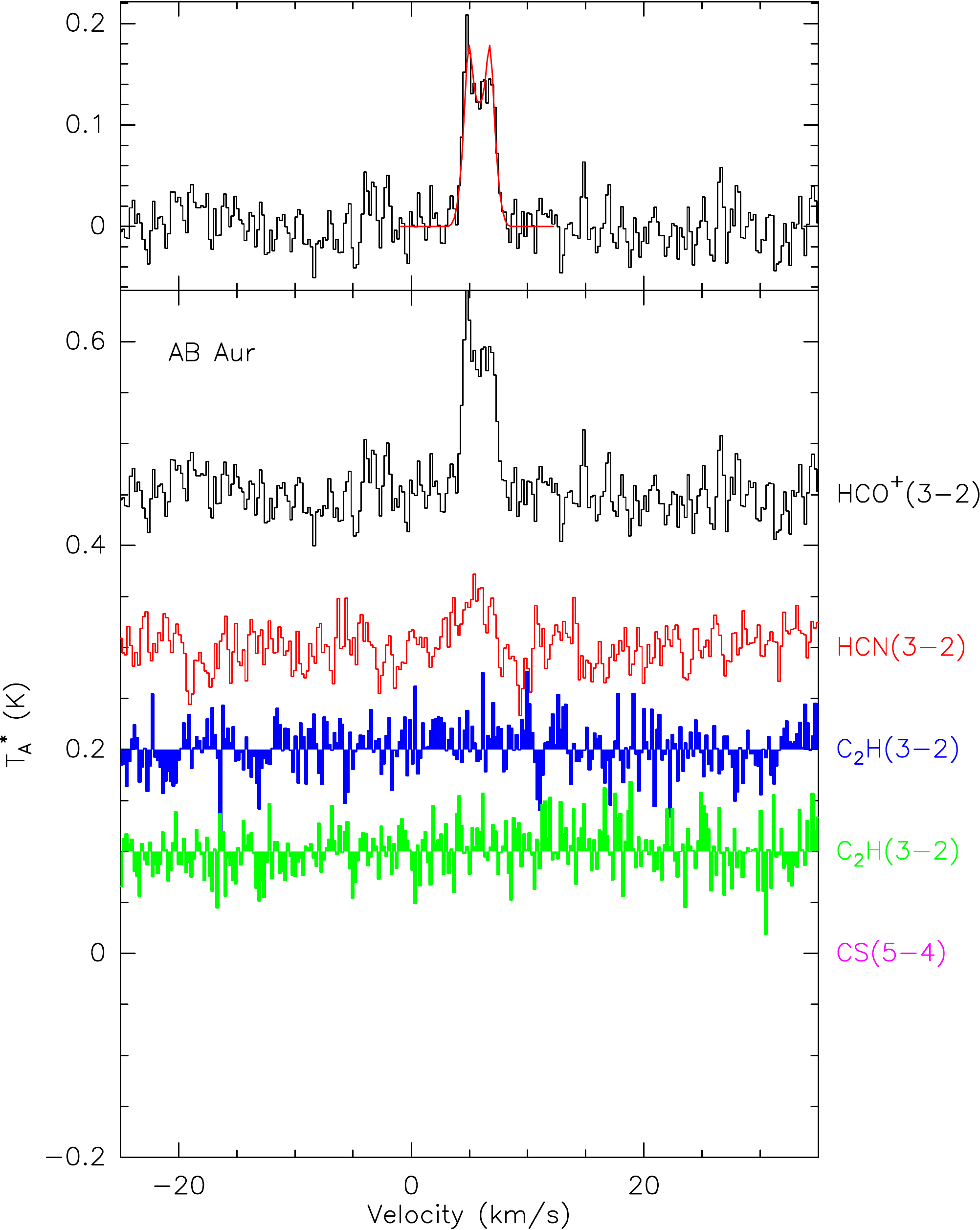}
\caption{Spectra of the observed transitions towards AB Aur}
\label{fig:AB_AUR}
\end{figure}
\begin{figure}
\includegraphics[height=11.5cm]{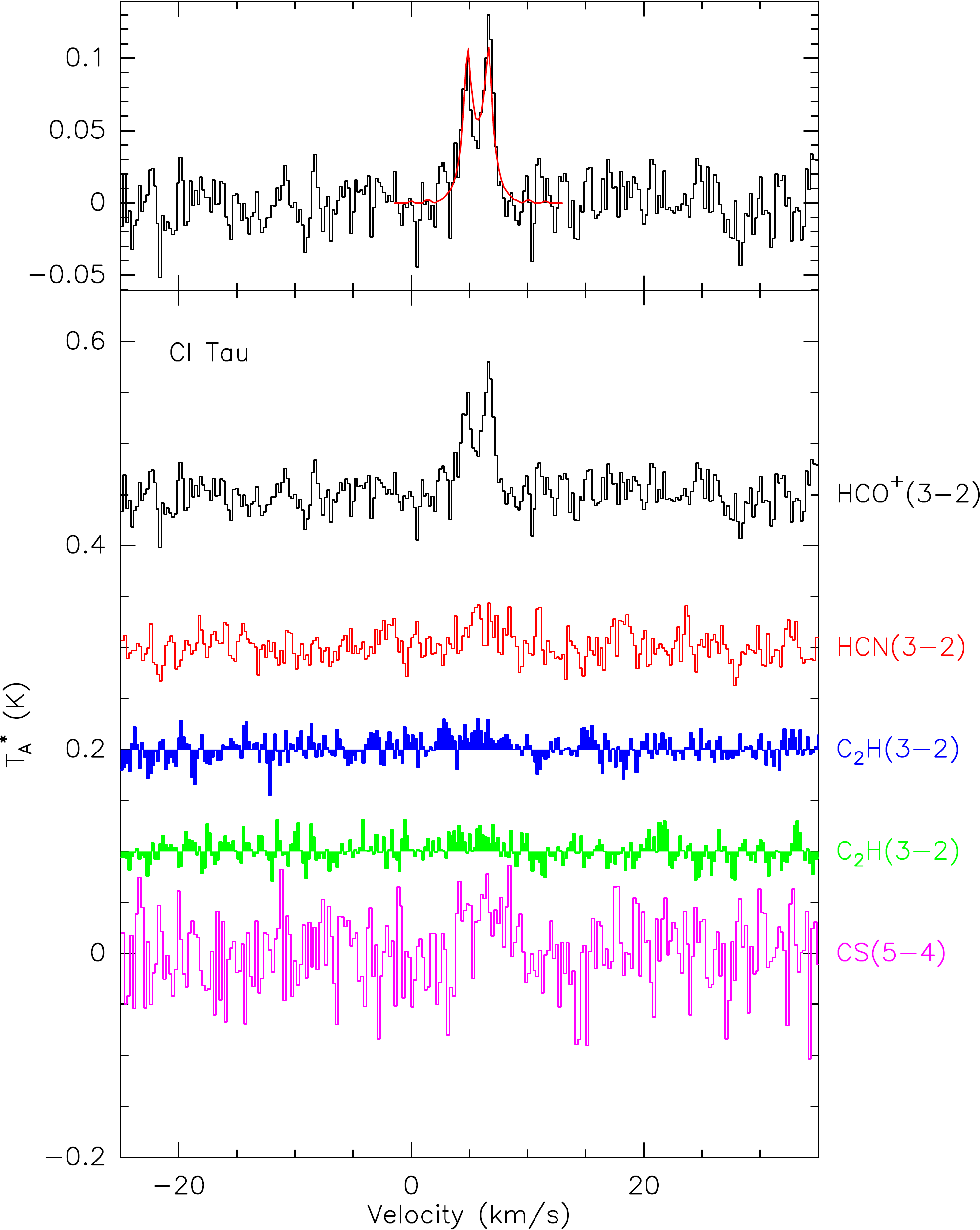}
\caption{Spectra of the observed transitions towards CI Tau}
\label{fig:CI_TAU}
\end{figure}
\begin{figure}
\includegraphics[height=11.5cm]{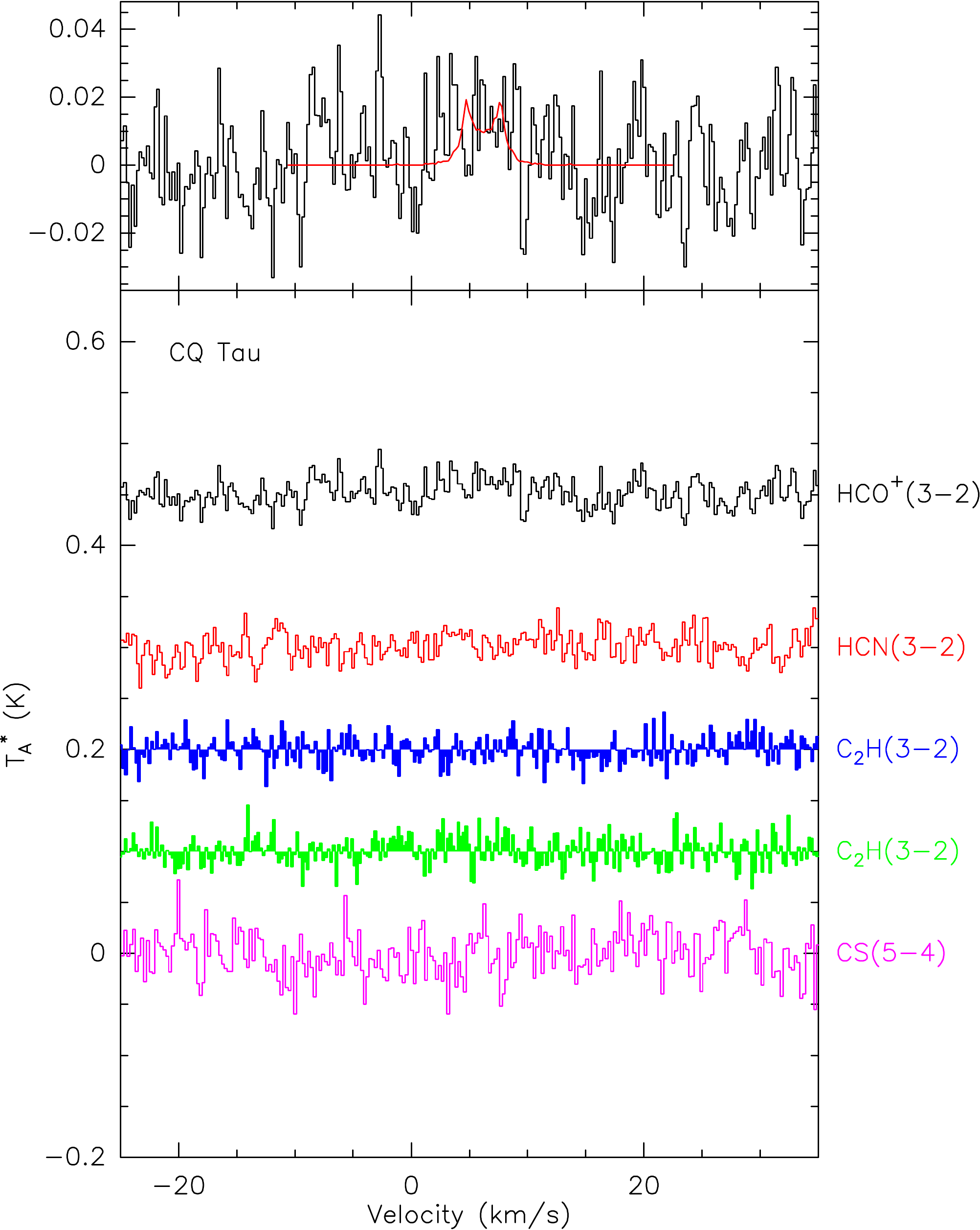}
\caption{Spectra of the observed transitions towards CQ Tau}
\label{fig:CQ_TAU}
\end{figure}
\begin{figure}[!t]
\includegraphics[height=11.5cm]{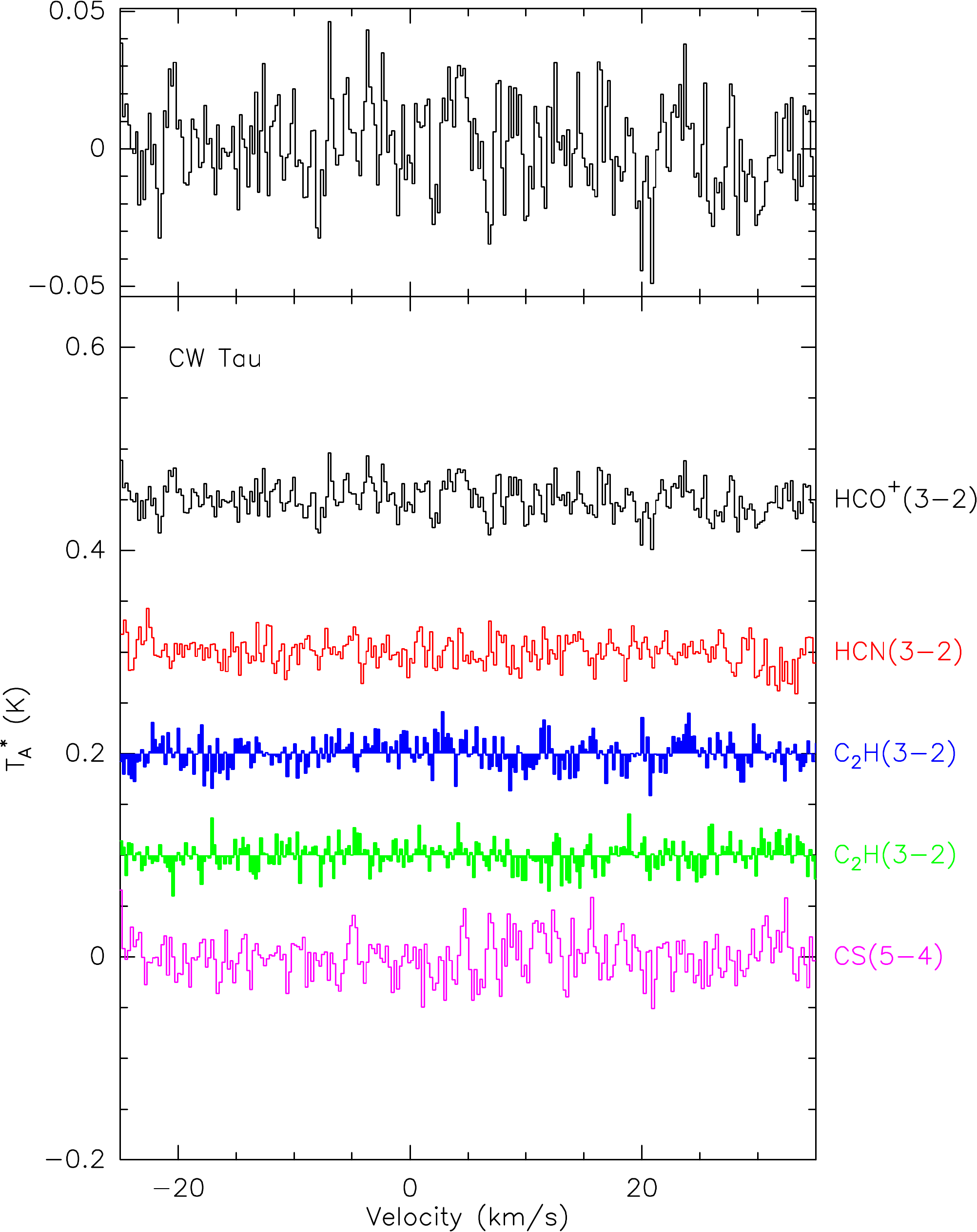}
\caption{Spectra of the observed transitions towards CW Tau}
\label{fig:CW_TAU}
\end{figure}
\clearpage
\begin{figure}[!t]
\includegraphics[height=11.5cm]{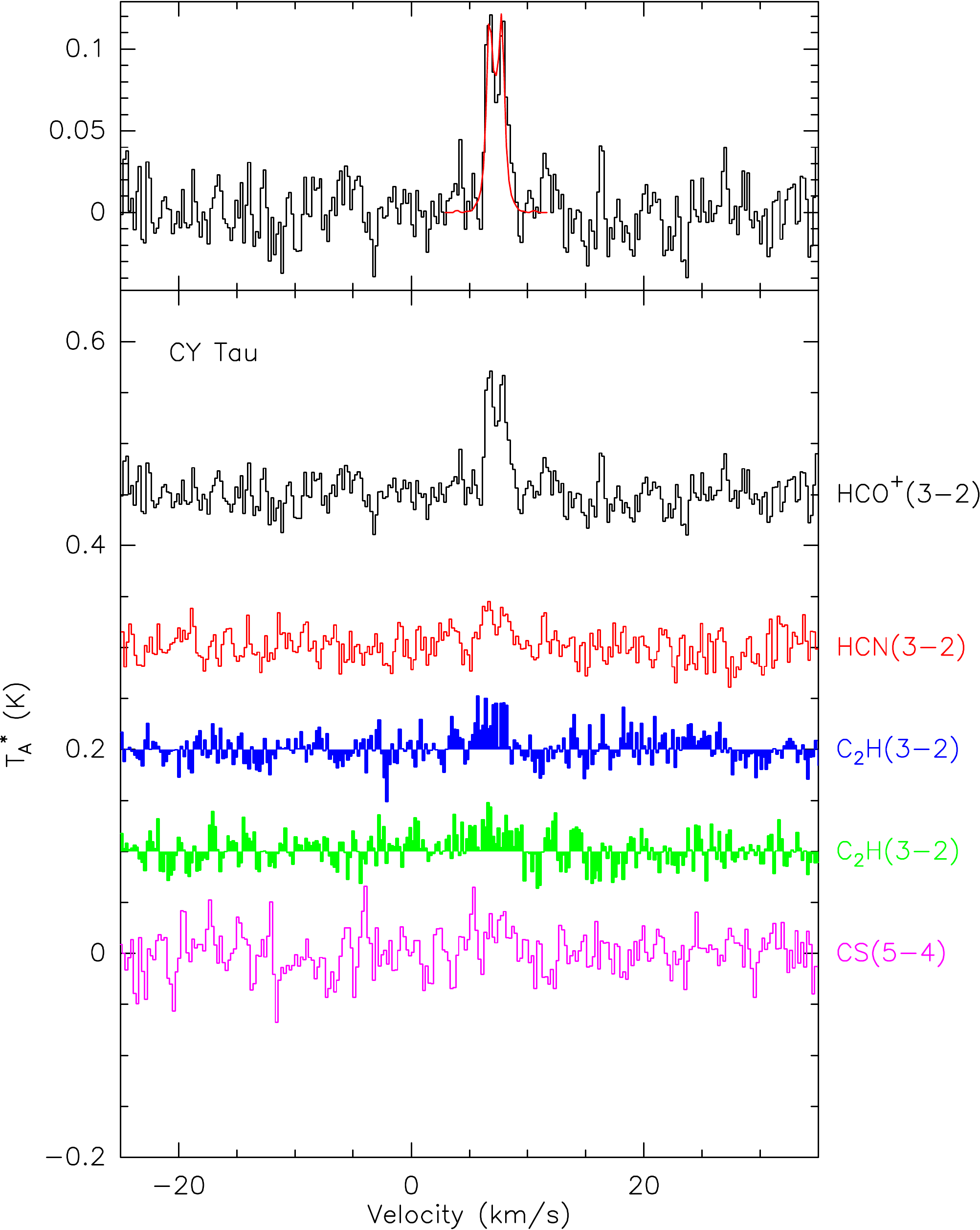}
\caption{Spectra of the observed transitions towards CY Tau}
\label{fig:CY_TAU}
\end{figure}
\begin{figure}
\includegraphics[height=11.5cm]{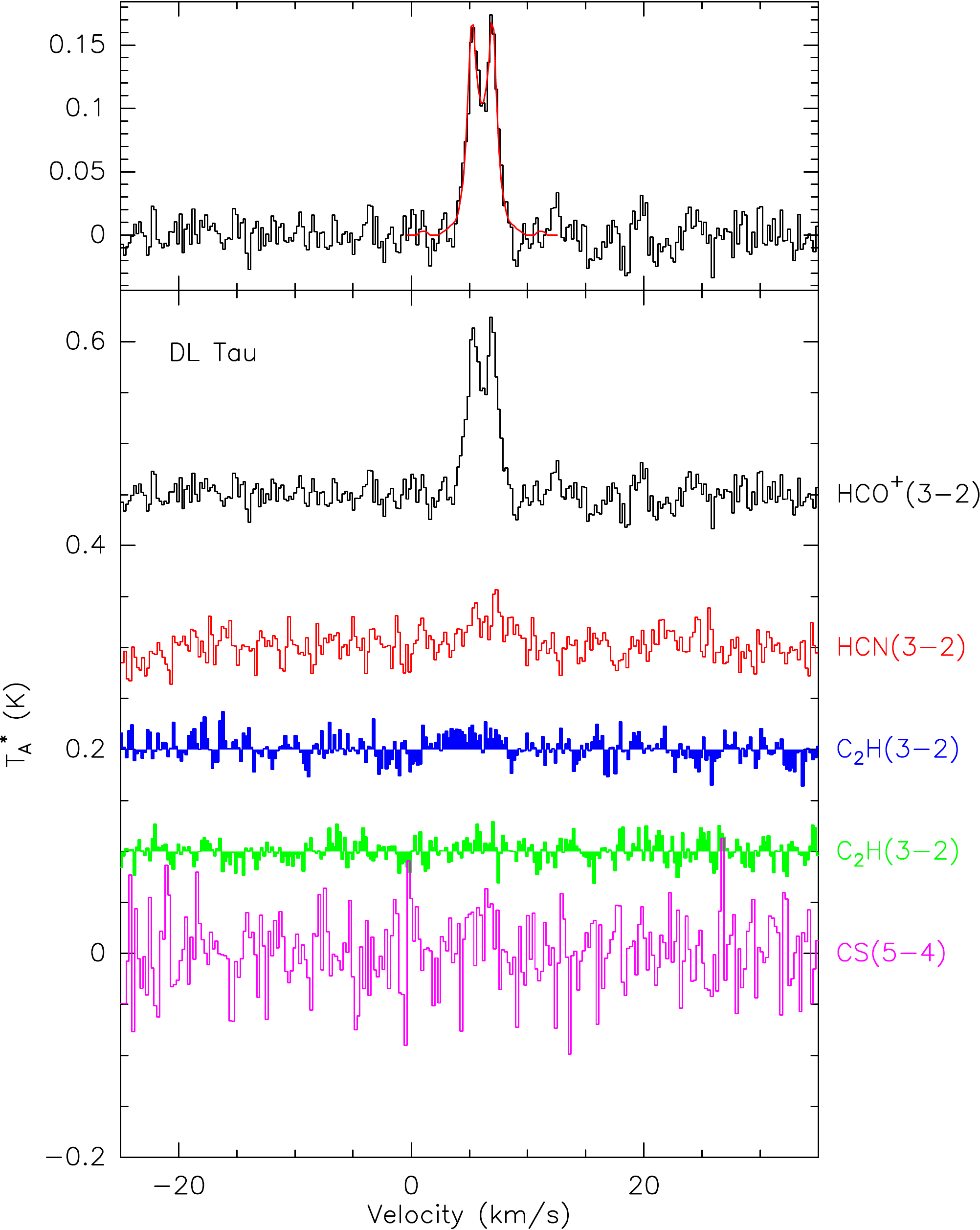}
\caption{Spectra of the observed transitions towards DL Tau}
\label{fig:DL_TAU}
\end{figure}
\begin{figure}
\includegraphics[height=11.5cm]{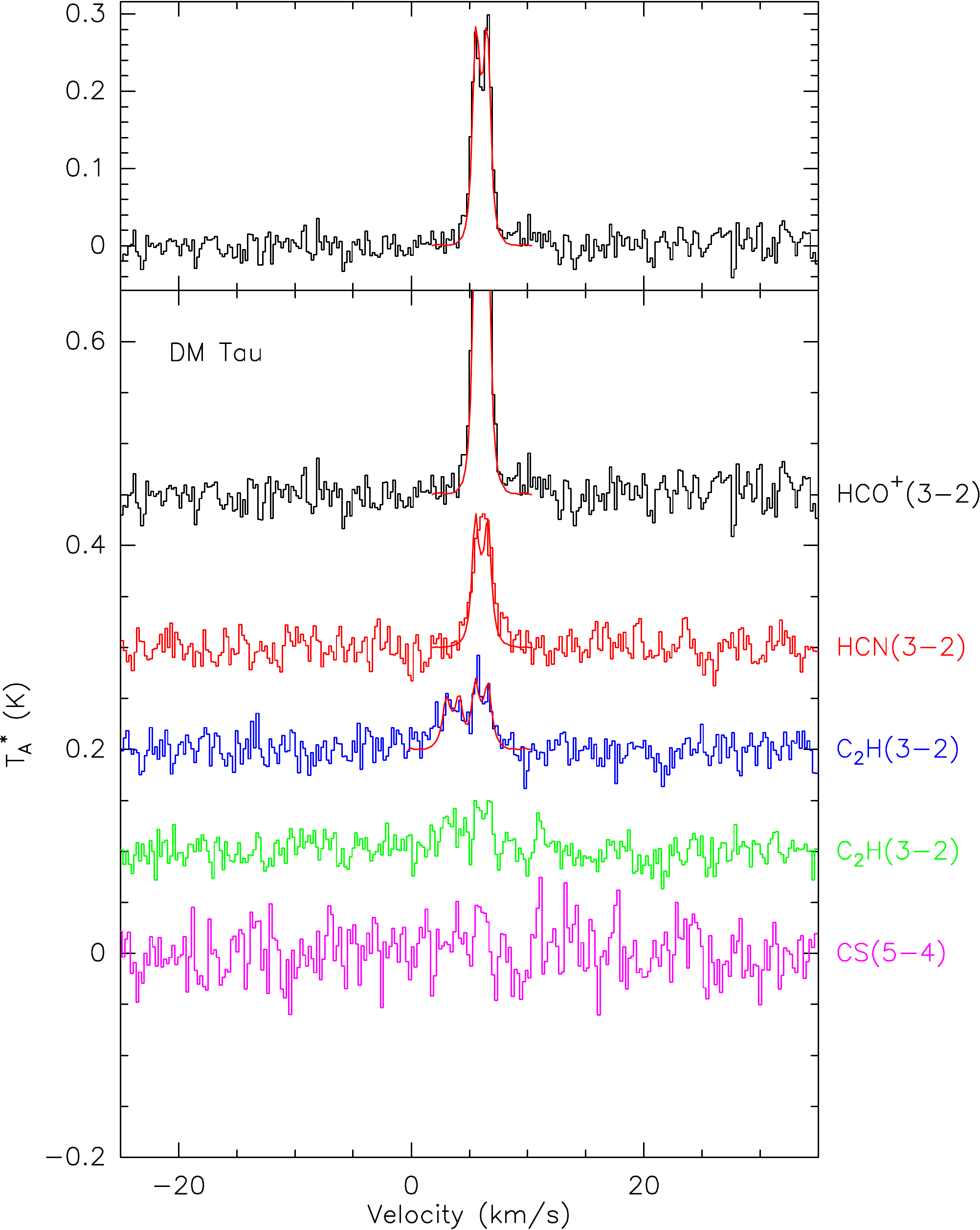}
\caption{Spectra of the observed transitions towards DM Tau}
\label{fig:DM_TAU}
\end{figure}
\begin{figure}
\includegraphics[height=11.5cm]{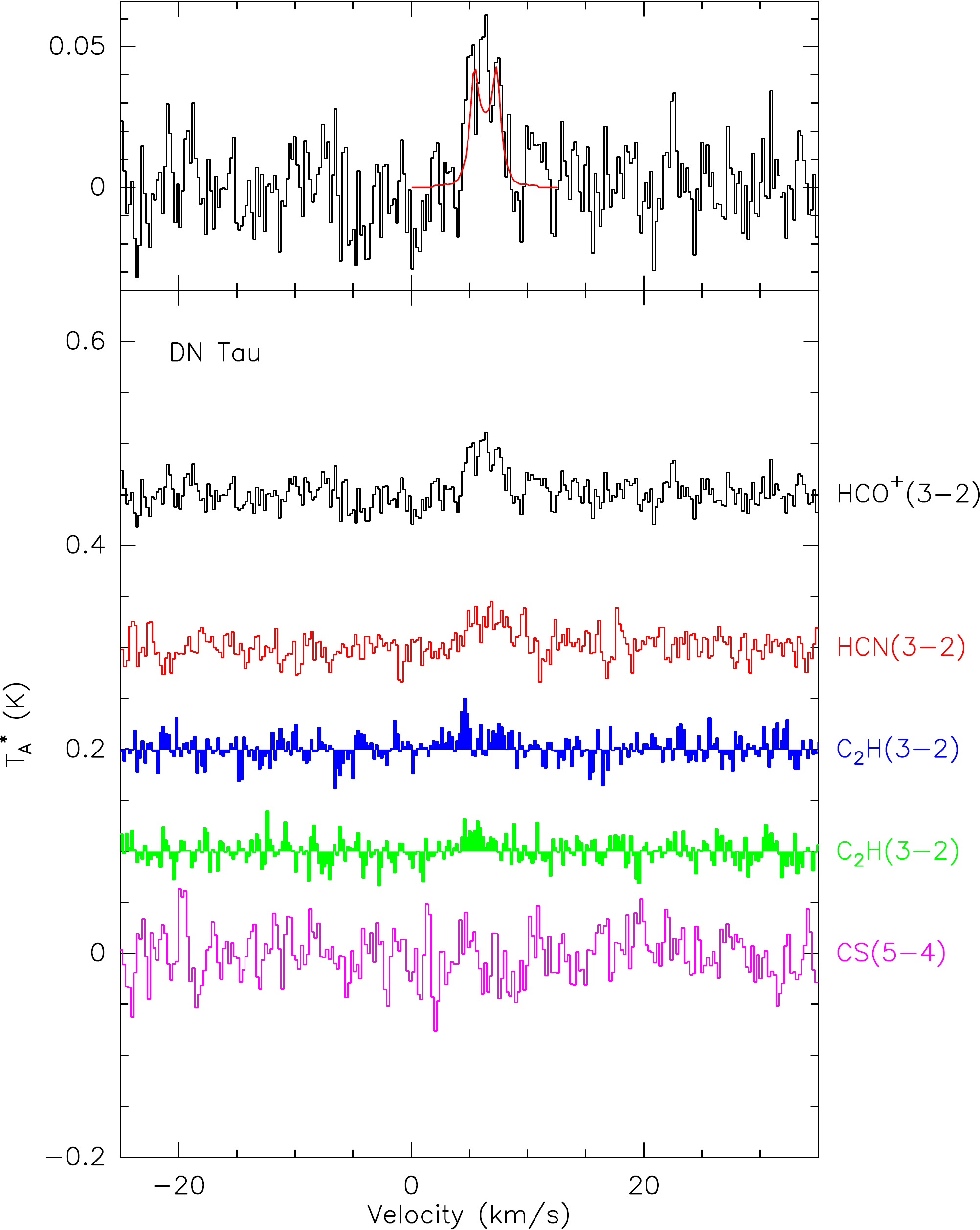}
\caption{Spectra of the observed transitions towards DN Tau}
\label{fig:DN_TAU}
\end{figure}
\clearpage
\begin{figure}
\includegraphics[height=11.5cm]{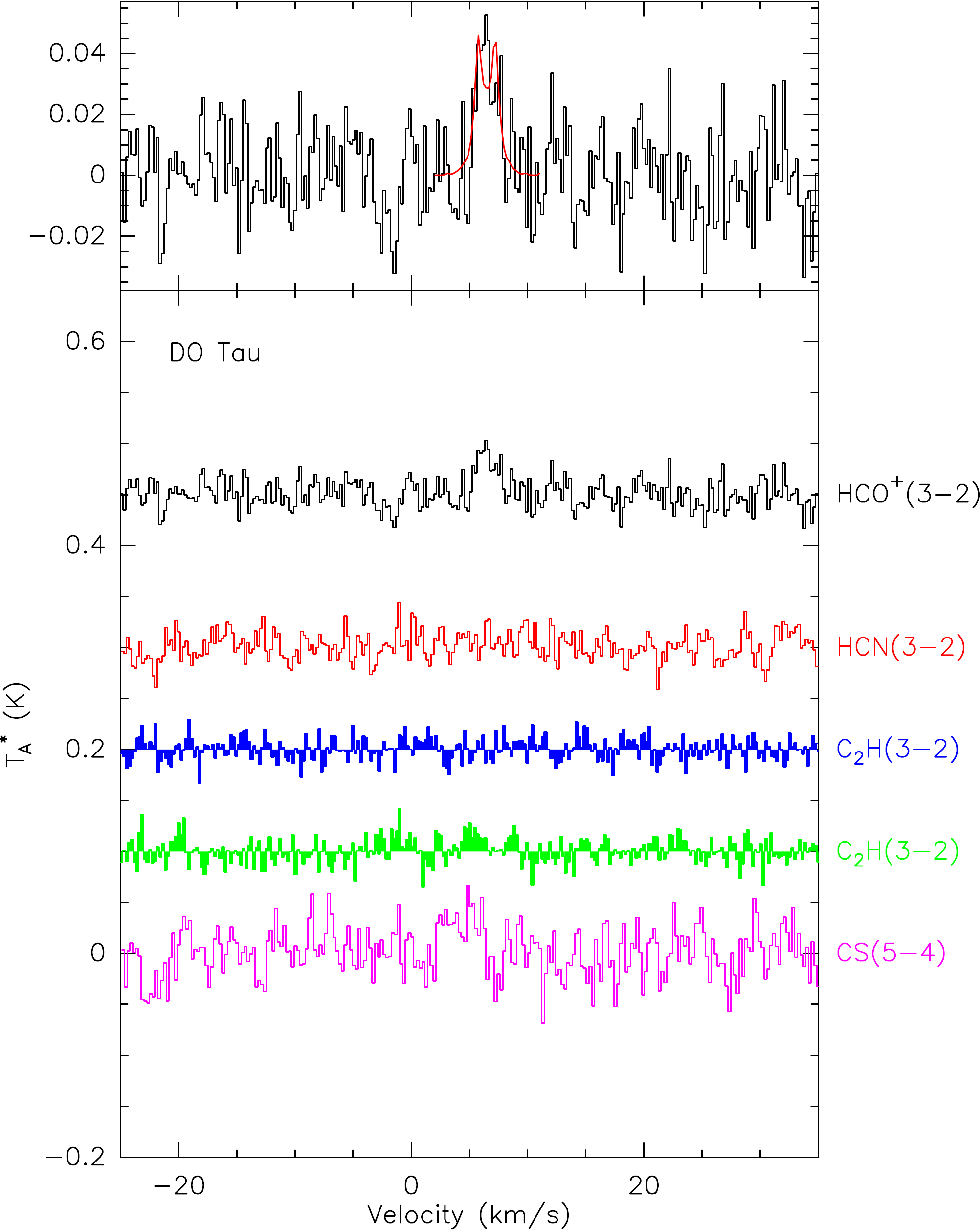}
\caption{Spectra of the observed transitions towards DO Tau}
\label{fig:DO_TAU}
\end{figure}
\begin{figure}
\includegraphics[height=11.5cm]{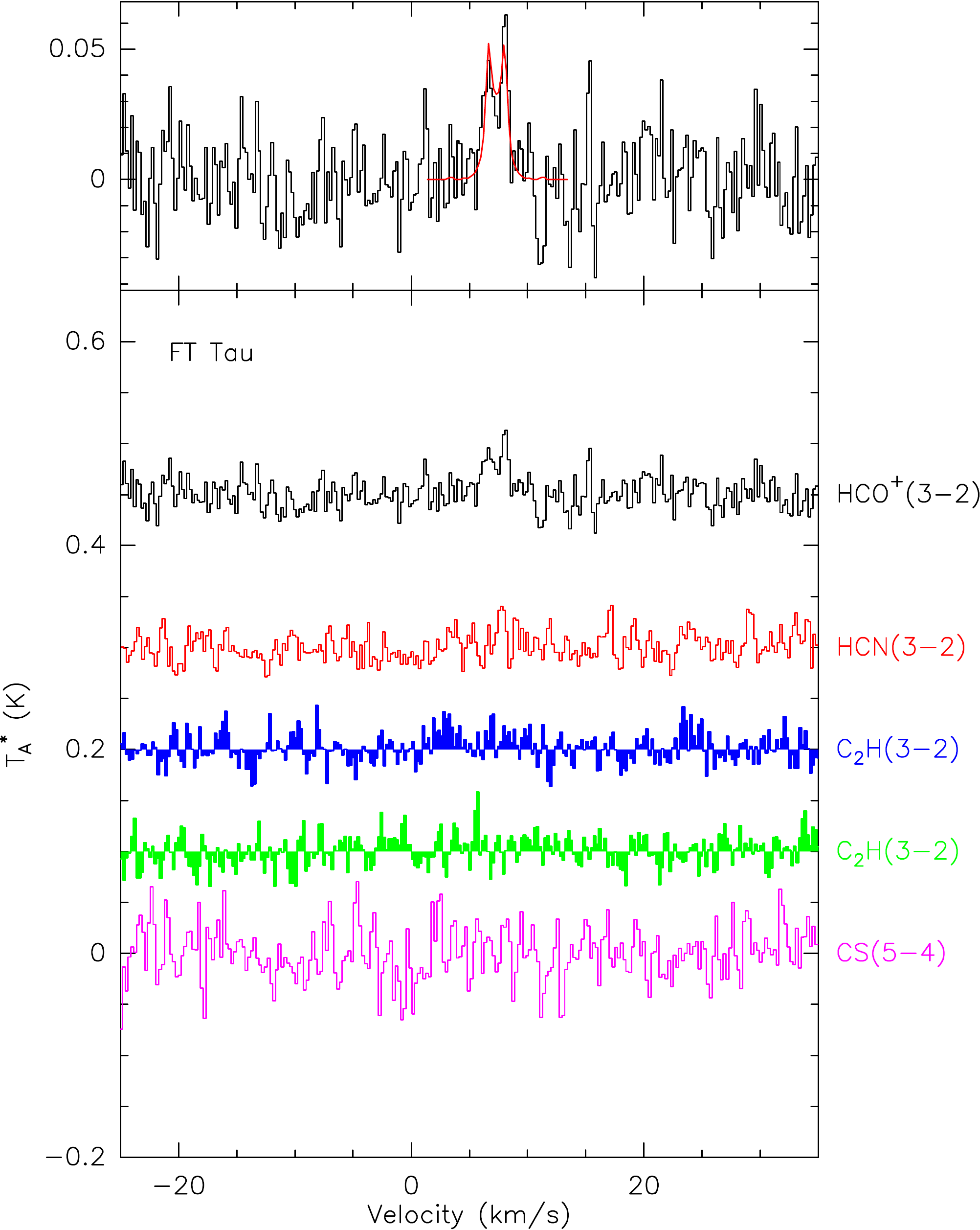}
\caption{Spectra of the observed transitions towards FT Tau}
\label{fig:FT_TAU}
\end{figure}
\begin{figure}
\includegraphics[height=11.5cm]{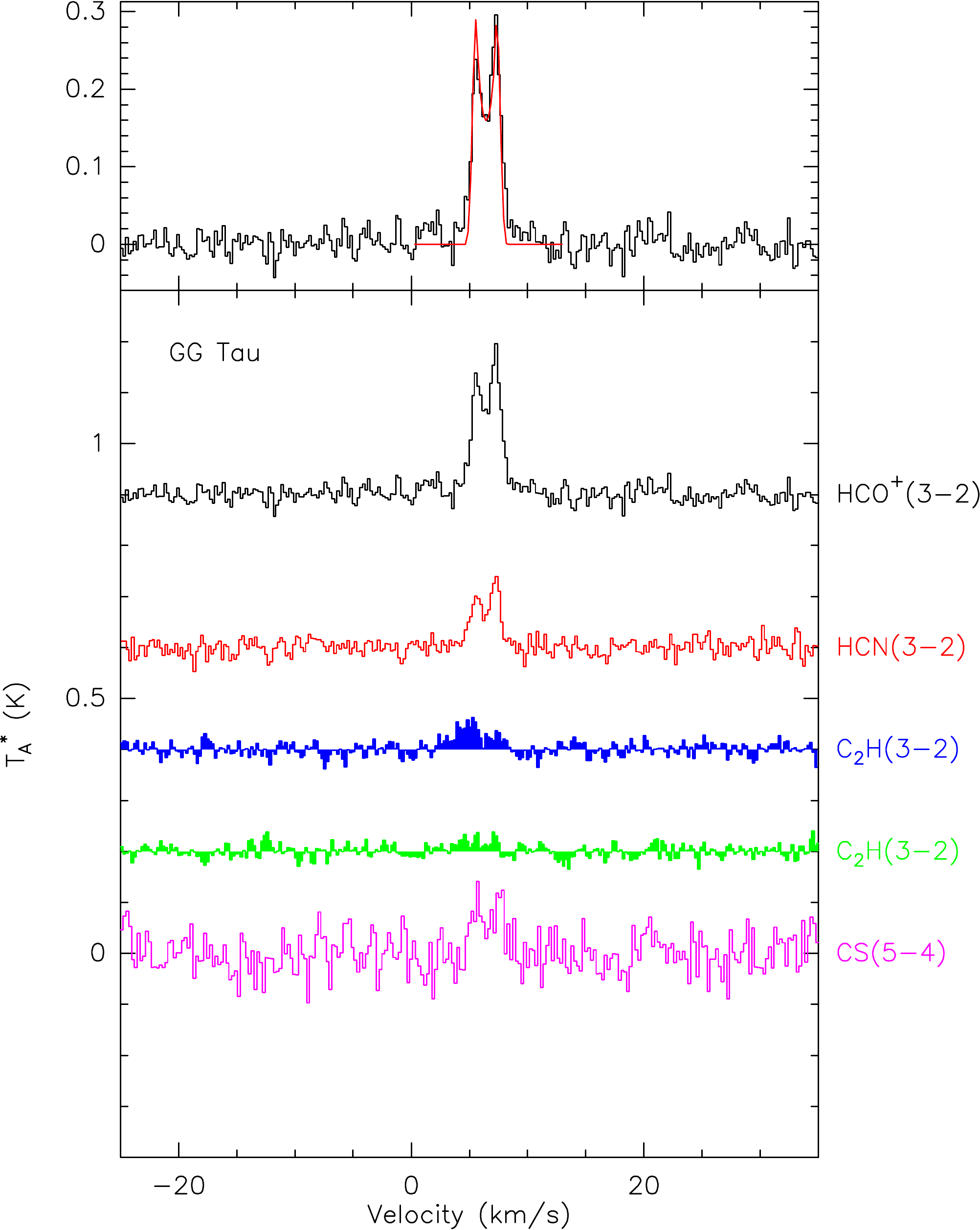}
\caption{Spectra of the observed transitions towards GG Tau}
\label{fig:GG_TAU}
\end{figure}
\begin{figure}
\includegraphics[height=11.5cm]{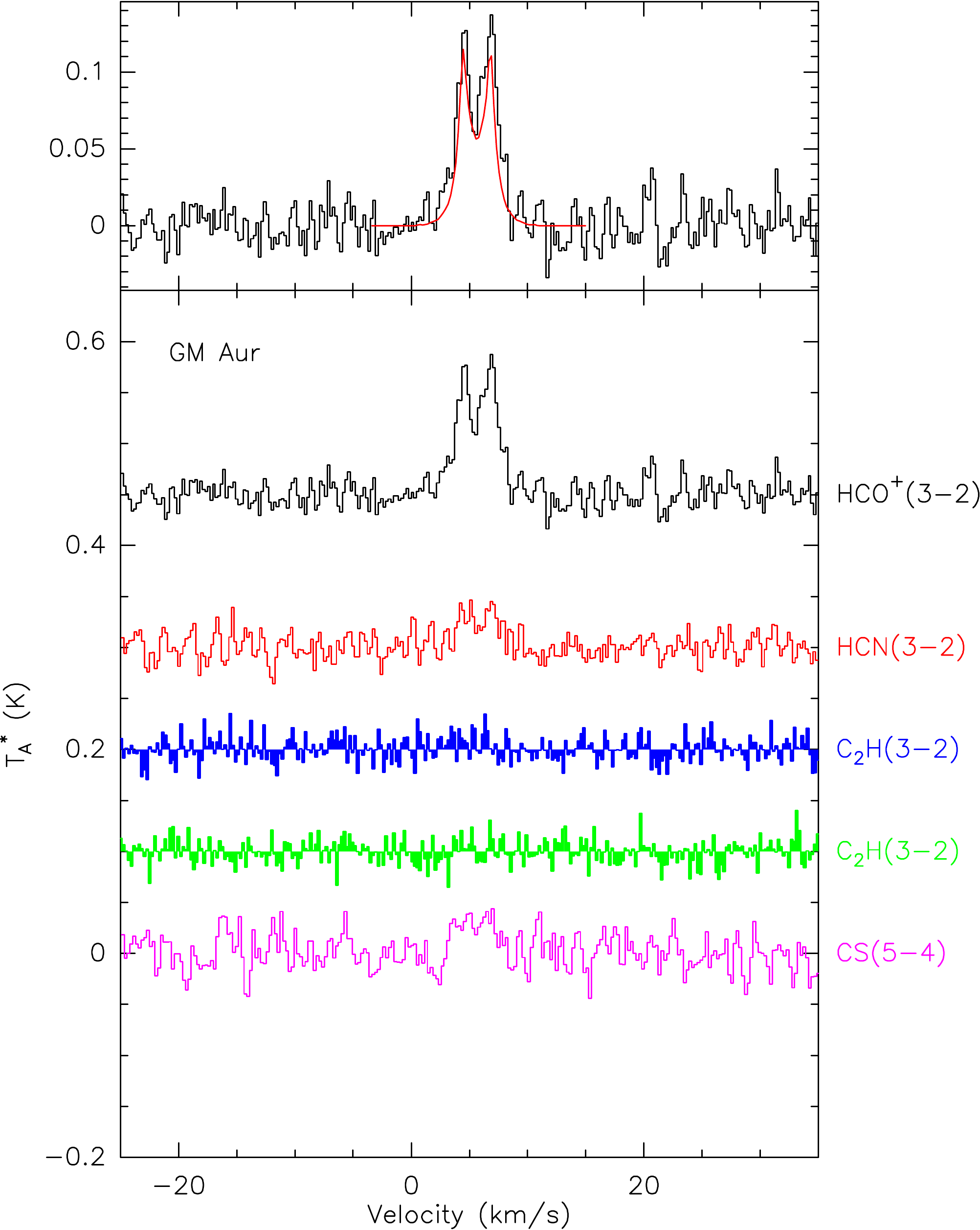}
\caption{Spectra of the observed transitions towards GM Aur}
\label{fig:GM_AUR}
\end{figure}
\clearpage
\begin{figure}
\includegraphics[height=11.5cm]{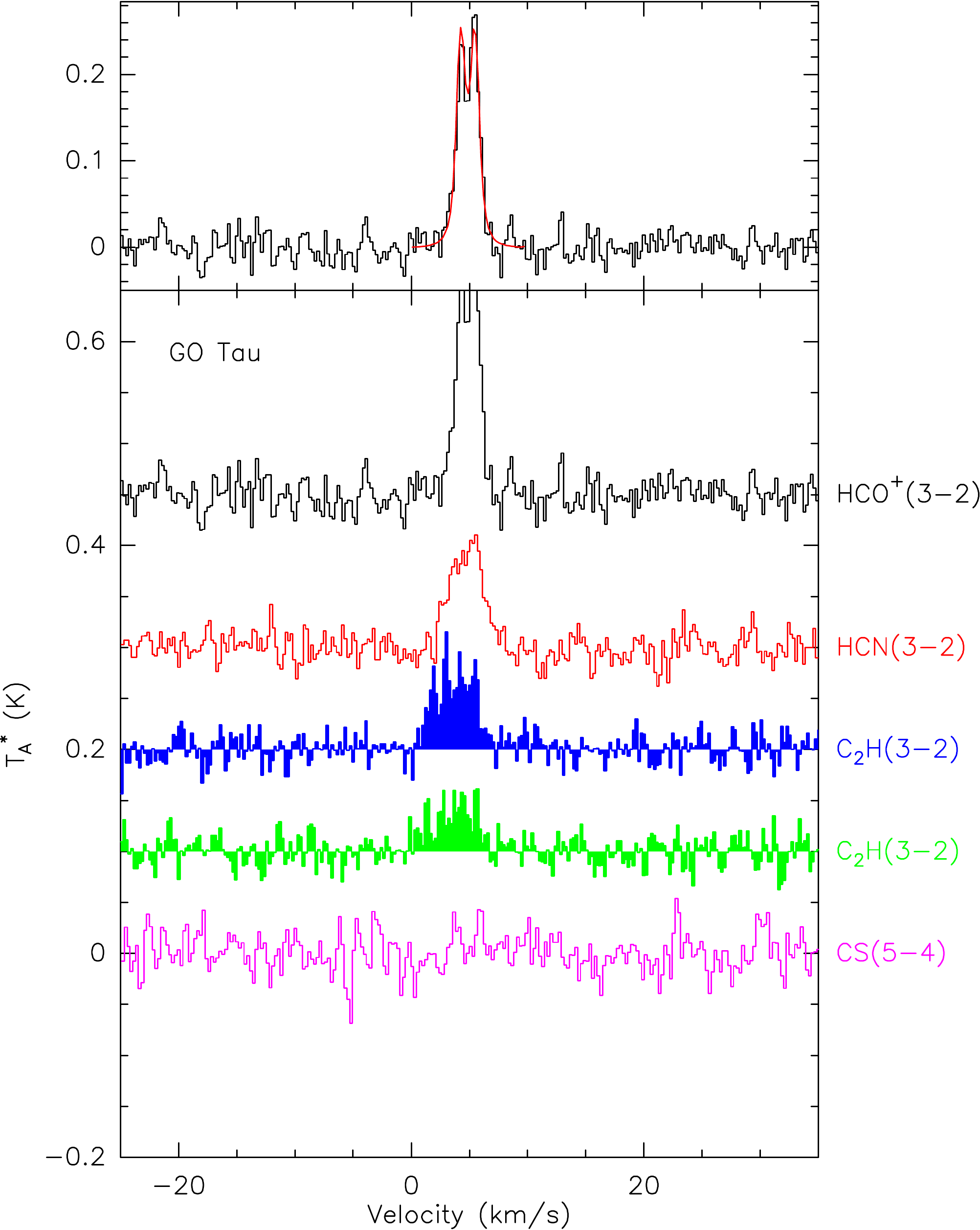}
\caption{Spectra of the observed transitions towards GO Tau}
\label{fig:GO_TAU}
\end{figure}
\begin{figure}
\includegraphics[height=11.5cm]{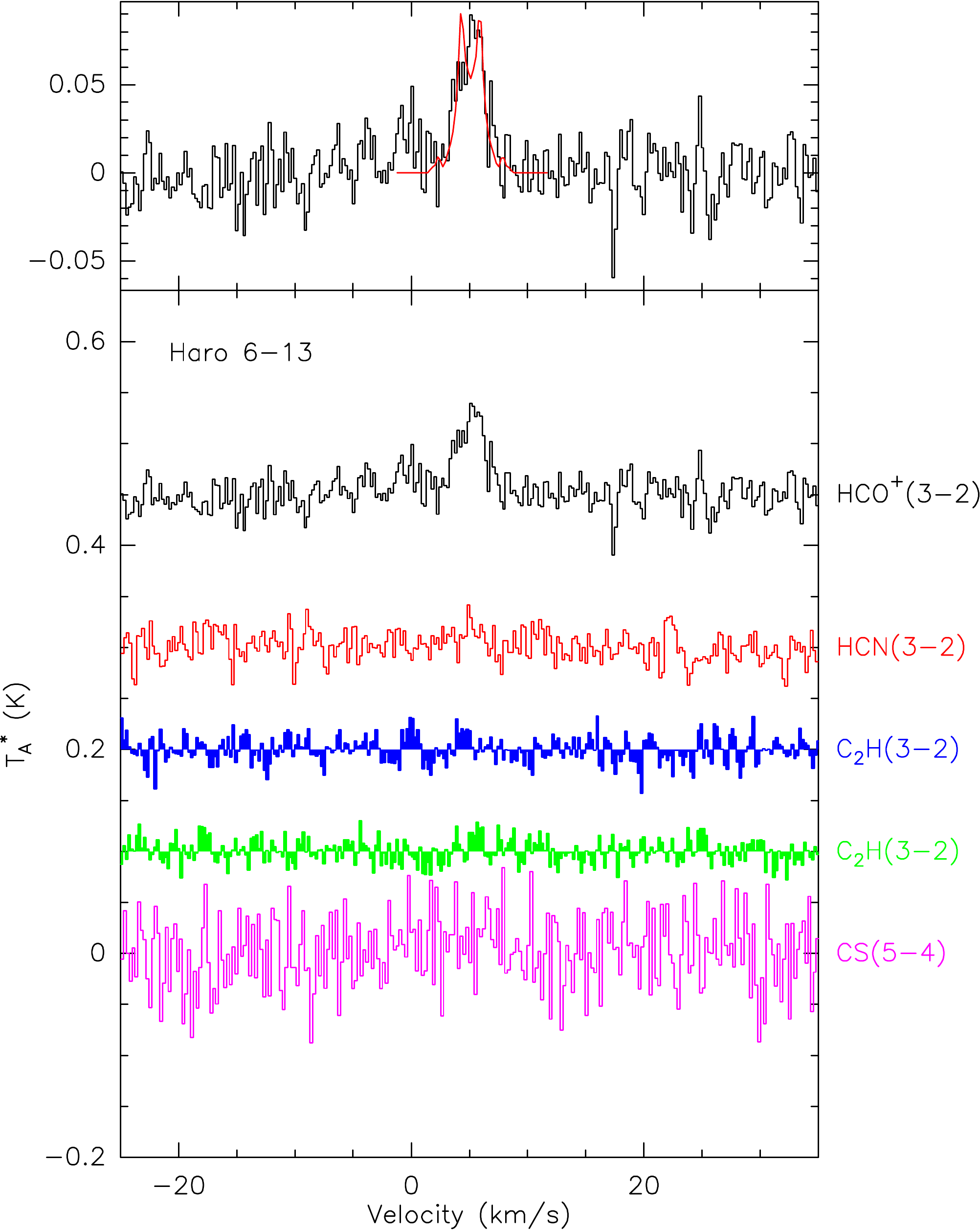}
\caption{Spectra of the observed transitions towards Haro 6-13}
\label{fig:HARO6-13}
\end{figure}
\begin{figure}
\includegraphics[height=11.5cm]{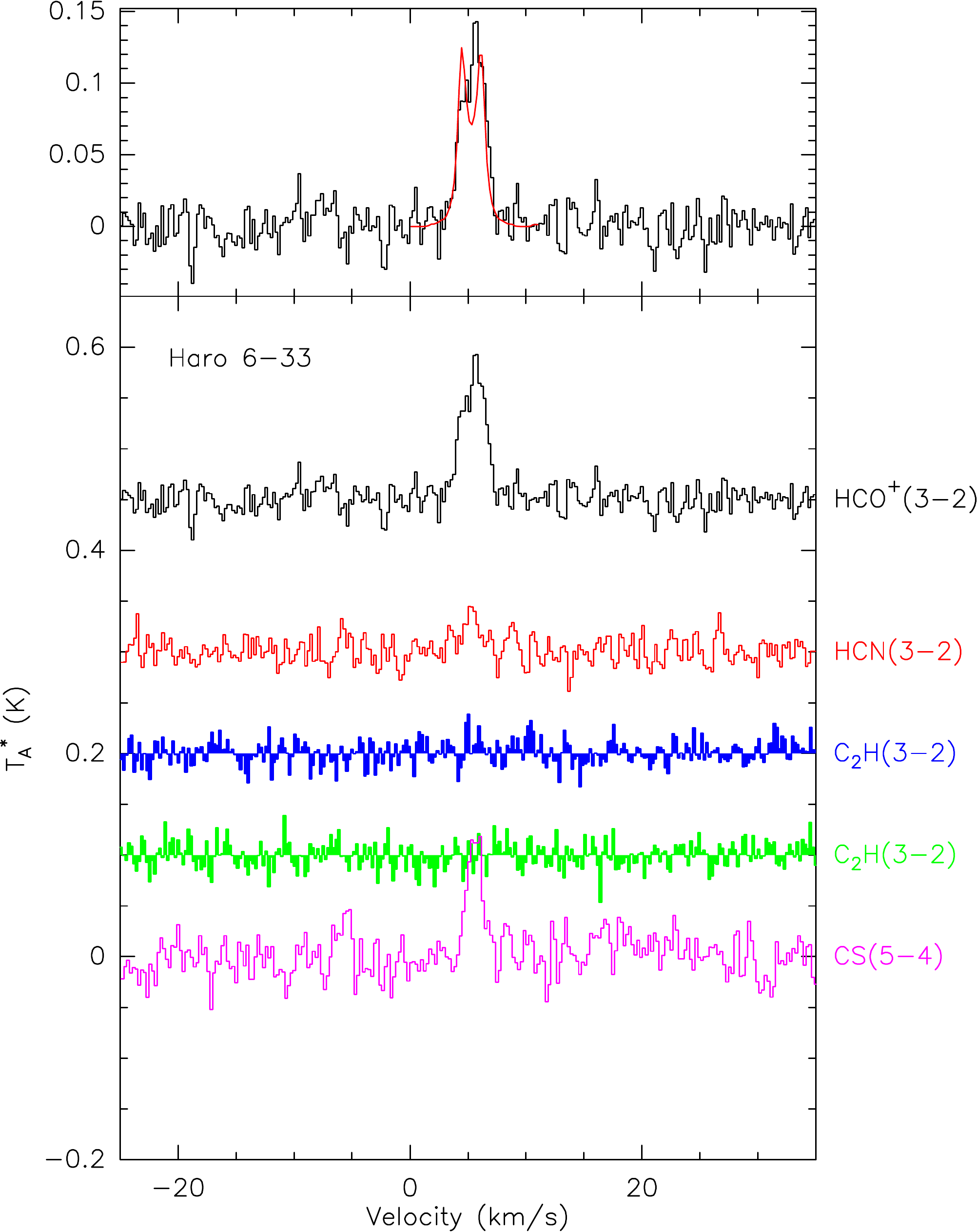}
\caption{Spectra of the observed transitions towards Haro 6-33}
\label{fig:HARO6-33}
\end{figure}
\begin{figure}
\includegraphics[height=11.5cm]{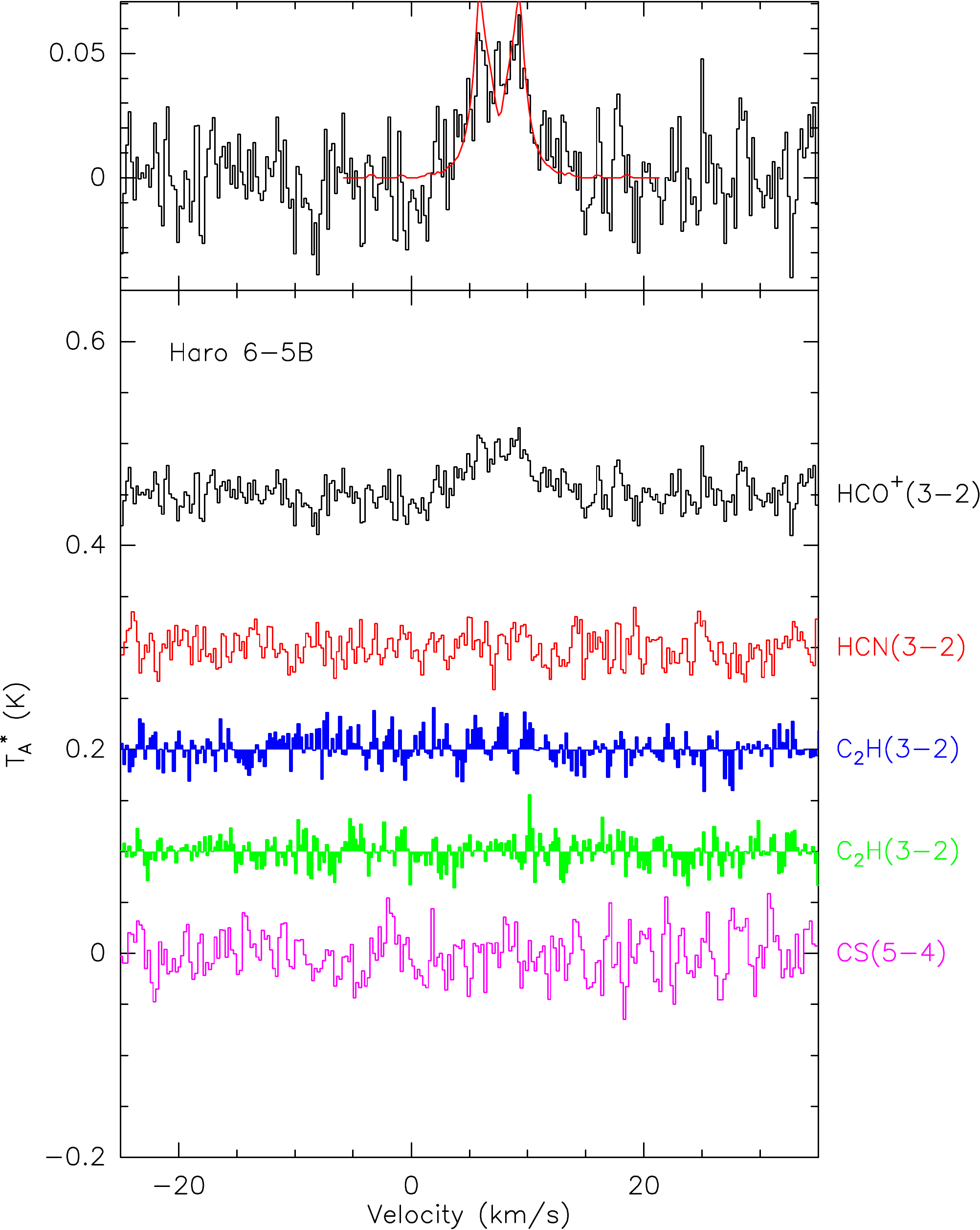}
\caption{Spectra of the observed transitions towards Haro 6-5B}
\label{fig:HARO6-5B}
\end{figure}
\clearpage
\begin{figure}
\includegraphics[height=11.5cm]{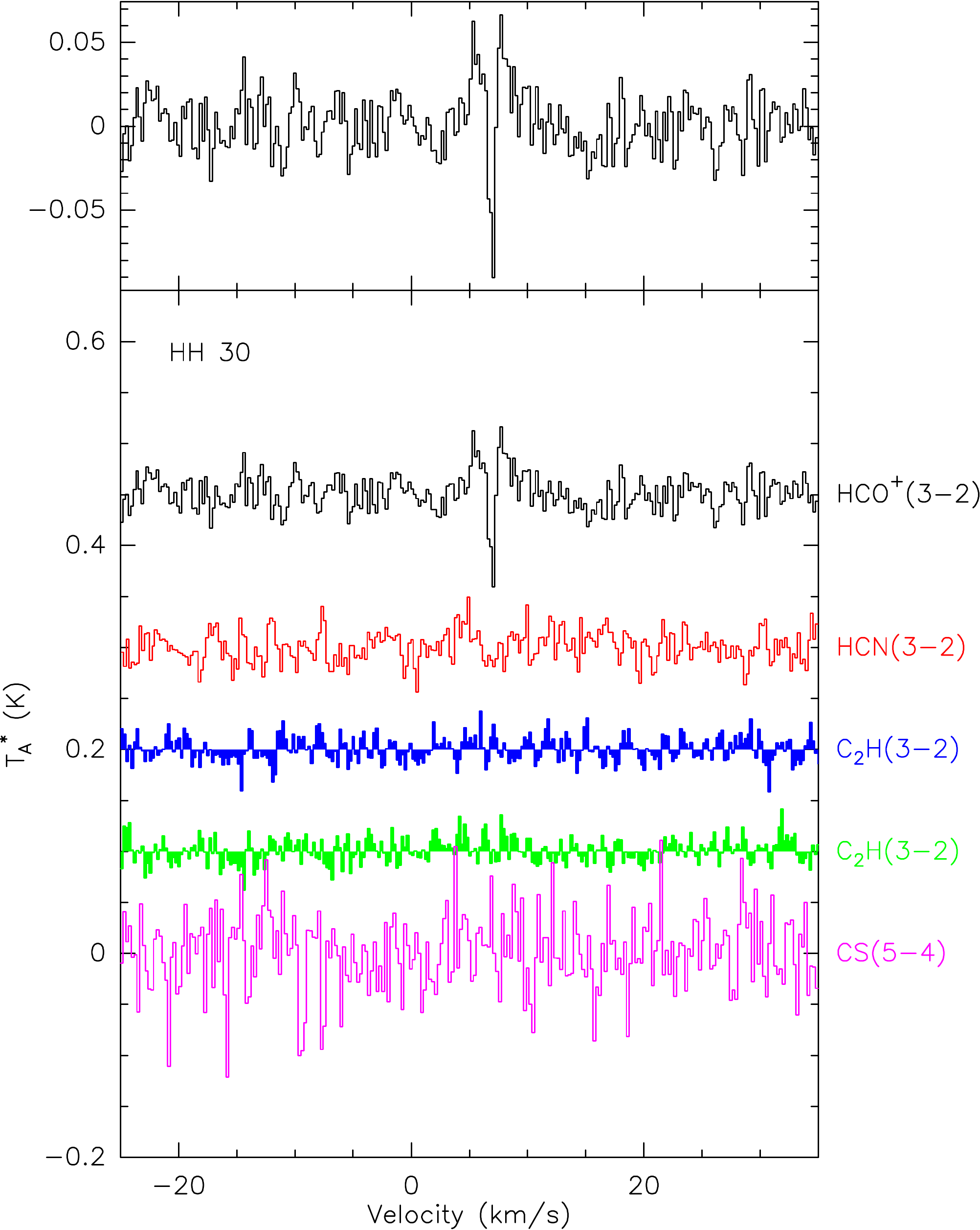}
\caption{Spectra of the observed transitions towards HH 30}
\label{fig:HH30}
\end{figure}
\begin{figure}
\includegraphics[height=11.5cm]{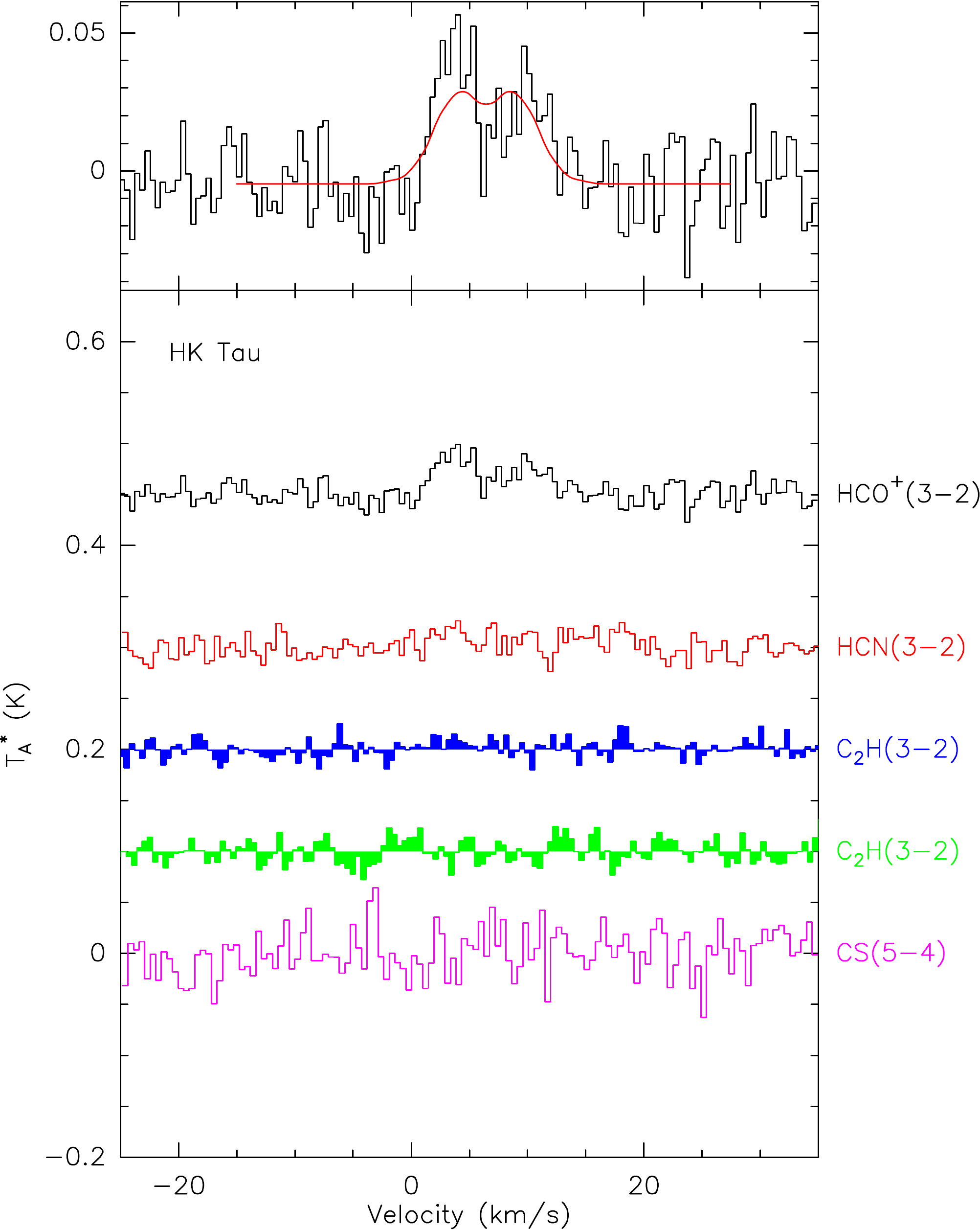}
\caption{Spectra of the observed transitions towards HK Tau}
\label{fig:HK_TAU}
\end{figure}
\begin{figure}
\includegraphics[height=11.5cm]{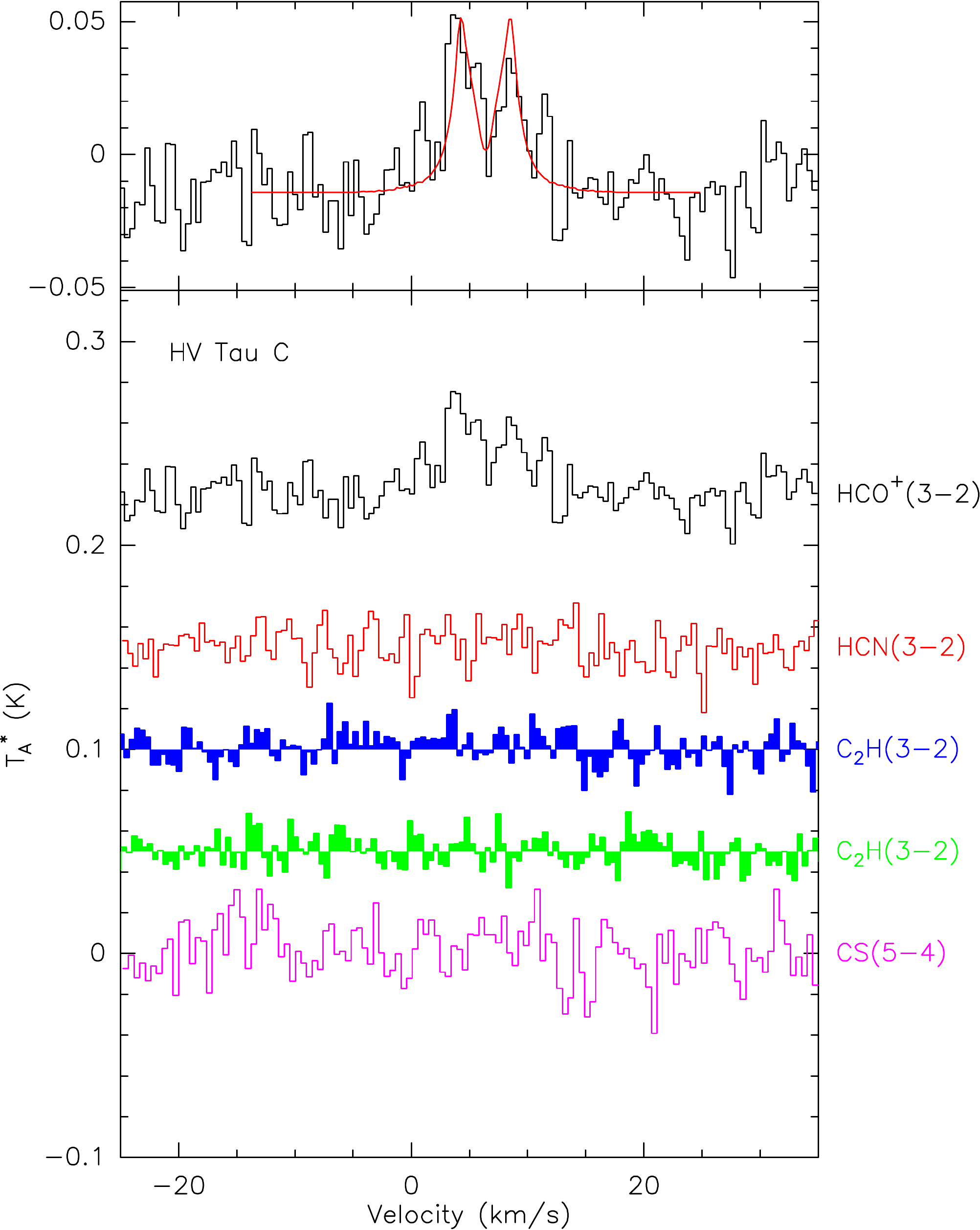}
\caption{Spectra of the observed transitions towards HV Tau C}
\label{fig:HV_TAU-C}
\end{figure}
\begin{figure}
\includegraphics[height=11.5cm]{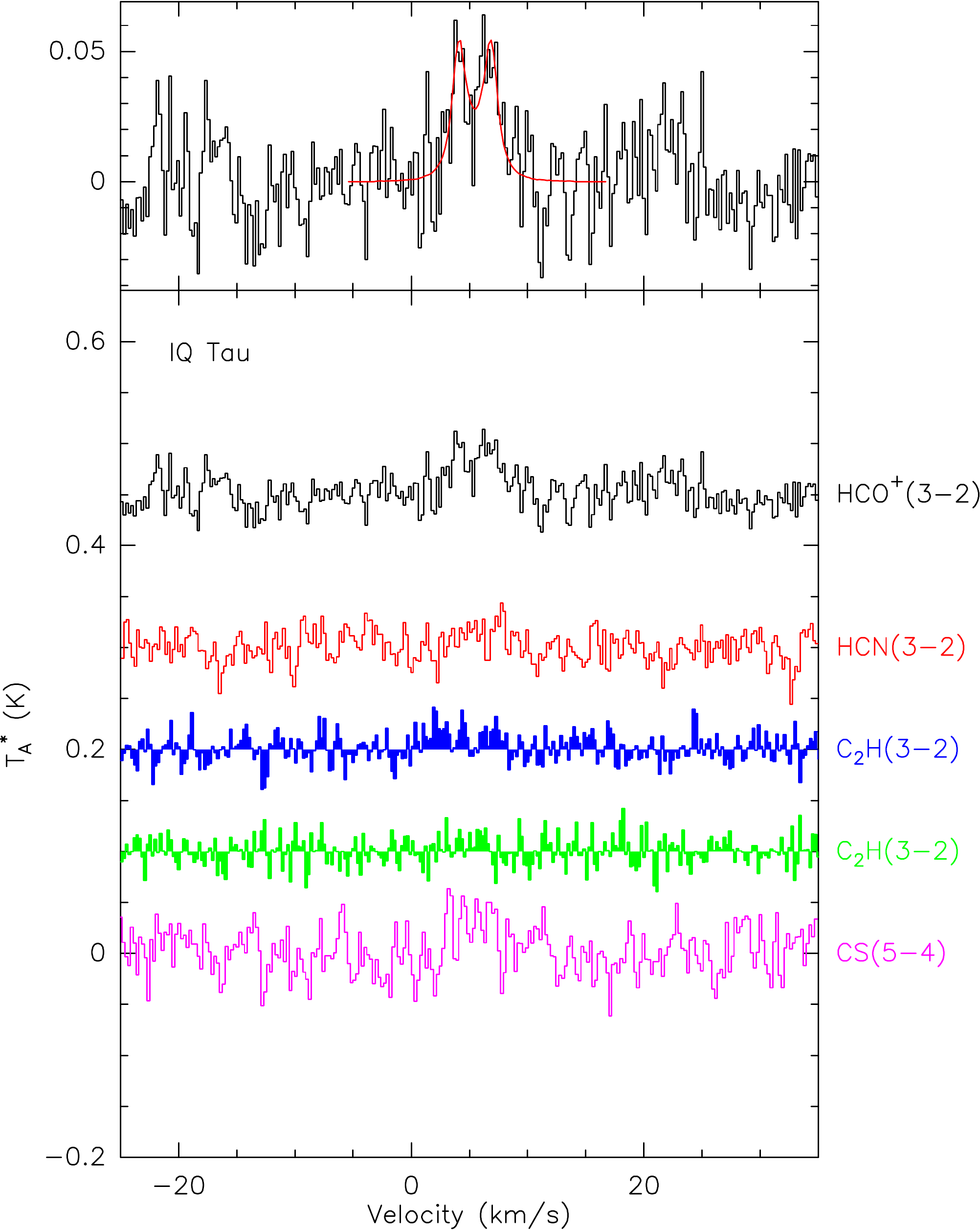}
\caption{Spectra of the observed transitions towards IQ Tau}
\label{fig:IQ_TAU}
\end{figure}
\clearpage
\begin{figure}
\includegraphics[height=11.5cm]{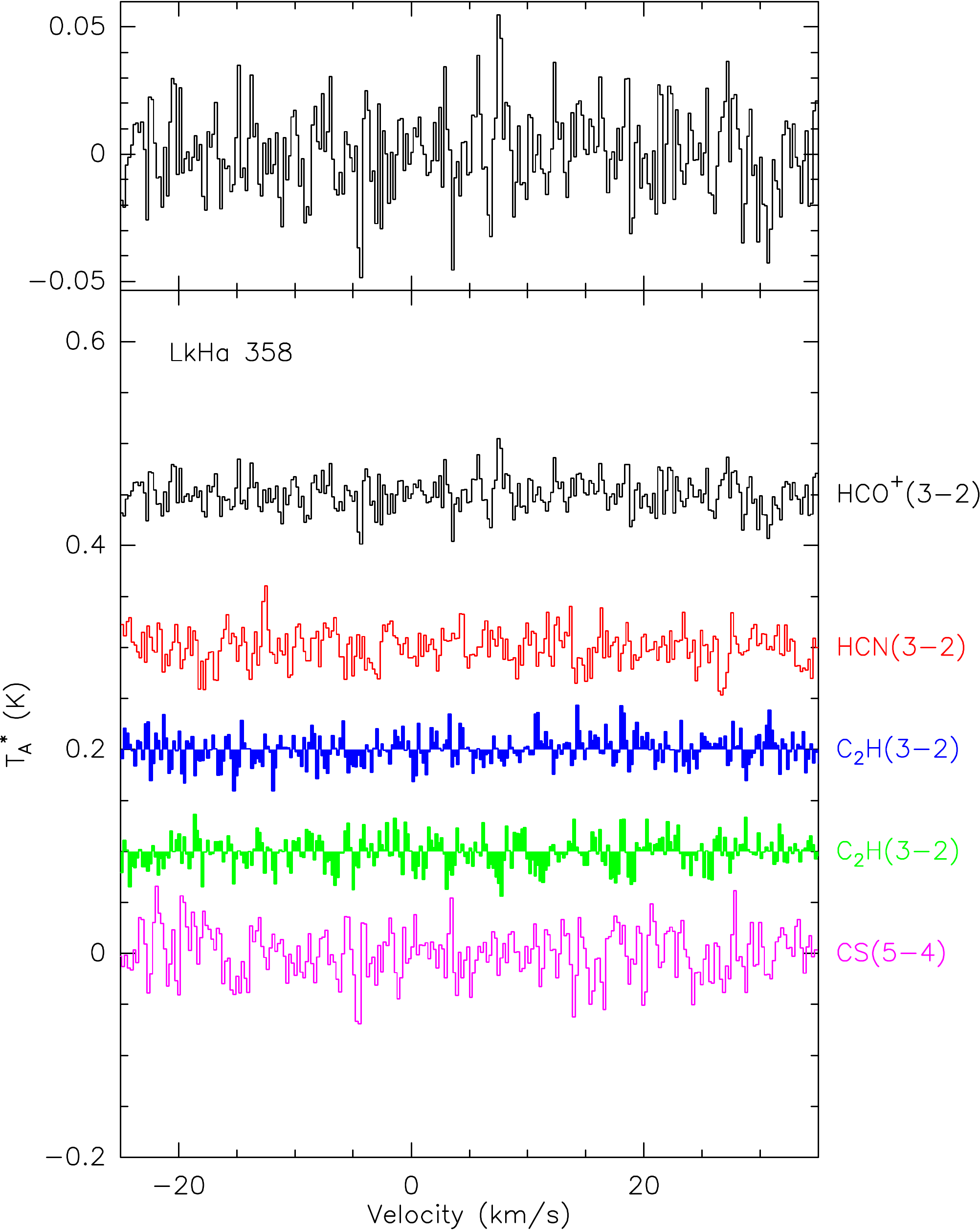}
\caption{Spectra of the observed transitions towards LkH$\alpha$ 358}
\label{fig:LKHA_358}
\end{figure}
\begin{figure}
\includegraphics[height=11.5cm]{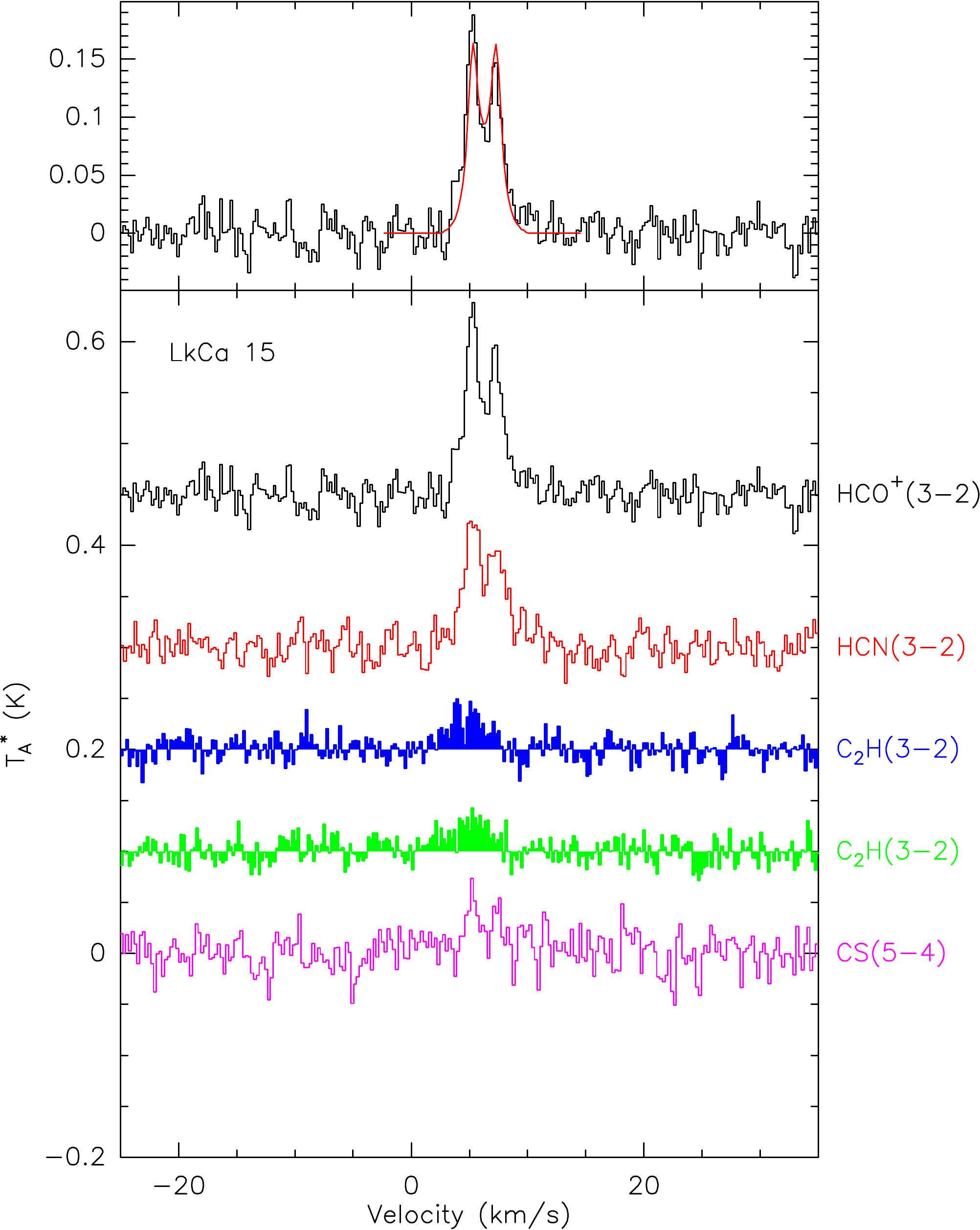}
\caption{Spectra of the observed transitions towards LkCa 15}
\label{fig:LKCA15}
\end{figure}
\begin{figure}
\includegraphics[height=11.5cm]{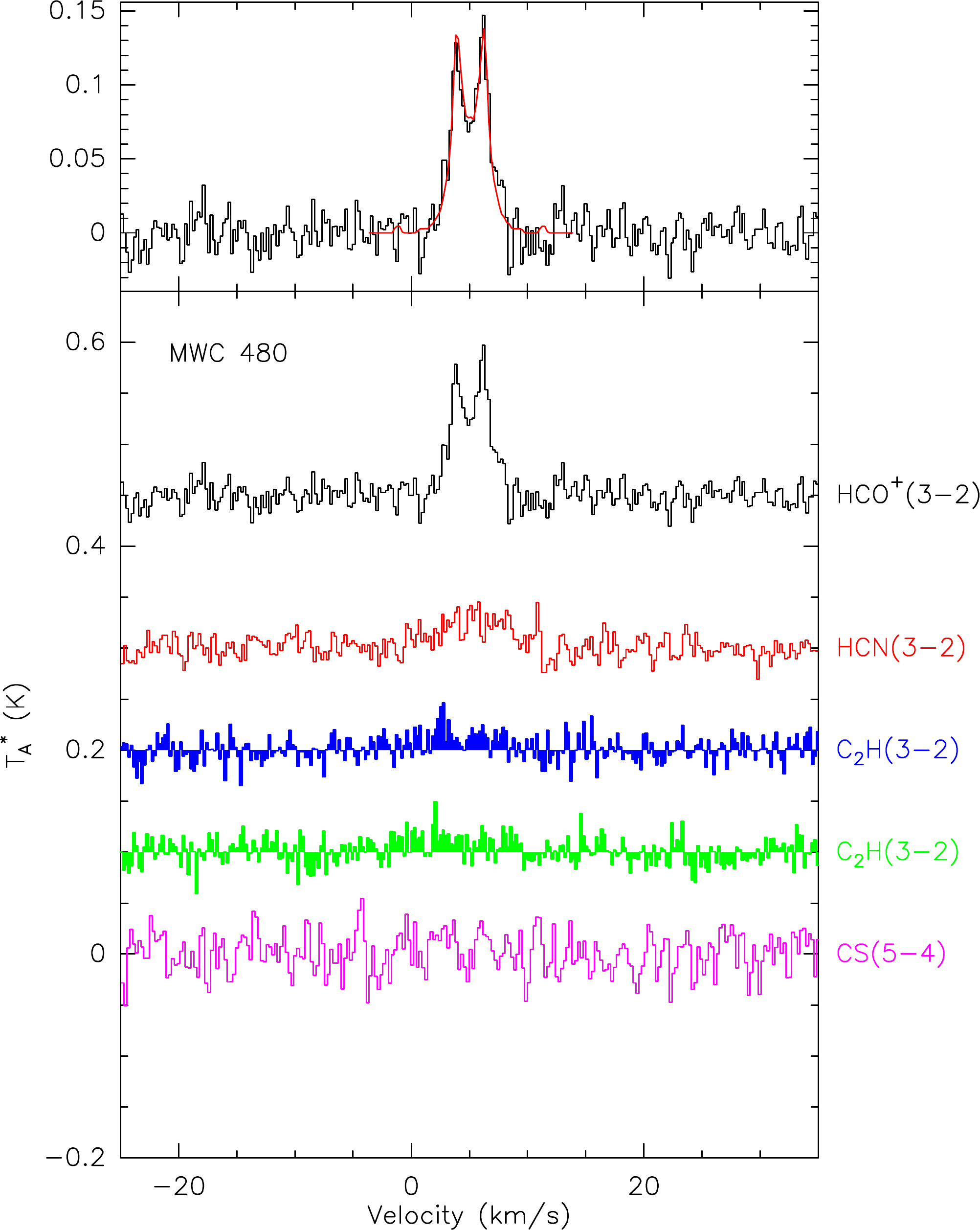}
\caption{Spectra of the observed transitions towards MWC 480}
\label{fig:MWC_480}
\end{figure}
\begin{figure}
\includegraphics[height=11.5cm]{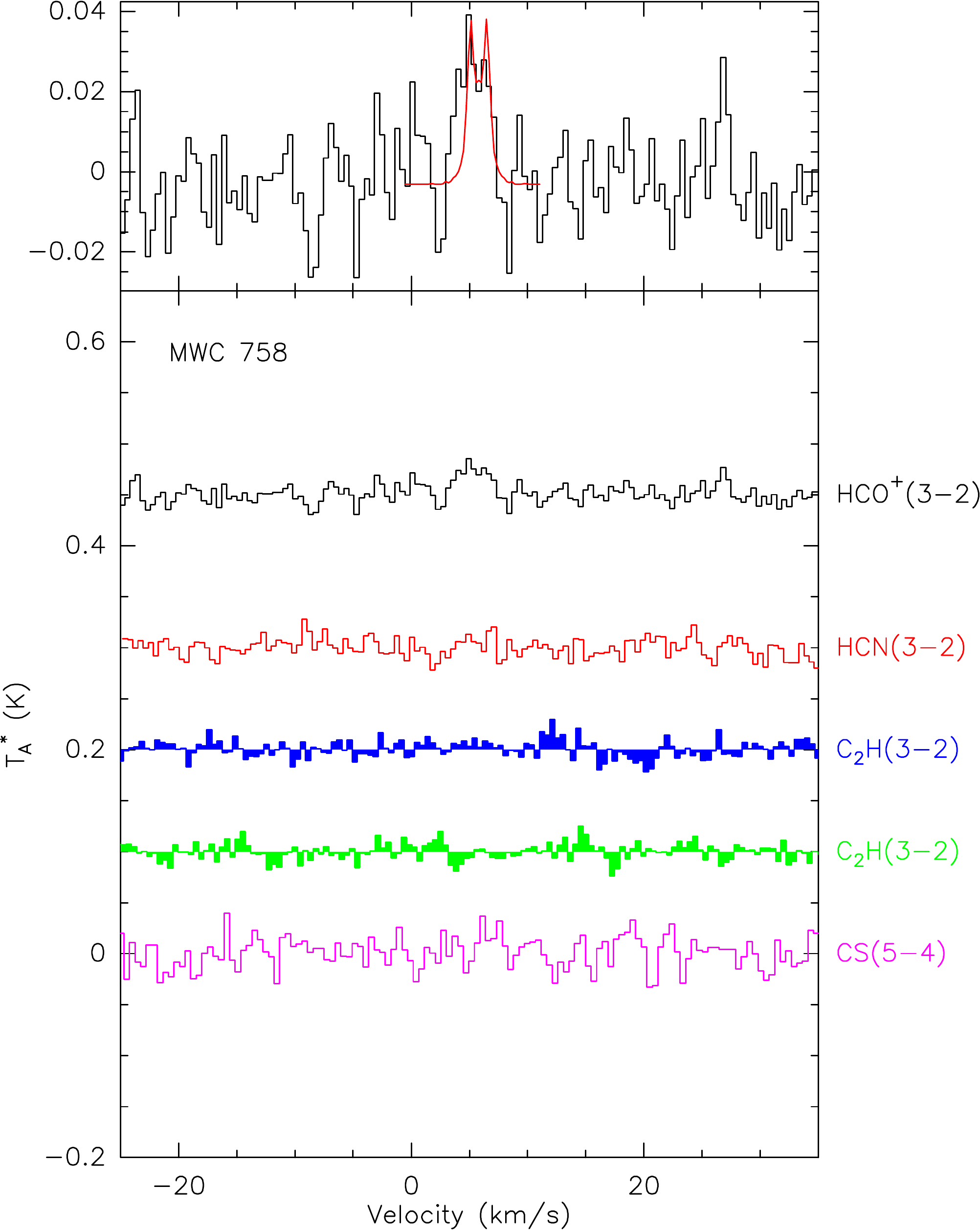}
\caption{Spectra of the observed transitions towards MWC 758}
\label{fig:MWC_758}
\end{figure}
\clearpage
\begin{figure}
\includegraphics[height=11.5cm]{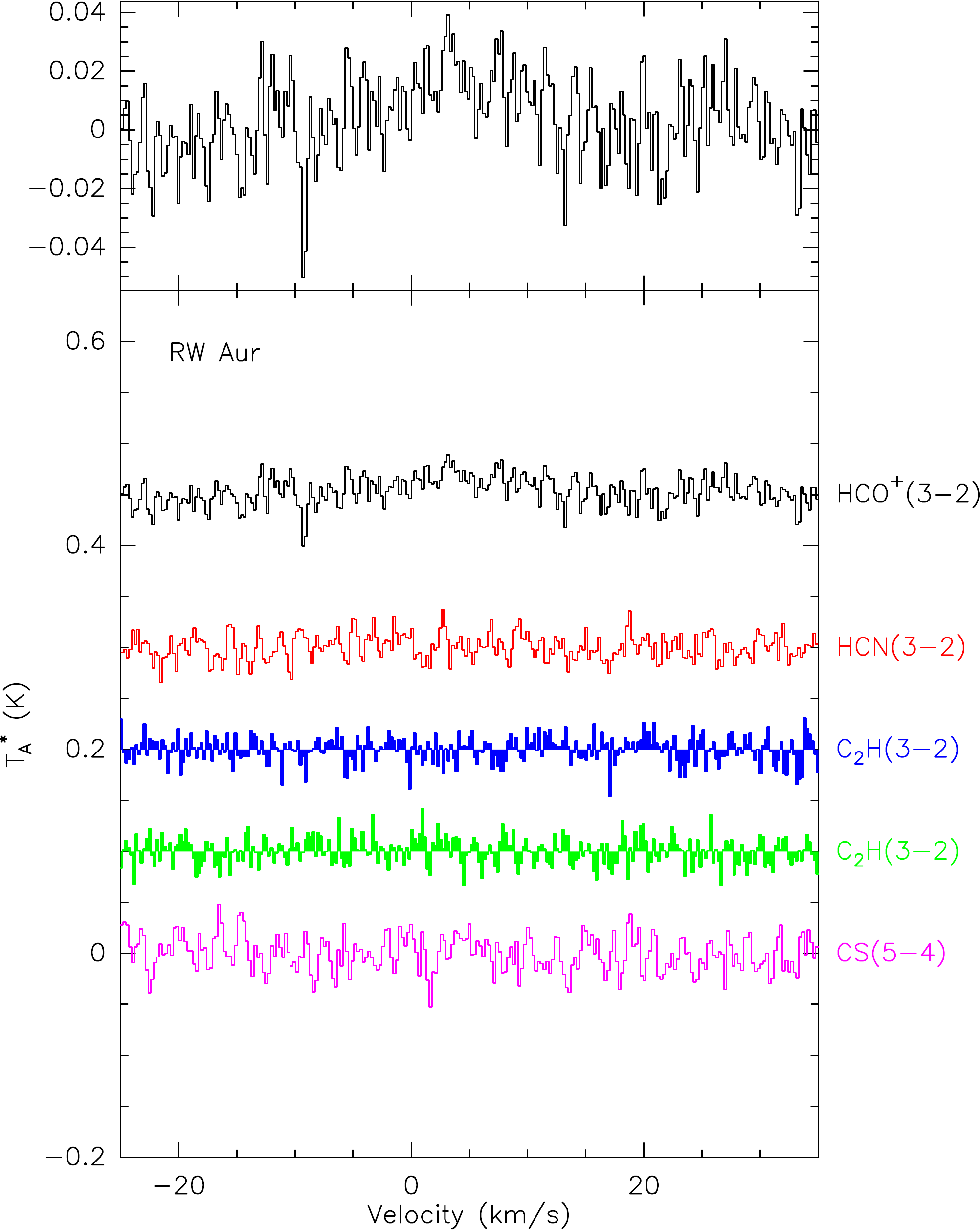}
\caption{Spectra of the observed transitions towards RW Aur}
\label{fig:RW_AUR}
\end{figure}
\begin{figure}
\includegraphics[height=11.5cm]{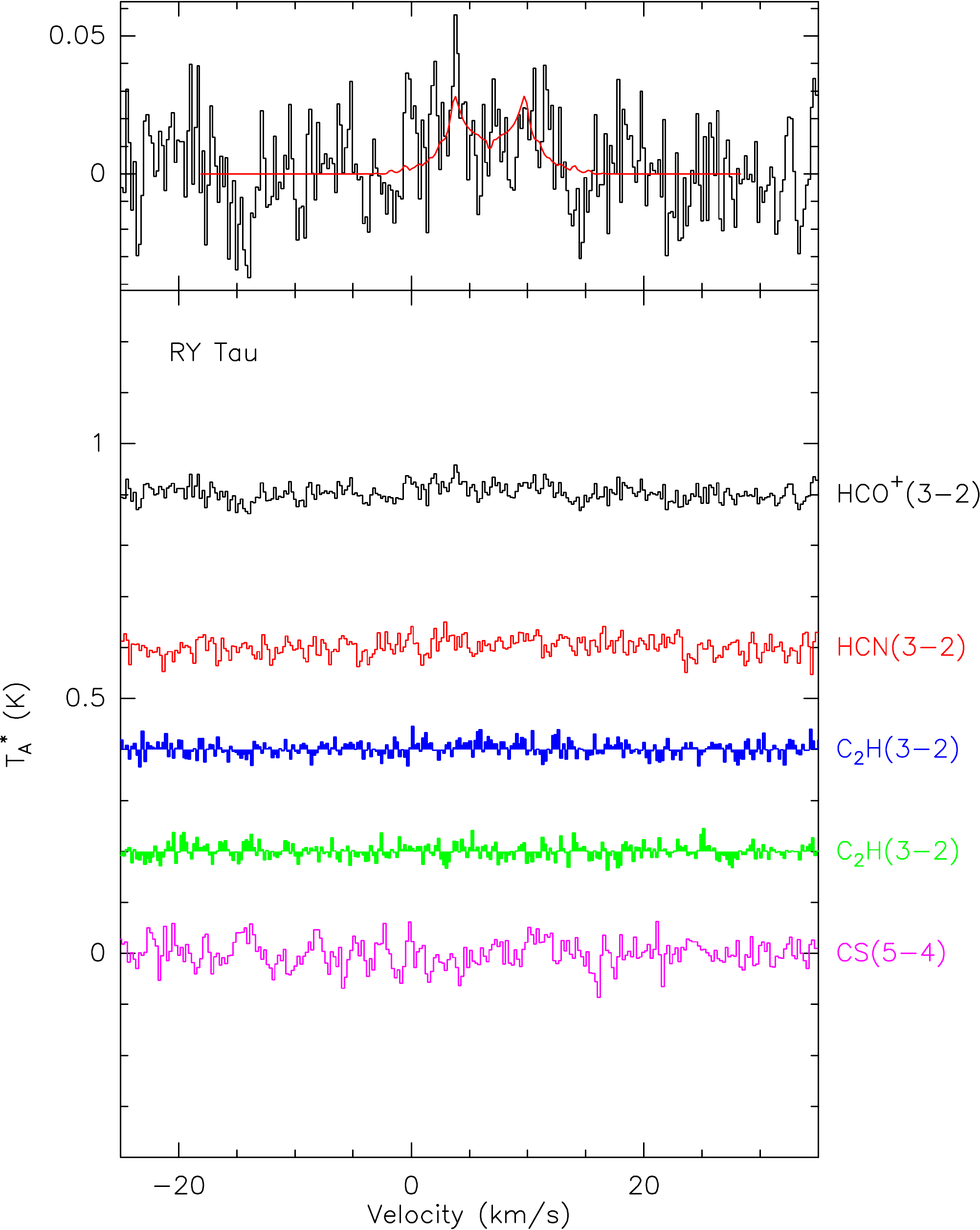}
\caption{Spectra of the observed transitions towards RY Tau}
\label{fig:RY_TAU}
\end{figure}
\begin{figure}
\includegraphics[height=11.5cm]{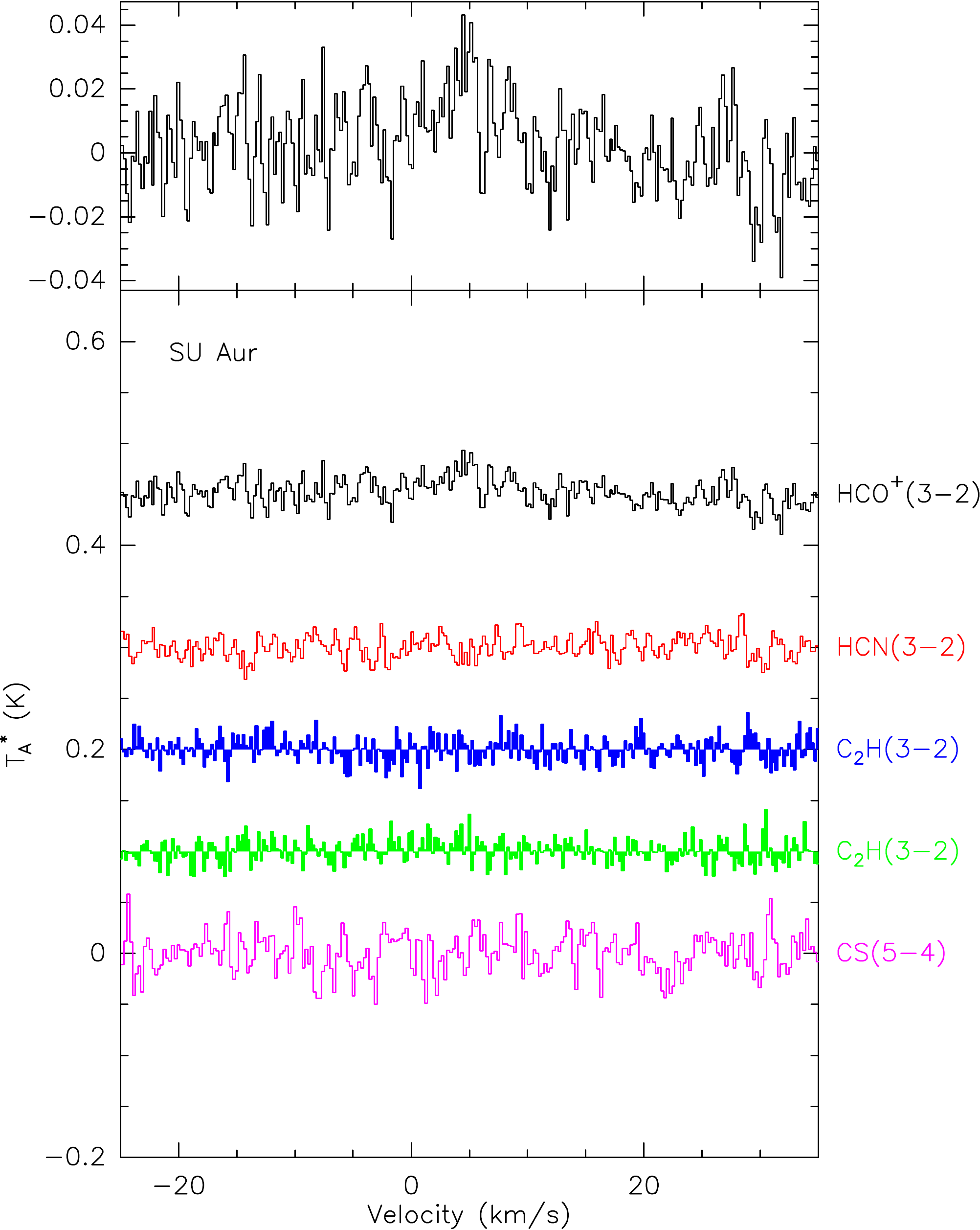}
\caption{Spectra of the observed transitions towards SU Aur}
\label{fig:SU_AUR}
\end{figure}
\begin{figure}
\includegraphics[height=11.5cm]{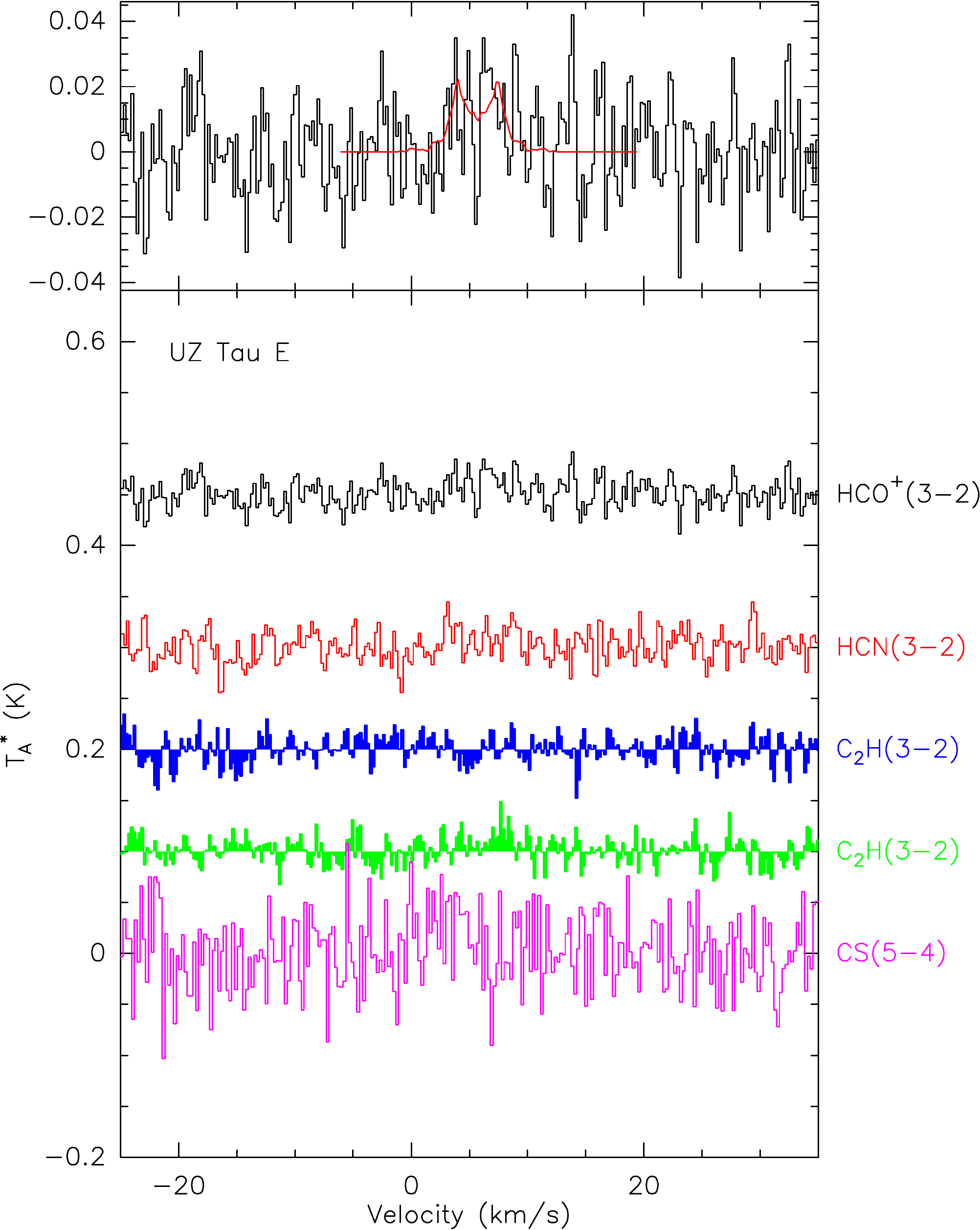}
\caption{Spectra of the observed transitions towards UZ Tau E}
\label{fig:UZTAU_E}
\end{figure}

\end{document}